\documentclass[aps,reprint,nofootinbib,nobibnotes,notitlepage,superscriptaddress,onecolumn,prd,
 amsmath,amssymb
]{revtex4-2}

\usepackage[caption=false]{subfig}
\usepackage{braket}
\usepackage{float}
\usepackage{lipsum}
\usepackage{graphicx}% include figure files
\usepackage{dcolumn}% align table columns on decimal point
\usepackage{bm}% bold math
\usepackage{natbib}
\usepackage{hyperref}
\hypersetup{
	colorlinks = true,
    linkcolor = Red,
    urlcolor  = Red,
    citecolor = Red
}

\usepackage{microtype}

\usepackage{verbatim}
\usepackage[amssymb]{SIunits}
\usepackage{tabularx}
\usepackage[dvipsnames]{xcolor}
\usepackage{wasysym}

\usepackage{tikz}
\usepackage{tikz-feynman}

\usepackage{color}
\definecolor{darkgreen}{RGB}{0,120,0}

\definecolor{darkgreen}{RGB}{0,120,0}

\newcommand{\delD}[1]{(2\pi)^3\delta_\mathrm{D}\left({#1}\right)}

\newcommand{\av}[1]{\left\langle{#1}\right\rangle} 

\newcommand{\vk}{\vec k}
\newcommand{\hk}{\hat{\vec k}}
\newcommand{\vK}{\vec K}

\newcommand{\C}{\mathsf{C}}
\newcommand{\Si}{\mathsf{S}^{-1}}
\newcommand{\F}{\mathcal{F}}

\newcommand{\A}{\mathsf{A}}
\newcommand{\Ai}{\mathsf{A}^{-1}}

\newcommand{\hr}{\hat{\vec r}}

\newcommand{\hK}{\hat{\vec K}}

\newcommand{\tj}[6]{\begin{pmatrix} {#1} & {#2} & {#3}\\ {#4} & {#5} & {#6}\end{pmatrix}}

\renewcommand{\vr}{\vec r}

\newcommand{\gnldotdot}{g_{\rm NL}^{\dot{\sigma}^4}}
\newcommand{\gnldotdel}{g_{\rm NL}^{\dot{\sigma}^2(\partial\sigma)^2}}
\newcommand{\gnldeldel}{g_{\rm NL}^{(\partial\sigma)^4}}
\newcommand{\fnl}{f_{\rm NL}^{\rm loc}}
\newcommand{\gnl}{g_{\rm NL}^{\rm loc}}
\newcommand{\taunl}{\tau_{\rm NL}^{\rm loc}}

\definecolor{darkgreen}{RGB}{0,120,0}
\newcommand{\resub}[1]{{#1}}%\color{darkgreen}{#1}}}

\DeclareSymbolFont{toneletters}{T1}{\familydefault}{m}{it}
\SetSymbolFont{toneletters}{bold}{T1}{\familydefault}{bx}{it}

\DeclareMathSymbol\edth{\mathord}{toneletters}{"F0}

\usepackage[skip=4pt plus1pt, indent=10pt]{parskip}

\usepackage{adjustbox}
\usepackage{dashbox}
\usepackage{xspace}

\def\beq{\begin{eqnarray}}
\def\eeq{\end{eqnarray}}

\let\vec\mathbf

\usepackage{listings}

\definecolor{dkgreen}{rgb}{0,0.6,0}
\definecolor{gray}{rgb}{0.5,0.5,0.5}
\definecolor{mauve}{rgb}{0.58,0,0.82}

\usepackage{empheq}

\def\mnras{\rm{MNRAS}}

\def\aap{\rm{A\&A}}

\def\jcap{\rm{JCAP}}

\defcitealias{Philcox4pt1}{Paper 1}
\defcitealias{Philcox4pt3}{Paper 3}

\newcommand{\paperone}{\citetalias{Philcox4pt1}\xspace}
\newcommand{\paperthree}{\citetalias{Philcox4pt3}\xspace}
\newcommand{\polyspec}{\textsc{PolySpec}\xspace}

\begin{document}

\setlength{\parskip}{0pt}

\title{\texorpdfstring{\Large Searching for Inflationary Physics with the CMB Trispectrum:\\
\large
2. Code \& Validation}{Searching for Inflationary Physics with the CMB Trispectrum: 2. Code \& Validation}
%Constraining Single-Field Inflation, Local Interactions, Spinning Massive Particle Exchange, Chiral Physics, Gravitational Lensing, and Beyond
}
%\date{\today}

\author{Oliver~H.\,E.~Philcox}
\email{ohep2@cantab.ac.uk}
\affiliation{Simons Society of Fellows, Simons Foundation, New York, NY 10010, USA}
\affiliation{Center for Theoretical Physics, Columbia University, New York, NY 10027, USA}
\affiliation{Department of Physics,
Stanford University, Stanford, CA 94305, USA}

\begin{abstract} 
    \noindent To unlock the vast potential of the CMB trispectrum, we require both robust estimators and efficient computational tools. In this work, we introduce the public code \href{https://github.com/oliverphilcox/PolySpec}{\polyspec}: a suite of quartic estimators designed to measure the amplitudes of a wide variety of inflationary templates, including local non-Gaussianity, effective field theory models, direction-dependent trispectra, spinning massive particle exchange, and weak gravitational lensing. \polyspec includes a \textsc{python}/\textsc{cython} implementation of each estimator derived in \paperone and has been carefully optimized to ensure efficient use of computational resources. We perform a broad range of validation tests, which demonstrate that the estimator is unbiased and minimum-variance, both in Gaussian and non-Gaussian regimes. In addition, we forecast constraints on various types of trispectra; this highlights the utility of CMB polarization and demonstrates that many models of primordial physics are poorly correlated with the simple templates considered in previous studies. This work lays the foundation for the \textit{Planck} trispectrum analyses performed in \paperthree.
\end{abstract}

\maketitle
\setlength{\parskip}{4pt}

\section{Introduction}
\noindent If new physics occurs in the early Universe, how can we detect it? One of the most attractive options is via `primordial non-Gaussianity' (PNG, \citep[e.g.,][]{Bartolo:2004if,Chen:2010xka,Komatsu:2003iq,Komatsu:2010hc}) -- distortions to the fiducial distribution of curvature perturbations induced by new particles, new interactions, or new vacuum states during inflation. As shown by various experiments including COBE, WMAP, and \textit{Planck} \citep[e.g.,][]{2006JCAP...05..004C,Senatore:2009gt,2014A&A...571A..24P,Planck:2015zfm,Planck:2019kim,WMAP:2003elm,Kunz:2001ym,Komatsu:2001wu}, the large-scale temperature and polarization fluctuations of the cosmic microwave background (CMB) appear consistent with Gaussian statistics, thus any deviations must be small. That said, they may still be non-zero, and there exist many non-trivial inflationary models capable of satisfying the current CMB bounds.

Whilst there is only one way for a distribution to be Gaussian, there are many ways for it to be non-Gaussian. The simplest signature of PNG is a non-vanishing skewness, or, in multiple dimensions, a non-trivial three-point function (or bispectrum). This statistic has been extensively studied in both theoretical and observational contexts, and has been used to place bounds on a wide variety of inflationary phenomena (see \citep{Chen:2010xka} for a review), such as additional fields (local PNG), self-interactions (equilateral and orthogonal PNG) and non Bunch-Davies vacua (folded PNG). Whilst the bispectrum is a very powerful tool, it is by no means the only statistic. Another option (that is the subject of this work) is the four-point function (or trispectrum), which is a higher-dimensional analog of the kurtosis. This describes the correlations between fluctuations at four points in space, and can be sourced by many effects such as primordial particle exchange, new particles uncoupled to the inflaton, chiral physics, gravitational lensing, and self-interactions protected by symmetries (as summarized in \citep{Philcox4pt1}). Of course, one can continue to higher-order; though the five- and six-point functions do not appear to contain much new information, some features such as rare particle production can only be probed by looking at the tails of the primordial curvature distribution, essentially probing $N$-point correlation functions at very high $N$ \citep[e.g.,][]{Flauger:2016idt,Munchmeyer:2019wlh,Kim:2021ida,Philcox:2024jpd}.

Despite its strong theoretical motivations, the inflationary trispectrum has received far less attention than the inflationary bispectrum. This is particularly true observationally: whilst more than 50 types of bispectra have been searched for in CMB data \citep[e.g.,][]{2014A&A...571A..24P,Planck:2015zfm,Planck:2019kim,Creminelli:2005hu,Senatore:2009gt,Fergusson:2010gn,2015arXiv150200635S,Marzouk:2022utf,2006JCAP...05..004C,Komatsu:2003iq,Liguori:2010hx,Babich:2004gb,Sohn:2024xzd,Cabass:2022wjy,Cabass:2022ymb,Cabass:2024wob,Feng:2015pva,Sekiguchi:2013hza,Fujita:2013qxa,Komatsu:2001wu,Smidt:2010ra,2003MNRAS.341..623S,Munchmeyer:2014nqa,Shiraishi:2012rm,Philcox:2023xxk,Philcox:2024wqx,Philcox:2023ypl,PhilcoxCMB,Philcox:2024jpd,Sohn:2023fte,Philcox:2023xxk,Philcox:2024wqx}, direct analyses of the four-point function have been restricted to just five models: two local shapes, one constant template \citep{Fergusson:2010gn}, and two Effective Field Theory of Inflation (EFT; \citep[e.g.,][]{Cheung:2007st,Senatore:2010wk}) interactions \citep{2015arXiv150200635S,Feng:2015pva,Fergusson:2010gn,Smidt:2010ra,2014A&A...571A..24P,Planck:2015zfm,Planck:2019kim}. Moreover, CMB polarization has been included in only one analysis \citep{Marzouk:2022utf}. The relative sparsity of trispectrum analyses is partly due to practical considerations: na\"ive computation of a four-point function from data has $\mathcal{O}(N_{\rm pix}^4)$ complexity, which is clearly infeasible given that current data have $N_{\rm pix}\approx 10^7$ pixels. As such, it is vital to develop efficient estimators for trispectrum non-Gaussianity. Whilst efficient estimators exist for computing the binned and  modal trispectrum \citep{Fergusson:2010gn,Regan:2010cn,Philcox:2023uwe,Philcox:2023psd} (which were used to place constraints on some primordial models in \citep{PhilcoxCMB,Philcox:2023ypl}), compressing the data in this fashion can be lossy and computationally inefficient \citep[cf.][]{Philcox:2024wqx}, thus it is more desirable to build direct estimators for the primordial amplitudes of interest. This can be highly non-trivial, thus previous works have considered only the simplest templates: local non-Gaussianity \citep{Kogo:2006kh,Munshi:2009wy,2011MNRAS.412.1993M,Izumi:2011di} and the EFTI shapes \citep{2015arXiv150200635S}.
% , and CMB lensing \citep[e.g.,][]{Okamoto:2003zw,Lewis:2006fu,Namikawa:2012pe,Maniyar:2021msb,Hanson:2010rp,Carron:2022edh}, all of which were included in the official \textit{Planck} analyses \citep{2014A&A...571A..24P,Planck:2015zfm,Planck:2019kim}.

In this triplet of works, our goal is to perform an in-depth analysis of the CMB trispectrum. This has multiple aims: (1) to connect recent developments in inflationary theory to observational data (in particular via the `cosmological collider' formalism \citep[e.g.,][]{Arkani-Hamed:2015bza,Lee:2016vti,Chen:2009zp,Flauger:2016idt}); (2) to define a suite of model trispectrum templates whose amplitudes can be constrained to probe a variety of physical processes; (3) to develop minimum-variance quartic estimators for each template, facilitating quasi-optimal parameter constraints; (4) to develop a fast and efficient public code to implement such estimators; (5) to forecast the detectability and distinguishability of our templates in realistic scenarios; (6) to constrain non-Gaussianity using the \textit{Planck} temperature and polarization dataset; (7) to bound physical models of inflation using these constraints. This builds on a number of previous works, in particular the wide variety of theoretical inflationary trispectrum studies (summarized in \paperone), as well as the optimal estimation procedure developed in \citep{2015arXiv150200635S} (building on \citep{2011MNRAS.417....2S,Komatsu:2003iq}). In \paperone, we tackled objectives (1), (2) and (3); here, we focus on (4) and (5), leaving (6) and (7) for \paperthree. 

This work presents the \href{https://github.com/oliverphilcox/PolySpec}{\polyspec} code: a fast and flexible package for performing quasi-optimal analyses of CMB temperature and polarization data. This heavily extends the \textsc{PolyBin} code presented in \citep{Philcox:2023uwe,Philcox:2023psd,PolyBin} (which estimates binned power spectra, bispectra and trispectra), and allows for minimum-variance estimation of eleven types of primordial trispectrum as well as two late-time effects. In particular, we build estimators for \resub{cubic and quadratic-squared} local non-Gaussianity (parametrized by $\gnl$ and $\taunl$), constant non-Gaussianity ($g_{\rm NL}^{\rm con}$), the three self-interaction EFTI templates ($\gnldotdot,\gnldotdel,\gnldeldel$), direction-dependent exchange non-Gaussianity assuming a parity-even ($\tau_{\rm NL}^{n,\rm even}$), parity-odd ($\tau_{\rm NL}^{n,\rm odd}$) or generalized ($\tau_{\rm NL}^{n_1n_3n}$) basis set, intermediate annd light spin-$s$ field exchange ($\tau_{\rm NL}^{\rm light}(s,\nu_s)$), heavy spin-$s$ field exchange ($\tau_{\rm NL}^{\rm heavy}(s,\mu_s)$), gravitational lensing $(A_{\rm lens}$) and unresolved point sources ($t_{\rm ps}$).\footnote{Precise template definitions can be found in \paperone and are summarized in the appendix of \paperthree; here, we note the rough form and features of each template when it is introduced in \S\ref{sec: validation}.} The \polyspec code is built in \textsc{python} for flexibility and makes extensive use of the \textsc{ducc} package for spherical harmonic transforms (which is an upgrade to \textsc{libsharp} \citep{Reinecke_2013}), with additional rate-limiting steps implemented in \textsc{c}. In addition to presenting the code, we perform an extensive battery of validation tests, affording us confidence that (a) the estimators are working as expected, (b) the various hyperparameters are well chosen, and (c) the code makes efficient use of computational resources.

The remainder of this work is as follows. In \S\ref{sec: estimators}, we give an overview of the trispectrum estimators introduced in \paperone, \resub{paying close attention} to their statistical properties and underlying approximations. Next, \S\ref{sec: code} presents the \polyspec code, including details of our numerical implementation and its basic usage. An extensive suite of forecasts and validation tests are performed in \S\ref{sec: validation}, assessing the bias and optimality of each class of estimator under various conditions, in addition to the dependence on a variety of hyperparameters. In \S\ref{sec: timings}, we assess the computational scalings and runtimes of \polyspec, before concluding in \S\ref{sec: summary} with a summary and a plan for the final installment of this series: application to the \textit{Planck} dataset. For reference, we summarize the hyperparameters used in the estimators in Tab.\,\ref{tab: hyperparameters} and demonstrate the basic usage of \polyspec in Fig.\,\ref{fig: sample-code}.

\section{Overview of the Optimal Estimators}\label{sec: estimators}

\subsection{Definition \& Properties}\label{subsec: estimator-def}
\noindent We begin with a brief summary of the trispectrum estimators discussed in \paperone, \resub{highlighting} the various numerical approximations that will be tested in this work. Our starting point is the four-point function of the primordial curvature perturbation, $\zeta$, expressed in terms of a set of amplitudes $\{A_\alpha\}$ and corresponding templates $T^{(\alpha)}_\zeta$:
\beq
    \av{\zeta(\vk_1)\zeta(\vk_2)\zeta(\vk_3)\zeta(\vk_4)}_c = \delD{\vk_1+\vk_2+\vk_3+\vk_4}\sum_\alpha A_\alpha\,T^{(\alpha)}_\zeta(\vk_1,\vk_2,\vk_3,\vk_4).
\eeq
We identify two classes of templates: (1) \textbf{contact} templates, which can be separated into factors depending only on a single $\vk_i$; (2) \textbf{exchange} templates, coupled by the intermediate momentum $\vK = \vk_1+\vk_2$ (or permutations). Roughly speaking, these correspond to self-interactions and exchange processes during inflation. 

The goal of the optimal estimator program is to obtain a measurement for the amplitudes $\{A_\alpha\}$ from a given dataset. Here, we work with the CMB data, denoted $d^i$, where the abstract index $i$ encodes either pixels and spins or harmonic coefficients and polarizations. The data can be related to an underlying CMB map via
\beq
    d^i = [\mathsf{P}a]^i+n^i,
\eeq
where $\mathsf{P}$ is the pointing matrix and $n$ is some noise field. When working with masked data, we will assume that $\mathsf{P}$ is the composition of a harmonic-space beam ($\mathsf{B}$), spherical harmonic synthesis ($\mathsf{Y}$) and a pixel-space mask ($\mathsf{W}$), \textit{i.e.}\ that $\mathsf{P} = \mathsf{W}\mathsf{Y}\mathsf{B}$. For unmasked data, we may use simply $\mathsf{P} = \mathsf{B}$, with all fields defined in harmonic-space. As shown in \paperone (and many other works, including \citep{Philcox:2024rqr,Philcox:2023uwe,Philcox:2023psd,2015arXiv150200635S,Regan:2010cn}) this
%, coupled with the known relation between $a^i$ and the primordial curvature, 
leads to the following estimator for $A_\alpha$ (assuming Einstein summation):
\beq\label{eq: general-estimator}
    \widehat{A}_\alpha &=& \sum_\beta \F^{-1}_{\alpha\beta}\widehat{N}^\beta\\\nonumber
    \widehat{N}_\alpha &\equiv& \frac{1}{4!}\frac{\partial\av{a^{i_1}a^{i_2}a^{i_3}a^{i_4}}_c}{\partial A_\alpha}\left[h_{i_1}h_{i_2}h_{i_3}h_{i_4}-6\,h_{i_1}h_{i_2}\av{h_{i_3}h_{i_4}}+3\av{h_{i_1}h_{i_2}}\av{h_{i_3}h_{i_4}}\right]^*\\\nonumber
    \mathcal{F}_{\alpha\beta} &\equiv& \frac{1}{4!}\left[\left(\frac{\partial\av{a^{i_1}a^{i_2}a^{i_3}a^{i_4}}_c}{\partial A_\alpha}\right)^*[\Si\mathsf{P}]_{i_1j_1}[\Si\mathsf{P}]_{i_2j_2}[\Si\mathsf{P}]_{i_3j_3}[\Si\mathsf{P}]_{i_4j_4}\frac{\partial\av{a^{j_1}a^{j_2}a^{j_3}a^{j_4}}_c}{\partial A_\beta}\right]^*,
\eeq
where $\F$ is known as the Fisher matrix, and $h[d] \equiv \Si d$ is a filtered version of the observed data, applying a linear weighting operator $\Si$. This is a quartic estimator with a variety of useful properties (which will be tested below):
\begin{itemize}
    \item \textbf{Non-Gaussian}: Under the null hypothesis (with $A_\alpha = 0$), the estimator returns zero. This is guaranteed by the expectation terms in \eqref{eq: general-estimator}, which arise from a Hermite expansion of the underlying likelihood, and include the mean-field and realization-dependent noise terms typically included in lensing analyses \citep[e.g.,][]{Carron:2022edh}. If $\av{hh}$ is replaced by an approximate covariance (\textit{i.e.}\ from an unknown true cosmology), the bias arises only at second order.
    \item \textbf{Unbiased}: The estimator recovers signals with arbitrary $A_\alpha$, \textit{i.e.}\ $\mathbb{E}[\widehat{A}_\alpha] = A_\alpha$ (up to higher-order effects). Practically, it is unbiased for all choices of weighting schemes, beams, and masks, as well as correlations between templates (as long as they are jointly analyzed).
    \item \textbf{Minimum Variance}: For a suitable choice of weighting scheme, \eqref{eq: general-estimator} is the lowest-variance unbiased estimator for $A_\alpha$ in the Gaussian limit ($A_\alpha \to 0$). Formally, this requires $\Si = \mathsf{P}^\dag \C^{-1}$, where $\C_{ij} = \av{d_id_j^*}$ is the covariance between pixels and spins in the dataset, \textit{i.e.}\ $\Si d$ is the beam-deconvolved inverse-variance-filtered data. This applies also to cut-sky data (as discussed in \paperone).
    \item \textbf{Known Covariance}: Assuming the $\Si$ weighting and working under null assumptions, the covariance of the estimator is given by $\mathrm{cov}\left(\widehat{A}_\alpha,\widehat{A}_\beta\right) = \F_{\alpha\beta}^{-1}$. Noting that $\F$ is the second derivative of the log-likelihood (\textit{i.e.}\ the Fisher matrix), this implies that the estimator saturates the Cram\'er-Rao bound, and is hence optimal. We caution that this does not imply that the distribution of $\widehat{A}_\alpha$ is Gaussian; as discussed in \citep{Smith:2012ty,2014A&A...571A..24P,Marzouk:2022utf}, the sampling distribution of $\widehat{\tau}_{\rm NL}^{\rm loc}$ can be highly skewed, for example.
\end{itemize}

\begin{table}[!t]
    \centering
    \begin{tabular}{c|l}
      \textbf{Parameter}  & \textbf{Description}\\\hline
      $\Si$ & Linear operator used to weight the data. This may include beam-deconvolution and inverse-variance weighting.\\
      $N_{\rm disc}$ & Number of simulations used to remove Gaussian contributions from the estimator numerator.\\
      $N_{\rm fish}$ & Number of Gaussian random fields used to estimate the Fisher matrix, $\F_{\alpha\beta}$. \\
      $N_{\rm opt}$ & Number of radial integration points in the estimators, after optimization (cf.\,\ref{eq: contact-optim}\,\&\,\ref{eq: exchange-optim}).\\
      $N_{\rm fish}^{\rm opt}$ & Number of Gaussian random fields used to estimate $\F$ when optimizing exchange trispectra.\\
      $f_{\rm thresh}$ & Target accuracy for the optimization algorithm.\\
      $\ell_{\rm min},\ell_{\rm max}$ & External $\ell$-range included in the analysis. We set the primordial template to zero outside these ranges. \\
      $L_{\rm min}, L_{\rm max}$ & Internal $L$-range included in the analysis. We set the primordial template to zero outside these ranges.\\
      $N_k$ & Number of points in $k$-space used to compute the transfer function integrals.\\
      $N_{\rm side}$ & \textsc{healpix} $N_{\rm side}$ parameter specifying the dimension of the pixelation grid.\\
      $k_{\rm coll}$ & Truncation scale used to restrict the collider estimators to the collapsed limit.\\
      $N_{\rm CPU}$ & Number of CPU threads used to run \polyspec.
    \end{tabular}
    \caption{Key hyperparameters used in the trispectrum estimators of this work and the \polyspec code. These control a variety of aspects of the estimator, such as weighting, scale cuts and Monte Carlo convergence. The impact of each parameter is assessed in \S\ref{sec: code}\,\&\,\ref{sec: validation}.}
    \label{tab: hyperparameters}
\end{table}

\subsection{Separable Implementation}\label{subsec: sep-impl}
\subsubsection{Monte Carlo Summation}

\noindent To apply \eqref{eq: general-estimator} to data, it must be heavily simplified. A na\"ive implementation scales as $\mathcal{O}(N_{\rm pix}^4)$ for a total of $N_{\rm pix}$ pixels and spins; given that $N_{\rm pix}\sim 10^7-10^8$, this is clearly infeasible. As discussed in \paperone (building on \citep{2015arXiv150200635S,Philcox:2023uwe,Philcox:2023psd,Philcox:2024rqr}), it can be simplified using a number of computational tricks. For the numerator, one can replace the expectations $\av{hh}$ with Monte Carlo averages:
\beq\label{eq: num-mc}
    \av{h_{i_1}h_{i_2}} &\to& \frac{1}{N_{\rm disc}}\sum_{n=1}^{N_{\rm disc}}h_{i_1}[\delta^{(n)}]h_{i_2}[\delta^{(n)}]\\\nonumber
    \quad\av{h_{i_1}h_{i_2}}\av{h_{i_3}h_{i_4}} &\to& \frac{2}{N_{\rm disc}}\sum_{n=1}^{N_{\rm disc}/2}h_{i_1}[\delta^{(n)}]h_{i_2}[\delta^{(n)}]h_{i_3}[\delta^{(N_{\rm disc}-n)}]h_{i_4}[\delta^{(N_{\rm disc}-n)}]
\eeq
where $\{\delta^{(n)}\}$ are some set of $N_{\rm disc}$ independent and identically distributed simulations whose covariance is assumed to match that of the data (\textit{i.e.}\ we assume $\av{\delta^{(n)}\delta^{(n')\dag}} = \delta^{\rm K}_{nn'}\av{dd^\dag}$).\footnote{In the language of CMB lensing estimators, this includes the `realization-dependent bias' term and the $\mathrm{N}^{(0)}$ contribution \citep[e.g.,][]{Namikawa:2012pe,Carron:2022edh}.} Formally, this gives an additional contribution to the estimator variance scaling as $1/N_{\rm disc}$. If the covariance of the simulations does not match that of the data, the estimator will be biased; for this reason, it is advisable to use realistic simulations (though we note that accurate higher-order statistics are not required). 

Next, the Fisher matrix $\F_{\alpha\beta}$ can be simplified using tricks derived from the mathematical field of stochastic trace estimation \citep{girard89,hutchinson90,2015arXiv150200635S,Oh:1998sr}. Inserting a set of $N_{\rm fish}$ random fields $\{a^{(n)}\}$ with known covariance $\A = \av{aa^\dag}$, this can be rewritten in terms of a second Monte Carlo sum:
\beq\label{eq: fisher-mc}
    \F_{\alpha\beta} \to \frac{1}{4!}\frac{1}{N_{\rm fish}}\sum_{n=1}^{N_{\rm fish}}Q^{i_1}_\alpha[\Si\mathsf{P} a^{(n)}]\,\times\,[\Si \mathsf{P}]_{i_1j_1}\,\times\,Q^{j_1}_\beta[\Ai a^{(n)}],
\eeq
where $Q^{i_1}_{\alpha}[x] = \partial_{A_\alpha}\av{a^{i_1}a^{i_2}a^{i_3}a^{i_4}}_cx^*_{i_2}y^*_{i_3}z^*_{i_4}$.\footnote{In practice, we utilize a modified version of \eqref{eq: fisher-mc} to ensure optimal use of Monte Carlo simulations; this is discussed in \citep{Philcox4pt1,2015arXiv150200635S}.} This is the outer product of two vectors, and can be efficiently computed without forming the full matrix. The underlying simulations do not have to closely represent the data. We simply require their covariance to be exactly known and invertible -- a good choice is unmasked Gaussian random fields generated with the beam-deconvolved power spectra.
%\footnote{For fast convergence, $\mathsf{A}^{-1}$ should approximate $\Si\mathsf{P}\approx \mathsf{P}\mathsf{C}^{-1}\mathsf{P}^\dag$.} 
This procedure does not induce bias in $\F$, but leads to a stochastic error scaling as $1/N_{\rm fish}$ (which can induce bias in $\F^{-1}$).

\subsubsection{Factorization \& Optimization}
\noindent We must also simplify the template derivatives, $\partial_{A_\alpha}\av{a^{i_1}a^{i_2}a^{i_3}a^{i_4}}$, which appear in the numerator and the Fisher matrix. Since $\av{a^{i_1}a^{i_2}a^{i_3}a^{i_4}}$ is proportional to the four-point function of $\zeta$, this is equivalent to requiring a simple representation of the primordial correlator. As discussed extensively in \paperone, all the templates considered in this work can be written in either contact- or exchange-factorizable form. For contact templates
\beq
\av{\zeta(\vk_1)\zeta(\vk_2)\zeta(\vk_3)\zeta(\vk_4)}_c = \delD{\vk_1+\vk_2+\vk_3+\vk_4}t^{(1)}(\vk_1)t^{(2)}(\vk_2)t^{(3)}(\vk_3)t^{(4)}(\vk_4)+\text{23 perms.},
\eeq
where we split the trispectrum into independent functions of a single $\vk_i$ and symmetrize over permutations of $\{\vk_1,\vk_2,\vk_3,\vk_4\}$. For the EFTI templates, this is realized as an integral over conformal time $\tau$; the resulting expression is equivalent but with and extra factor of $\int_{-\infty}^0 d\tau$. Additionally writing the Dirac delta as an integral gives
\beq
    \av{\zeta(\vk_1)\zeta(\vk_2)\zeta(\vk_3)\zeta(\vk_4)}_c = \int_0^\infty r^2dr\,\int d\hr\,\prod_{i=1}^4\left[e^{i\vk_i\cdot\vr}t^{(i)}(\vk_i)\right]+\text{23 perms.},
\eeq
which is explicitly integral-separable. Transforming to harmonic space, inserting \eqref{eq: general-estimator} and performing the various $\vk$ integrals and $\ell,m$ sums, the quartic term of the general estimator takes the form
\beq\label{eq: contact-optim}
    \widehat{N}_{\rm contact} \sim \int_0^\infty r^2dr\,\int d\hr\,f^{(1)}[d](\hr,r)f^{(2)}[d](\hr,r)f^{(3)}[d](\hr,r)f^{(4)}[d](\hr,r).
\eeq
This depends on a set of scalar functions $f^{(i)}$, which are linear in the data. Schematically, 
\beq\label{eq: g-r-def}
    f^{(i)}[x](\hr,r) \sim \sum_{\ell m}g^{(i)}_\ell(r)[\Si d]_{\ell m}Y_{\ell m}(\hr),
\eeq
where the harmonic-space weights $g_\ell^{(i)}(r)$ are integrals over the CMB transfer functions whose form can be found in \paperone. The key message is that the estimator can be computed as a sum over pixels (\textit{i.e.}\ $\int d\hr$) and a numerical integral over $r$ (and possibly $\tau$). All operations can be performed using spherical harmonic transforms (hereafter SHTs) and direct summation, leading to an $\mathcal{O}(N_{\rm pix}\log N_{\rm pix})$ scaling.

For exchange-factorizable templates, we start from the definition
\beq
    \av{\zeta(\vk_1)\zeta(\vk_2)\zeta(\vk_3)\zeta(\vk_4)}_c &=& \int_{\vK}\delD{\vk_1+\vk_2-\vK}\delD{\vk_3+\vk_4+\vK}\\\nonumber
    &&\,\times\,\left[t^{(1)}(\vk_1)t^{(2)}(\vk_2)t^{(3)}(\vk_3)t^{(4)}(\vk_4)t^{(5)}(\vK)+\text{23 perms.}\right],
\eeq
introducing the exchange momentum $\vK$. In this case, the estimator involves two radial integrals (corresponding to the two Dirac delta functions); after some simplifications, we find the schematic form
\beq\label{eq: exchange-optim}
    \widehat{N}_{\rm exchange} &\sim& \sum_{L}\int_0^\infty r^2dr\,\int_0^\infty r'^2dr'\,F_{L}(r,r')\\\nonumber
    &&\,\times\,\left(\int d\hr\,f^{(1)}[d](\hr,r)f^{(2)}[d](\hr,r)Y_{LM}(\hr)\right)\left(\int d\hr'\,f^{(3)}[d](\hr',r')f^{(4)}[d](\hr',r')Y_{LM}(\hr')\right)^*,
\eeq
which involves two quadratic terms that are combined in harmonic-space via some $F_{L}(r,r')$ coupling. This can be efficiently estimated in $\mathcal{O}(N_{\rm pix}\log N_{\rm pix})$ time using SHTs and a two-dimensional radial integral.

Computation of the above estimators is limited by the numerical integrals, since the integrand must be computed using SHTs. To ameliorate this, one can use optimization algorithms to define a low-dimensional approximation to the integral that retains accuracy whilst reducing computation time \citep{2011MNRAS.417....2S,Philcox4pt1}. This first rewrites the radial integrals as finely sampled discrete sums, e.g., 
\beq
    \int_0^\infty r^2dr \to \sum_{i=1}^{N_s} r_i^2\delta r_i, \qquad\qquad  \int_0^\infty r^2dr\int_{-\infty}^0d\tau\to\sum_{i=1}^{N_s} r_i^2\delta r_i\delta\tau_i
\eeq
given some initial grid of $N_s\gg1$ sampling points. We then search for an accurate approximation to this sum (or more specifically, the high-dimensional Fisher matrix) using a length-$N_{\rm opt}\ll N_s$ subset of the sampling points, in addition to a set of weights. The optimized representation depends on both the template and analysis settings (e.g., fiducial cosmology and scale-cuts), but can be efficiently computed by minimizing an appropriate convex loss, asserting convergence when the true and approximated Fisher matrices agree within $f_{\rm thresh}$ (usually set to $10^{-3}$ or $10^{-4}$). As shown in \paperone, this induces a multiplicative error $\sim \sqrt{f_{\rm thresh}}$ in $\F$, and can be efficiently implemented for both contact and exchange trispectra (though is more efficient for the former). In this work, we include an optimization step in all analyses: as verified in \S\ref{subsubsec: optim-test} (and throughout \S\ref{sec: validation}), this yields accurate estimators using just $N_{\rm opt}\approx 10-100$ sampling points. 

\section{The PolySpec Code}\label{sec: code}
\noindent In this work, we introduce the \polyspec code:\footnote{Publicly available on GitHub: \href{https://github.com/oliverphilcox/PolySpec}{GitHub.com/OliverPhilcox/PolySpec}.} a fast and flexible implementation of the above trispectrum estimators in \textsc{python} and \textsc{cython}. The code is built around the binned power spectrum, bispectrum, and trispectrum estimator \textsc{PolyBin} \citep{Philcox:2023uwe,Philcox:2023psd} (which is subsumed into \polyspec), and makes extensive use of the \textsc{healpix} and \textsc{healpy} packages  \citep{Gorski:2004by,Zonca:2019vzt}, as well as \textsc{ducc} for fast SHTs (itself based on \textsc{libsharp} \citep{Reinecke_2013}).\footnote{The SHTs could be expedited using a GPU-accelerated code \citep[e.g.,][]{Belkner:2024vor}, the resulting algorithm will likely be limited by memory and the speed of transfers onto and off the GPU.} Though many experiments are shifting from \textsc{healpix} to more modern pixelation schemes, \polyspec works with \textsc{healpix} for consistency with the full-sky \textit{Planck} dataset. This choice could be fairly easily altered in the future.

\begin{figure}
\centering
\noindent\begin{minipage}{.9\textwidth}
\begin{lstlisting}{Name}
import polyspec as ps, numpy as np

# Load base class
base = ps.PolySpec(Nside, fiducial_Cl, beam, backend="ducc")

# Load the trispectrum template class, specifying the templates to analyze
tspec = ps.TSpecTemplate(base, smooth_mask, applySinv, ["gNL-loc","tauNL-direc:1,1,0"], 
                            lmin, lmax, k_array, transfer_array, Lmin, Lmax)

# Perform optimization to compute the radial integration points
tspec.optimize_radial_sampling_1d()

# Compute the Fisher matrix as a Monte Carlo sum
fish = np.mean([tspec.compute_fisher_contribution(seed) for seed in range(Nfish)],axis=0)

# Compute the trispectrum estimator
tspec.generate_sims(Ndisc)
estimate = np.linalg.inv(fish)@tspec.Tl_numerator(data)

# Print diagnostics
tspec.report_timings()
\end{lstlisting}
\end{minipage}
\caption{Sample \textsc{python} code demonstrating the \polyspec package. Here, we jointly estimate the $\gnl$ and $\tau_{\rm NL}^{110}$ amplitudes from an observed map `data`, using Gaussian random fields to remove the disconnected contributions. Detailed code tutorials are available on GitHub, which include extensive discussion of the various code inputs and outputs.}
\label{fig: sample-code}
\end{figure}

\subsection{Outline}
\noindent \polyspec is comprised of four main segments: precomputation, optimization, estimation of the Fisher matrix, and computation of the trispectrum numerator. As shown in Fig.\,\ref{fig: sample-code}, these can be invoked using \textsc{python}, and depend on a number of inputs, such as the fiducial power spectrum, the beam, the mask and a suite of simulations (for the disconnected terms). Below, we summarize each part of the code (which are not necessarily run in order):
\begin{itemize}
    \item \textbf{Precomputation}: We compute all required $k$-space integrals, including the $\ell$-space weights, $g_\ell(r)$, used in \eqref{eq: g-r-def} and the coupling matrices, $F_L(r,r')$, entering the exchange trispectrum estimators \eqref{eq: exchange-optim}. These are obtained using numerical quadrature, given grids of $k$ and $r$ (or $r,\tau$) and the numerical transfer functions computed using \textsc{camb} or \textsc{class} \citep{Lewis:1999bs,Blas:2011rf}. The integrands involve $j_\ell(kr)$ Bessel functions and their derivatives; to compute these, we adopt a similar procedure to \citep{2015arXiv150200635S}, first precomputing an array of $j_\ell(x)$ values (with $\Delta x = 0.01$ for $x<2\ell_{\rm max}$ and $0.1$ else), then interpolating the result to the two-dimensional grid of $k$ and $r$. We employ Steed's algorithm \citep{BARNETT1974377} to recursively compute all Bessel functions up to order $L$, such that $|j_L(x)|>10^{-150}$ (avoiding underflow error), setting values outside this regime to zero. For $x\gg L$, we replace the Bessel functions by their large-argument limits for efficiency. These operations are all performed in \textsc{cython} making use of \textsc{gsl}.
    \item \textbf{Optimization}: We compute the optimized radial integration grids and corresponding weights using the algorithms described in \paperone. This has the following steps:
        \begin{enumerate}
            \item Generate some initial grid of $r$ (and, if EFTI templates are used, $\tau$) points, and run the precomputation step. We adopt an initial grid similar to \citep{2015arXiv150200635S}, preferentially sampling regions where the integrand peaks.\footnote{Explicitly, we use a linear grid with spacing $\Delta r = 50$ for $r\in [1,0.95r_\star]$ and $r\in[1.05r_{\rm hor},r_{\rm hor}+5000]$, and $\Delta r = 5$ for $r\in [0.95r_\star,1.05r_{\rm hor}]$, where $r_\star$ ($r_{\rm hor}$) is the distance to last scattering (the horizon) and we work in $\mathrm{Mpc}$ units. For EFTI templates, we use a logarithmically-spaced grid of $-\tau\in[10/\ell_{\rm max},10^6]$, and set the radial spacings to the maximum of the above and $-0.1\tau$, with $r_{\rm max} = r_{\rm hor}+5000-5\tau$. Minor variation in these choices gives negligible changes to our results.}
            \item Compute the contributions to the Fisher matrix from each integraiton point, \textit{i.e.}\ $\F(r,r')$, working under idealized conditions of a unit mask and translation-invariant noise. For contact trispectra, this is performed analytically using Gauss-Legendre quadrature with $(2\ell_{\rm max}+1)$ points; for exchange trispectra, we use the Monte Carlo algorithm described below using $N_{\rm fish}^{\rm opt}\gtrsim 1$ Monte Carlo iterations (as described below).
            \item Generate a low-dimensional set of $N_{\rm opt}$ radial integration points and corresponding weights: $\{r_i,w_i\}$ (or $\{r_i,\tau_i,w_i\}$), using the greedy algorithm presented in \paperone, stopping when the Fisher matrix converges to $f_{\rm thresh}$. For exchange trispectra, the optimization algorithm is not guaranteed to converge;\footnote{This usually occurs only for templates that cannot be meaningfully constrained from the CMB, as discussed in \S\ref{subsec: valid-direc}.} in this case, the algorithm exits if the loss function does not (significantly) improve after a specified number of iterations.
            \item Rerun the precomputation step with the optimized set of basis points.
        \end{enumerate}
    \item \textbf{Fisher Matrix}: The Fisher matrix is computed as a Monte Carlo average over $N_{\rm fish}$ (unmasked) Gaussian realizations, as in \eqref{eq: fisher-mc}. For each choice of random seed, we do the following:
        \begin{enumerate}
            \item Draw two Gaussian random fields, $a^{(k)},a^{(k')}$, and apply the $\mathsf{S}^{-1}\mathsf{P}$ and $\Ai$ filters. A typical $\Si$ involves inpainting small holes in the mask, transforming to harmonic-space and multiplying by the inverse power spectrum, $C_{\ell,\rm tot}^{-1,XY}$ for fields $X,Y$ (which includes the beam and noise), then multiplying by the beam (to yield deconvolved spectra). After applying the filters, we drop all modes with $\ell>\ell_{\rm max}$, which leads to significant expedition.
            \item Compute the required one-dimensional maps $f[a](\hr,r_i)$ for each radial integration point, $r_i$. Each template requires a small number of maps (typically between one and five, see Tab.\,1 of \paperone), and can be computed using chained harmonic transforms and harmonic-space products as in \eqref{eq: g-r-def}.
            \item Construct the $Q_{\alpha}$ derivatives for each template of interest using both random fields. This requires multiplication in map- and/or harmonic-space as well as (possibly spin-weighted) SHTs. 
            \item Using SHTs, apply the $\Si \mathsf{P}$ weighting to $Q_\alpha$ following \eqref{eq: fisher-mc}, and compute the Fisher matrix estimate as an outer product of $\{Q_\alpha\}$, summing over modes in harmonic-space.
            \end{enumerate}
    \item \textbf{Numerator}: The numerator involves similar operations to the Fisher matrix. Given the data and a set of $N_{\rm disc}$ realistic simulations $\{\delta^{(n)}\}$, we perform the following:
        \begin{enumerate}
            \item Apply the $\Si$ filter to the data and each simulation, dropping modes outside the desired $\ell$-range.
            \item Compute the required one-dimensional maps $f(\hr,r_i)$, as before. When analyzing multiple simulations, we can hold $\{f[\delta^{(n)}]\}$ in memory to avoid recomputation.
            \item For contact templates, we compute the trispectrum numerator via \eqref{eq: contact-optim}, which involves a sum over pixels. For exchange templates, we compute the quadratic term in the estimators: $\int d\hr\,f(\hr,r)f'(\hr,r)Y_{LM}(\hr)$ for each $r$, dropping any modes outside the desired $L$-range. These are then combined pairwise, and summed over $r,r
            $, weighted by $F_L(r,r')$ as in \eqref{eq: exchange-optim}. 
            \item Repeat step 3 to subtract the disconnected trispectrum component, following \eqref{eq: general-estimator}, 
        \end{enumerate}
\end{itemize}
Notably, the lensing and point-source estimators require only the Fisher and numerator segments. Given the numerator, $\widehat{N}_\alpha$, and the data-independent Fisher matrix $\F_{\alpha\beta}$, the estimated templates amplitudes are given as $\sum_{\beta}\F^{-1}_{\alpha\beta}\widehat{N}^\beta$, as in \eqref{eq: general-estimator} (as shown in Fig.\,\ref{fig: sample-code}). This performs a joint estimate of all templates included in the analysis; if one wishes to instead analyze the templates independently, we can set $\widehat{A}_\alpha^{\rm indep.} = \widehat{N}^\alpha/\F_{\alpha\alpha}$. Finally, we note that the \polyspec code requires some $\ell$ and $L$ range to be specified; in our implementation, we assume that the trispectrum template is zero outside of these ranges. This allows the scale-dependence of our constraints to be tested.

\subsection{Performance}
\noindent To minimize computation time, the rate-limiting steps of \polyspec are written in \textsc{c} (via \textsc{cython}). These include: computation of the Bessel functions, $k$-space integrals, harmonic-space filtering and convolution operations, summation over $r,r'$, summation over $\ell,m$, multiplication of pixel-space maps, Fisher matrix assembly, and computation of the idealized contact Fisher matrices (used in the optimization procedure). Each of these steps is additionally parallelized using \textsc{openmp} to take maximal advantage of a given computing node. Whilst we do not incorporate MPI parallelism, we note that computation of the Fisher matrix from $N_{\rm fish}$ realizations can be trivially parallelized across nodes. Overall, the runtime is a balance between (a) the time spent computing SHTs via the (parallelized) \textsc{ducc} code, (b) \textsc{cython} summation and convolution operations (particularly those involving direction-dependent trispectra), and (c) \textsc{numpy} memory management and array processing. The last component is hard to reduce without rewriting the entire pipeline in \textsc{c}, which would require direct interface with an SHT code and reduce flexibility. 

In general, the runtime of \polyspec depends significantly on the template being analyzed. For contact trispectra, the numerators involve $\mathcal{O}(N_{\rm opt}N_{\rm disc})$ SHTs (each of which is paired with a multiplication and sum), which sets the dominant scaling. For exchange trispectra, we again require $\mathcal{O}(N_{\rm opt}N_{\rm disc})$ SHTs, then an $\mathcal{O}(N_{\rm opt}^2N_{\rm disc})$ summation, as in \eqref{eq: exchange-optim}.\footnote{The scalings with $N_{\rm opt}$ clearly demonstrate the importance of our optimization routines.} In the contact case, we could limit memory overhead by analyzing each radial point in turn; in contrast, the exchange estimators require all $N_{\rm opt}$ maps to be held in memory, else the scaling reduces to $\mathcal{O}(N_{\rm opt}^2N_{\rm disc})$ SHTs. For the Fisher matrix, the scalings are similar, except that $N_{\rm disc}$ is replaced with $N_{\rm fish}$. The prefactor is also important: exchange templates with $n$-th order direction-dependence (e.g., a trispectrum depending on $\mathcal{L}_{n}(\hk_1\cdot\hk_3)$ for Legendre polynomial $\mathcal{L}_n$), require $\sim(2n+1)\times$ more SHTs and summations. Furthermore, the EFTI shapes require considerably larger $N_{\rm opt}$ due to the coupled $r,\tau$ integral. As a result, computation of simple templates such as $\gnl$ and $\taunl$ is around an order of magnitude faster than for the EFTI shapes or, for example, collider templates with large spin. This is explicitly demonstrated in \S\ref{sec: timings}.

\section{Validation}\label{sec: validation}

\subsection{General Principles}
\noindent Before applying the estimators to observational data, we must first perform validation tests to ensure that our implementation is free from numerical (and human) error and to set fiducial values for the hyperparameters (summarized in Tab.\,\ref{tab: hyperparameters}). By necessity, the \polyspec code is long and complex, thus this is a somewhat arduous task. Here, we principally validate our pipeline by checking whether the estimators obey the theoretical properties outlined in \S\ref{subsec: estimator-def}. In particular, we ask the following questions: 
\begin{enumerate}
    \item \textbf{Does the estimator lead to false detections?} This tests whether we have correctly subtracted the disconnected (\textit{i.e.}\ Gaussian) contributions to the estimator, and can be answered by applying the estimator to unmasked simulations.
    \item \textbf{Can the estimator recover a true input signal?} Given some suite of non-Gaussian simulations, we can perform an end-to-end validation test by comparing the estimator expectation, $\mathbb{E}[\widehat{A}_\alpha]$, to the input value, $A_\alpha^{\rm fid}$. Whilst it is difficult to create simulations with generic non-Gaussianity, previous works have generated $\fnl$, $\gnl$ and $A_{\rm lens}$ simulations that we can use to test our local and lensing estimators. 
    \item \textbf{Is the estimator biased by a mask?} In theory, a non-trivial mask will increase the variance of the estimator but not change its mean; this can be tested by analyzing masked Gaussian or non-Gaussian simulations.
    \item \textbf{Is the estimator optimal when applied to Gaussian data?} As discussed in \S\ref{subsec: estimator-def}, the estimator should satisfy $\mathrm{cov}(\widehat{A}_\alpha,\widehat{A}_\beta) \to \F^{-1}_{\alpha\beta}$ in the limit of optimal $\Si$ if the dataset is Gaussian. We can test this explicitly for all templates by generating synthetic data with and without a mask and measuring the variances and correlation structure.
    \item \textbf{Does the Fisher matrix recover the analytical expectation?} For contact trispectra, the Fisher matrix can be computed exactly under idealized conditions (see \paperone and \citep{2015arXiv150200635S}). This can be used to validate the (unmasked) Monte Carlo estimator for $\F_{\alpha\beta}$.
    \item \textbf{Does the estimator scale as expected with scale cuts?} For some templates, previous works \citep[e.g.,][]{Kalaja:2020mkq,Shiraishi:2013oqa} have forecast the dependence of $\mathrm{var}(\widehat{A}_\alpha)$ on $\ell_{\rm max}$ or $L_{\rm max}$. These can be tested explicitly.
    \item \textbf{Are the numerical integrals converged?} We can test for convergence by increasing the number of points in the $k$, $r$ and $\tau$ arrays and checking for variations in $\mathbb{E}[\widehat{A}_\alpha]$ and $\F_{\alpha\beta}$. This is not required for the lensing estimators.
    \item \textbf{Is the optimization procedure sufficiently accurate?} As discussed in \S\ref{sec: estimators}), the error induced by the optimization procedure should scale as $\sqrt{f_{\rm thresh}}$. This is important to validate, particularly when one is analyzing masked templates (given that optimization is performed under idealized conditions).
    \item \textbf{How many simulations do we need for the Fisher matrix \& numerator?} Formally, the estimators require $N_{\rm disc}, N_{\rm fish}\to \infty$. As such, we must check that the induced errors in $\widehat{A}_\alpha$ are subdominant to cosmic variance.
\end{enumerate}

In the remainder of this section, we will test the above questions using synthetic data. To limit unnecessary computational expense, we mainly work at relatively low $\ell_{\rm max}\leq  512$; this is appropriate since almost all of the above questions scale with constraining power, e.g., if $N_{\rm fish}$ is too small or the optimization is unconverged, we will source a fractional error in the estimator, not an absolute one. We will perform the most validation tests for $\gnl$ and $\taunl$, since (a) these are the simplest and most well-studied templates, (b) we have non-Gaussian realizations of these models, and (c) all other estimators are built around these (with, for example, directional and collider $\tau_{\rm NL}$ just adding additional additional angular and/or scale dependence).

As a first step, we have checked that the \polyspec estimators pass a variety of sanity checks. These include: (a) checking that the Fisher derivative reproduces the estimator numerator with $d^\dag Q_\alpha[d]/4! = \tau_\alpha[d,d,d,d]$; (b) replacing the $\delta^{(n)}$ maps used in the estimator numerator with the data, $d$, and ensuring that the four-, two- and zero-field terms agree exactly; (c) checking that the various types of estimator agree exactly in the various limits (e.g., $\taunl = \tau_{\rm NL}^{0,\rm even}=(4\pi)^{-3/2}\tau^{000}_{\rm NL}=\tau_{\rm NL}^{\rm light}(0,3/2)$, $\tau_{\rm NL}^{\rm light}(s,3/2)\propto \tau_{\rm NL}^{ss0}$, $\tau_{\rm NL}^{\rm light}(s,0) = \tau_{\rm NL}^{\rm heavy}(s,0)$, \textit{et cetera}); (d) checking that the various symmetry relations are satisfied, for example, ensuring that $Q_{\ell m}$ behaves as a spin-$0$ field. Many of these are non-trivial since the numerator and Fisher matrix involve different code, different orders of operations, and different permutation structures, and one has to use different approaches to capture, for example, the complex oscillations in the heavy templates compared to the simple $\taunl$ form. All checks are passed to per-mille accuracy.

\subsection{Set-Up}\label{subsec: setup}
\noindent To perform the validation tests, we use the following default settings (several of which are varied below). Gaussian random field data are generated using the \textit{Planck} 2018 cosmology, specified by $\{h = 0.6732, \omega_b = 0.02238, \omega_c = 0.1201, \tau_{\rm reio} = 0.0543, \sum m_\nu = 0, n_s = 0.9660, A_s = 2.101\times 10^{-9}, k_{\rm pivot} = 0.05\,\mathrm{Mpc}^{-1}, r = 0\}$ \citep{2020A&A...641A...6P}. CMB temperature and polarization transfer functions are computed by running \textsc{camb} \citep{Lewis:1999bs} at high resolution for all $\ell\in[2,500]$, and $k\in[10^{-6},0.5]\,\mathrm{Mpc}^{-1}$, using $N_k = 3425$ sampling points. When simulating masked data, we apply the \textit{Planck} common component-separation mask \citep{Planck:2018yye}, which retains $\approx 77\%$ of the sky, and inpaint small holes, reducing the sky fraction by $3\%$. The CMB power spectrum, $C_\ell^{XY}$, is evaluated using \textsc{camb}, ignoring lensing contributions. We add a \textit{Planck}-like noise spectrum $N_{\ell}^{XY} = \delta_{\rm K}^{XY}\Delta^2_X \mathrm{exp}(\ell(\ell+1)\theta_{\rm FWHM}^2/(8\log 2))$ with $\Delta_T = \Delta_{E,B}/\sqrt{2} = 60\,\mu$K-arcmin and $\theta_{\rm FWHM} = 5\,$arcmin, but do not otherwise include a beam. The data is weigted by the following filter, defined by its action on some map $x_{\ell m}^X$ (retaining the beam for generality):
\beq
    [\Si x]^X_{\ell m} = B_\ell^X\sum_Y\left[B_\ell C_{\ell}B_\ell+N_\ell\right]^{-1,XY}[\mathbb{I}\,x]^Y_{\ell m},
\eeq
where $\mathbb{I}\,x$ is the map processed by a diffusive inpainting scheme. In the presence of a mask, this is not strictly optimal, but is found to be close in practice \citep{2011MNRAS.417....2S}. We will discuss optimal weighting schemes in \paperthree.

By default, we work with $N_{\rm side}=256$ \textsc{healpix} maps at $\ell_{\rm max}=512$, though we carefully consider the scaling of our constraints with $\ell_{\rm max}$. In the absence of a mask, the results are unchanged for any $N_{\rm side}>\ell_{\rm max}/3$ (since the theoretical trispectra are band-limited); in its presence, aliasing effects to apply to both the numerator and Fisher matrix and thus cancel at leading order (as we demonstrate below). The precise hyperparameters adopted for each test can be found below. All computations are performed on high-performance cluster nodes with $N_{\rm CPU} = 64$, using \textsc{cython} instructions generated for a Broadwell processor.

\subsection{Local Non-Gaussianity}
\noindent Our first templates are those generated by cubic and quadratic local transformations of the primordial potential, parametrized by $\gnl$ and $\taunl$ respectively. These generate contact and exchange inflationary trispectra with the schematic form (neglecting all numerical factors and permutations)
\beq
    T_\zeta \sim \gnl P_\zeta(k_1)P_\zeta(k_2)P_\zeta(k_3) + \taunl P_\zeta(k_1)P_\zeta(k_3)P_\zeta(K),
\eeq
for $K = |\vk_1+\vk_2|$. Note that $\taunl$ can be sourced by $\fnl$, with $\taunl \geq (\tfrac{6}{5}\fnl)^2$ \citep{Suyama:2007bg}. %These templates are of particular interest since they have both been used to construct non-Gaussian CMB realizations, which can beuse to perform recovery tests at the end of this section. 

\subsubsection{(Non-)Gaussian Simulations}
\noindent When analyzing $\gnl$, we utilize 100 full-sky CMB temperature and polarization simulations developed during the tailed cosmology study of \citep{Coulton:2024vot}. These were generated with $\ell_{\rm max}=499$ and the fiducial cosmology stated above, and are supplemented with noise and (optionally) a mask. We consider both $\gnl=0$ (\textit{i.e.}\ Gaussian) and $\gnl = 6\times 10^5$, with the latter value chosen to ensure detectability. We analyze $50$ simulations with the default parameters $\{N_{\rm fish}=100, \ell_{\rm min}=2, f_{\rm thresh} = 10^{-4}, N_{\rm side}=256\}$, using the other $N_{\rm disc}=50$ simulations to subtract the disconnected contributions, adopting the same value of $\gnl$ in all cases.\footnote{In realistic settings, one would typically use disconnected simulations without primordial non-Gaussianity. However, large $\fnl$ and $\gnl$ can alter the CMB power spectra, thus we here use non-Gaussian simulations to isolate the trispectrum effects.}

For $\taunl$, we use the simulations are described in \citep{Elsner:2009md}, which were generated with the WMAP5 cosmology $\{h = 0.701, \omega_b = 0.02265, \omega_c = 0.1143, \tau_{\rm reio} = 0.084, \sum m_\nu = 0, n_s = 0.96, A_s = 2.457\times 10^{-9}, k_{\rm pivot} = 0.002\,\mathrm{Mpc}^{-1}, r=0\}$. This suite was originally developed for validating $\fnl$ estimators, but can be similarly used to for four-point validation, setting $\taunl = (\frac{6}{5}\fnl)^2$.\footnote{This carries an important caveat here. When constructing the simulations, the authors of \citep{Elsner:2009md} first generated the harmonic coefficients of a Gaussian primordial potential, $\Phi^{\rm G}_{\ell m}(k)$, which were transformed to pixel-space and used to form a non-Gaussian $\Phi$ field. The $\ell=1$ modes in $\Phi^{\rm G}_{\ell m}(k)$ were set to zero, since they are not relevant for bispectrum analyses -- however, these modes also generate an $L=1$ trispectrum, which \textit{is} observationally accessible (and highly constraining). To avoid bias arising from this discrepancy, we omit the $L=1$ mode in our analysis, setting $L_{\rm min}=2$.} We restrict to $\ell_{\rm max}=512$, and consider both $\taunl = 0$ and $\taunl = 5.76\times 10^4$ (corresponding to $\fnl=200$). We analyze $100$ simulations using the same hyperparameter values as above and use $N_{\rm disc}=100$ simulations to subtract disconnected contributions from the estimator.

\subsubsection{Optimization}\label{subsubsec: optim-test}
\noindent We first discuss the (data-independent) optimization routine used to reduce the number of radial integration points in the trispectrum estimators. Starting from a finely sampled set of $N_s = 686$ sampling points (defined in \S\ref{sec: code}), we compute an optimized subset using a distance metric defined by the idealized Fisher matrix (computed over $5$ Monte Carlo realizations for $\taunl$), as described in \S\ref{sec: estimators}. Including both $T$- and $E$-modes with $\ell_{\rm max}=499$ and $L_{\rm max}=10$, we obtain $f_{\rm thresh}<10^{-4}$ with just $N_{\rm opt} = 30$ integration points for both $\taunl$ and $\gnl$, which corresponds to sub-percent error in $\F$. This represents a speed-up by $\approx 20\times$, and is stable to variations in the starting radial grid (see below). We caution that the optimization algorithm is tuned to a specific analysis and must be re-run for any change in scale-cuts, fields, beams or fiducial spectra.

\begin{figure}[!t]
    \centering
    \includegraphics[width=0.47\linewidth]{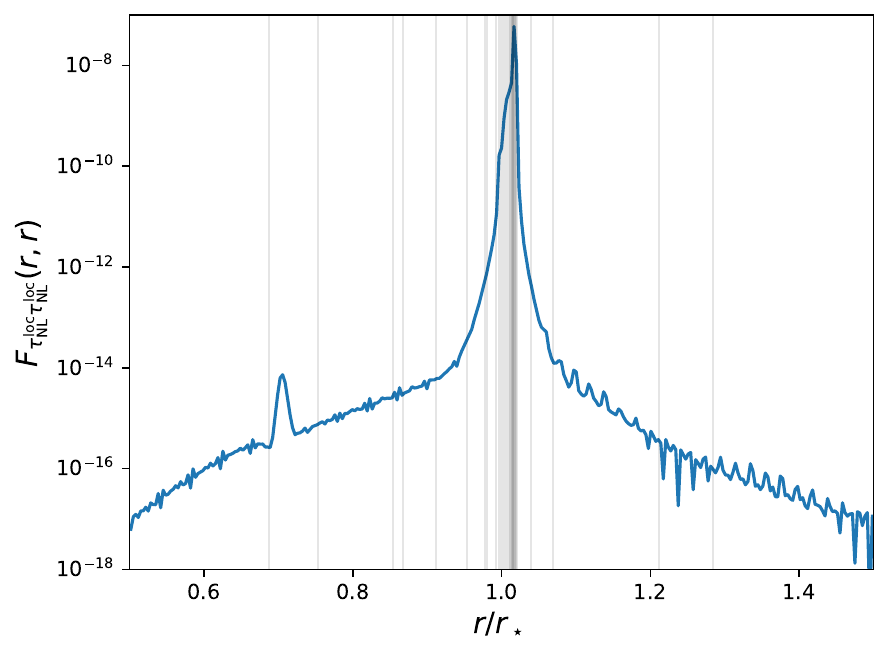}
    \includegraphics[width=0.47\linewidth]{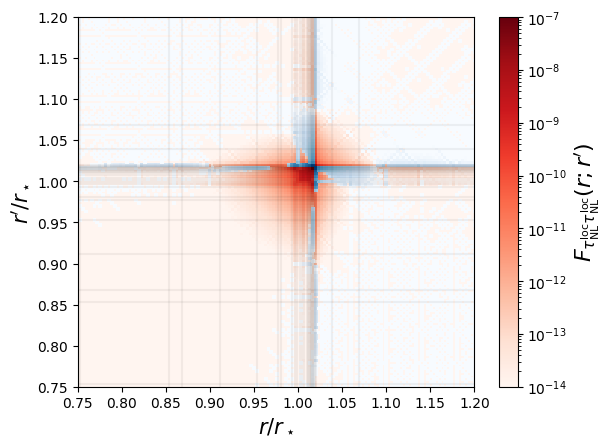}
    \caption{Contributions to the ideal $\taunl$ Fisher matrix as a function of the radial parameters $r,r'$, relative to the distance to last scattering, $r_\star\approx 14\,000\,\mathrm{Mpc}$. %The full Fisher matrix, which defines the optimal inverse variance, is given by the two-dimensional integral over this quantity. 
    The left panel shows the diagonal elements, which are sharply peaked at $r=r_\star$, whilst the right panel shows the full matrix, which exhibits strong positive (red) and negative (blue) correlations around recombination. Grey lines show the sampling points returned by our optimization algorithm; using just these points to define the numerator and Fisher matrix (as opposed to integrating over the full $r,r'$ plane) incurs an error below $1\%$. These results are computed using $\ell\in[2,499]$, $L\in[2,10]$, including both temperature and polarization anisotropies.}
    \label{fig: optim1d}
\end{figure}

To make sense of the optimization procedure, we plot the contributions to the Fisher matrix as a function of radius in Fig.\,\ref{fig: optim1d}, focusing on the $\taunl$ template (which is more difficult to optimize, as discussed in \paperone).\footnote{For exchange templates, the Fisher matrix is an outer product of two $Q$ derivatives, each of which involves two-dimensional radial integrals (cf.\,\ref{eq: fisher-mc}): here, we take the derivative with respect to one $r$ in each $Q$. This is the two-dimensional matrix used to perform optimization, and closely approximates the Hessian of $\F$.} We find that matrix diagonal is strongly peaked around $r = r_\star\pm5\%$ (where $r_\star$ is the distance to last scattering), with contributions outside this regime suppressed by a factor $\gtrsim 10^6$ (although note the reionization bump at $r\approx 0.7r_\star$). In the two-dimensional $r,r'$ plane, the results are similar: we find strong correlations around recombination, which quickly decay (and switch sign around $r=r_\star$). Our optimization algorithm reconstructs the dominant contributions to $\F(r,r')$: given its sharply-peaked nature, it is unsurprising that it can be well-approximated with only a small number of sampling points (whose locations are shown as gray lines in the Figure).

\subsubsection{Gaussian Bias}
\noindent Next, we validate the estimator numerator. As in \eqref{eq: general-estimator}, this contains three piece: a four-field term ($\widehat{N}_4 \propto h^4$, dropping indices), a two-field term ($\widehat{N}_2\propto -6h^2\av{h^2}$), and a zero-field term ($\widehat{N}_0 \propto 3\av{h^2}^2$). In the Gaussian limit (\textit{i.e.}\ with vanishing $\gnl$ and beyond), the expectation of each term should be related by
\beq
    \mathbb{E}[\widehat{n}_4] = -\frac{1}{2}\mathbb{E}[\widehat{n}_2] = \mathbb{E}[\widehat{n}_0],
\eeq
such that the sum vanishes, and the estimator is unbiased. In the presence of non-Gaussianity, $\mathbb{E}[\widehat{N}_4]$ picks up an additional contribution proportional to the primordial amplitude of interest, whilst $\mathbb{E}[\widehat{N}_{0,2}]$ are unchanged (ignoring variations in the power spectrum).

\begin{figure}[!t]
    \centering
    \includegraphics[width=0.49\linewidth]{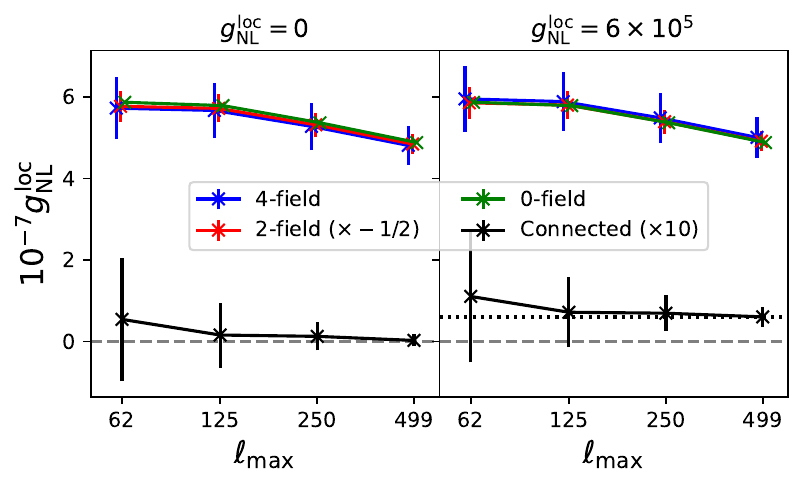}
    \includegraphics[width=0.49\linewidth]{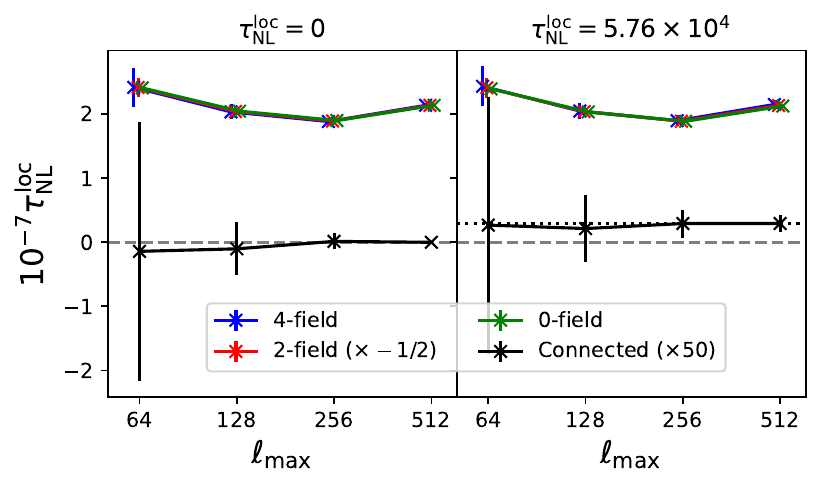}
    \caption{Breakdown of contributions to the $\gnl$ (left) and $\taunl$ (right) trispectrum estimators applied to Gaussian (left sub-panels) and non-Gaussian (right sub-panels) simulations, with fiducial values indicated by the titles. We show the \resub{quartic (blue), quadratic (red)}, and constant (green) numerator contributions normalized by the inverse Fisher matrix, as well as their sum (black), which gives the full estimator. As expected, the three terms precisely cancel if the fields are Gaussian, but reveal subtle signals consistent with the injected values (shown in dotted lines) in the non-Gaussian case. Results are generated from the mean of 50 ($\gnl$) and $100$ ($\taunl$) masked temperature and polarization simulations for four choices of $\ell_{\rm max}$ and errorbars showing the error in a single realization. We perform a detailed analysis of the injected signals in Fig.\,\ref{fig: g-tau-injection}.}
    \label{fig: breakdown}
\end{figure}

In Fig.\,\ref{fig: breakdown}, we validate these predictions for the $\gnl$ and $\taunl$ estimators, as applied to masked Gaussian and non-Gaussian simulations. We find excellent agreement between the three terms for Gaussian simulations, with the inclusion of non-zero $\gnl$ or $\taunl$ leading to a slight increase in $\widehat{N}_4$ (consistent with the true value, as discussed below). Despite the large injected signal, these differences are very small, with the disconnected term dominating by two orders of magnitude. This demonstrates the need for accurate simulations to avoid (second-order) bias from an incorrectly modelled two-point function. Notably, the variance of the connected trispectrum is much smaller than that of the disconnected piece; this is due to the inclusion of two-field terms and implies that the simpler estimator $\widehat{N}_4-\widehat{N}_0$ is suboptimal. Finally, we observe that the non-Gaussian error-bars reduce only slowly with $\ell_{\rm max}$; this is due to the large non-Gaussian contributions to the covariance. 

\subsubsection{Optimality \& Scale-Dependence}

\begin{figure}[!t]
    \centering
    \begin{minipage}{0.54\linewidth}
        \centering
        \includegraphics[width=0.8\linewidth]{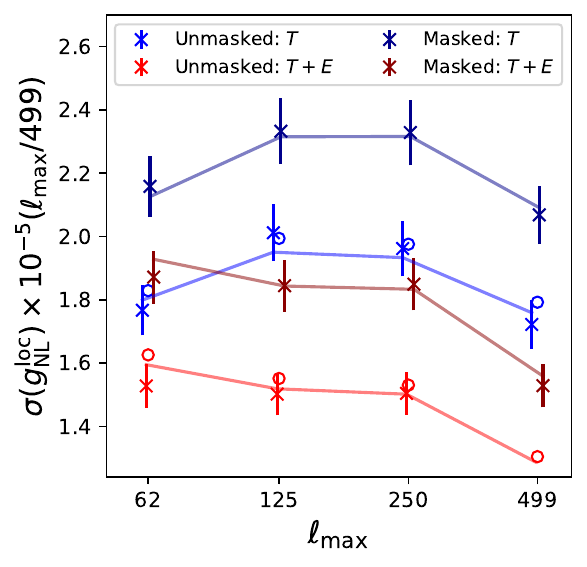}
        \caption{Optimality test for the $\gnl$ estimator. We compare the empirical standard deviations (errorbars) obtained from $250$ Gaussian simulations to those predicted from the Monte Carlo Fisher matrix (lines) and idealized analytic predictions (circles). Results are shown for various choices of $\ell_{\rm max}$ (scaling by $1/\ell_{\rm max}$), for both temperature (blue) and temperature and polarization (red) analyses, optionally including a \textit{Planck}-like mask and inpainting scheme (indicated by dark colors). Note that the various datasets are correlated, since they are constructed from the same Gaussian realizations. Up to noise fluctuations, we find excellent agreement between the three sets of estimators, implying that (a) the estimators are close to minimum-variance and (b) the Fisher matrix is unbiased.}
        \label{fig: g-err}
    \end{minipage}
    \hfill
    \begin{minipage}{0.44\linewidth}
        \centering
        \includegraphics[width=\linewidth]{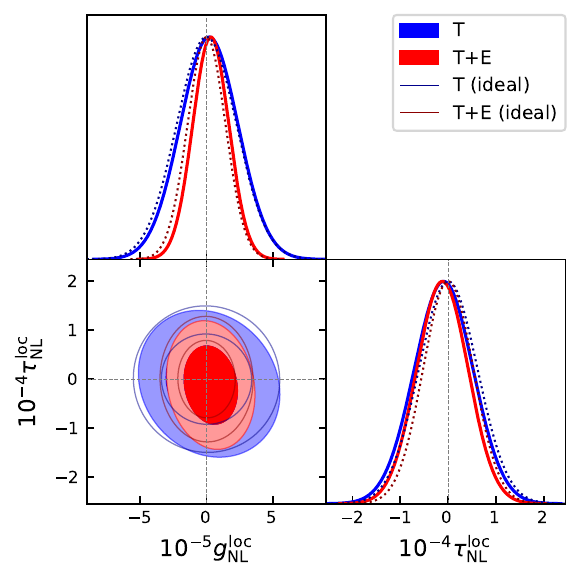}
        \caption{Joint constraints on $\gnl$ and $\taunl$ obtained by applying \polyspec to $250$ $\gnl=\taunl=0$ simulations and assuming a Gaussian likelihood. Results are derived using $\ell_{\rm max}=499$ and $L_{\rm max}=10$, and include a \textit{Planck}-like mask. Dotted lines show the idealized constraints, setting the mean to zero and the covariance to the inverse Fisher matrix. The two parameters are almost uncorrelated, and we obtain close-to-optimal parameter errors.}
        \label{fig: joint-g-tau}
    \end{minipage}
\end{figure}

\noindent If the estimators are optimal, their covariance should be equal to the inverse Fisher matrix.
%, \textit{i.e.}\ $\mathrm{cov}(\widehat{A}_\alpha,\widehat{A}_\beta) = \F^{-1}_{\alpha\beta}$ (assuming zero fiducial signal). 
This does not require contaminated simulations and is a powerful validation test (particularly given that the relevant code is mostly independent). In Fig.\,\ref{fig: g-err}, we compare the empirical variance of $\widehat{g}_{\rm NL}^{\rm loc}$ computed from $250$ Gaussian simulations (setting $N_{\rm disc}=100$ to reduce noise) alongside the Fisher matrix predictions. We find excellent agreement for all values of $\ell_{\rm max}$, both including and excluding polarization, and with and without a mask. This implies that estimator is close to minimum variance and that our masked $\Si$ weighting is close to optimal. Furthermore, the unmasked Fisher matrices are in excellent agreement with the idealized analytic predictions (computed with a very different method), which provides further validation of the estimator.

Fig.\,\ref{fig: g-err} additionally shows the dependence of $\sigma(\gnl)$ on scale-cuts and field content. Adding $E$-modes improves constraints by up to $40\%$ (particularly at larger $\ell_{\rm max}$); this suggests that the excision of polarization in the official \textit{Planck} $\gnl$ analyses led to significant information loss. When adding a mask, the constraints inflate by $20-30\%$, which is broadly consistent with the loss of survey area. On these scales, we find $\sigma(\gnl)\propto \ell_{\rm max}^{-1}$ to good accuracy, though some variation due to the transfer functions and differing noise properties of temperature and polarization; this agrees with the theoretical analysis of \citep{Kalaja:2020mkq}.

\begin{figure}[!t]
    \centering
    \includegraphics[width=0.4\linewidth]{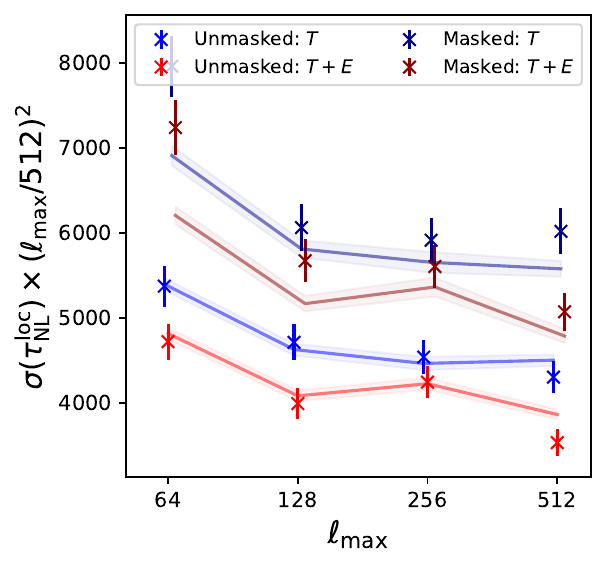}
    \includegraphics[width=0.4\linewidth]{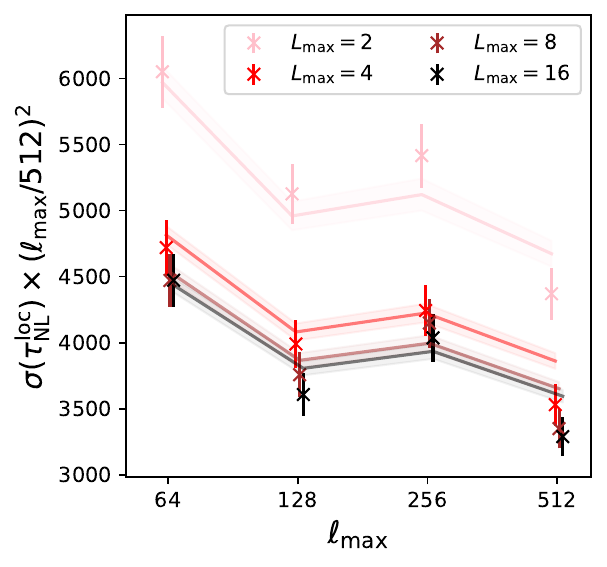}
    \caption{Optimality test for the $\taunl$ estimator. The left panel is analogous to Fig.\,\ref{fig: g-err}; we find a close-to-quadratic scaling of the errors with $\ell_{\rm max}$ and up to $20\%$ tighter errorbars when including polarization data. In most cases, the observed variances agree well with the Fisher matrix predictions, though we find some evidence for a mismatch in the masked dataset at $\ell_{\rm max}=64$, most likely due to the suboptimal weighting scheme. All results are obtained using $250$ Gaussian simulations with $N_{\rm disc}=100$ and $L_{\rm max}=4$. In the right panel, we show the effect of varying $L_{\rm max}$, restricting to unmasked $T+E$ simulations. As expected, the estimator is dominated by the lowest $L$-modes, with the removal of $L>8$ changing $\sigma(\taunl)$ by less than $2\%$.}
    \label{fig: tau-err}
\end{figure}

The left panel of Fig.\,\ref{fig: tau-err} shows the corresponding results for $\taunl$. Though analytic predictions cannot be easily obtained in this regime (as discussed in \paperone), we find generally good agreement between the Fisher matrix prediction and the empirical variances obtained from $250$ simulations (again setting $N_{\rm disc}=100$). An exception is the masked dataset at $\ell_{\rm max}=64$; 
%\ell_{\rm max}=64$, , though there is a slight evidence for a weak inflation of errorbars (as seen in previous analyses \citep[e.g.,][]{Philcox:2023xxk,Philcox:2024wqx}) 
this could indicate suboptimality in our approximate $\Si$ filtering and the increased dependence of the estimator on large-scales, which are strongly affected by the mask. In contrast to $\gnl$, the utility of $E$-modes in the $\taunl$ estimator is relatively minor, with the errorbar shrinking by at most $20\%$ (agreeing with \citep{Kalaja:2020mkq}). 

For all choices of field content and mask, we find $\sigma(\taunl)\sim \ell_{\rm max}^{-2}$ to good approximation -- this agrees with \citep{Kamionkowski:2010me,Kalaja:2020mkq,Kogo:2006kh} and beats the $\ell_{\rm max}^{-1}$ scaling expected for measuring $\fnl$ from the CMB bispectrum (recalling that $\taunl\geq (6/5 \fnl)^2$ \citep{Suyama:2007bg}). This occurs since the trispectrum contains two `hard modes' instead of one, and implies that the four-point function contains complementary information on $\fnl$ \citep{Kamionkowski:2010me} (though in practice the gain is marginal \citep{Philcox4pt3}). Turning to the internal leg, we find very different behavior. From the right panel of Fig.\,\ref{fig: tau-err}, it is clear that most of the signal-to-noise is sourced by the largest internal modes (\textit{i.e.} \ $L_{\rm max}=2$, since we drop $L=1$), with $\sigma(\taunl)$ tightening by $25\%$, $8\%$ and $2\%$ when increasing to $L_{\rm max}=4,8,16$ sequentially. This matches the findings of \citep{Kogo:2006kh,Kalaja:2020mkq}, and implies that we can restrict our analysis to highly collapsed configurations
%(essentially correlating the locally-measured power spectrum and two points in space, as in \citep{Feng:2015pva,Munshi:2009wy}) 
without appreciable loss of signal-to-noise.

Finally, we perform a joint analysis of $\gnl$ and $\taunl$ using masked Gaussian simulations (assuming the \textit{Planck} fiducial cosmology). As shown in Fig.\,\ref{fig: joint-g-tau}, we find good agreement between idealized and empirical covariances, validating our pipeline. Note that this assumes a Gaussian likelihood specified by the mean and covariance; in practice, the $\taunl$ posterior can be non-Gaussian, as we discuss in \paperthree. 
%\footnote{We note that $\F^{-1}$ is somewhat asymmetric here, which likely indicates suboptimality in the $\Si$ weights, although we find excellent agreement with simulations when considering the diagonal elements.} 
For this choice of $\ell_{\rm max}$ and cosmology, the addition of $E$-modes significantly tightens the $\gnl$ constraint but only minorly affects $\taunl$; we will find a similar result from the full high-resolution dataset in \paperthree. The two non-Gaussianity parameters are weakly correlated: this is expected since the corresponding trispectra peak in different regimes (squeezed and collapsed configurations), and matches previous work \citep[e.g.,][]{Smidt:2010ra}

\subsubsection{Injection Test}

\begin{figure}[!t]
    \centering
    \includegraphics[width=0.44\linewidth]{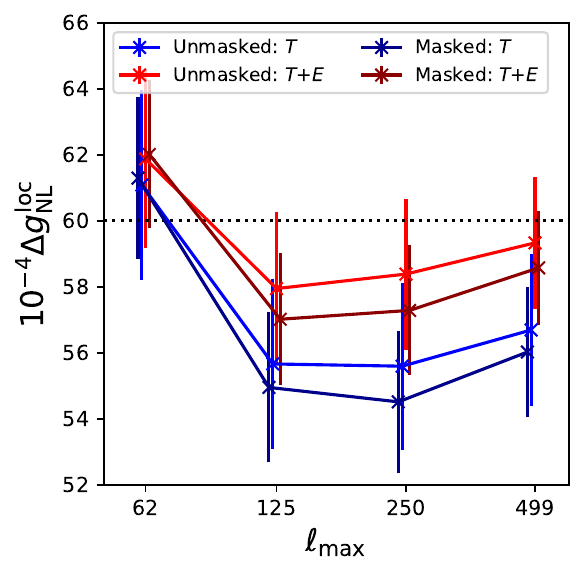}
    \includegraphics[width=0.44\linewidth]{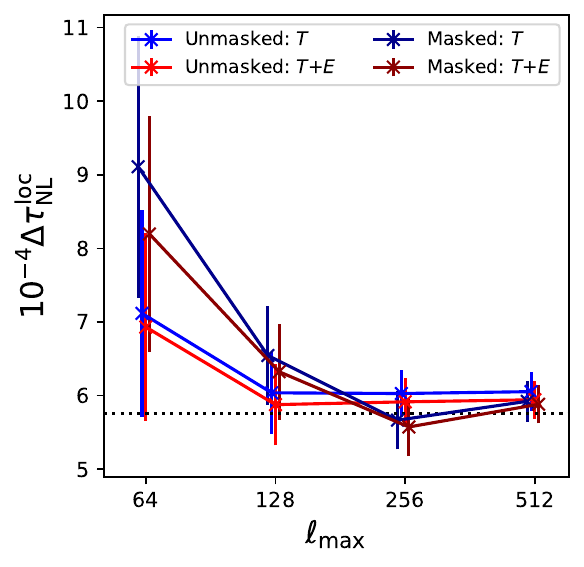}
    \caption{Parameter recovery test for $\gnl$ (left) and $\taunl$ (right). We plot the mean of $\gnl$ and $\taunl$ estimates obtained from \polyspec, taking the difference between estimates from simulations with and without injected non-Gaussianity. The fiducial values are shown with dotted lines, and errorbars represent the error in the mean from $50$ ($\gnl$) and $100$ ($\taunl$) simulations. Up to noise fluctuations, we find unbiased results in all cases, validating the estimators.}
    \label{fig: g-tau-injection}
\end{figure}

\noindent In Fig.\,\ref{fig: g-tau-injection} we verify that our estimators are unbiased in the presence of a signal. This is obtained by analyzing the non-Gaussian datasets discussed above, reducing noise by (a) averaging over $50$ ($100$) $\gnl$ ($\taunl$) simulations, and (b) taking the difference between non-Gaussian and Gaussian measurements (\textit{i.e.}\ $\widehat{A}[d_{\rm nG}]-\widehat{A}[d_{\rm G}]$ for data $d$), This reduces errors by $\sim 10\times$, which mimics the statistical power of a full high-resolution analysis. Although the results are noisy at low $\ell_{\rm max}$, they are unbiased in all cases. For the masked temperature-plus-polarization analysis at $\ell_{\rm max}=512$, we find $(10^{-5}\widehat{g}_{\rm NL}^{\rm loc},10^{-4}\widehat{\tau}_{\rm NL}^{\rm loc}) = (6.14\pm0.37, 5.86\pm0.17)$ in excellent agreement with the fiducial values $(6.00,5.76)$. This implies that our local estimators are able to recover an input signal at high fidelity. Lastly, we note that the variances appear to saturate by $\ell_{\rm max}=256$; this is due to the non-Gaussian noise contributions sourced by $\gnl$ and $\taunl$ and is not seen in Figs.\,\ref{fig: g-err}\,\&\,\ref{fig: tau-err} \citep[cf.][]{Elsner:2009md}.

\subsubsection{Consistency Tests}

\noindent Finally, we assess the dependence of the \polyspec $\gnl$ and $\taunl$ estimators on their various hyperparameters (which are summarized in Tab.\,\ref{tab: hyperparameters}). Firstly, we halve the number of Monte Carlo realizations used to measure the Fisher matrix, $N_{\rm fish}$. From the temperature-plus-polarization analysis at the largest $\ell_{\rm max}$, we find a $0.06\sigma$ ($0.08\sigma$) shift in the mean of $\gnl$ ($\taunl$), and a $1\%$ ($2\%$) change in the theoretical errorbars. This demonstrates that our choice of $N_{\rm fish}=100$ is conservative.\footnote{As shown in \paperthree, the shifts at $\ell_{\rm max}=2048$ are significantly smaller.} Halving the number of simulations used in the estimator numerator ($N_{\rm disc}$) gives a larger effect; we find a $0.26\sigma$ ($0.02\sigma$) shift in the mean, as well as a $2\%$ ($5\%$) change in the error-bars. This highlights the importance of carefully subtracting the disconnected contributions, which are particularly important for $\gnl$. We find that the estimators are insensitive to the $k$-integration grid, with a $0.003\sigma$ ($0.002\sigma$) change to the mean and $0.08\%$ ($0.03\%$) variation in the Fisher matrix obtained from doubling $N_k$ (and hence the number of points in the \textsc{camb} transfer function). Finally, we vary the optimization parameters; we find negligible ($<0.1\%$ in the Fisher matrix, $<0.01\sigma$ in the mean) effects from (a) doubling the number of points in the pre-optimized radial integration grid (and thus the $r$-resolution), (b) loosening the optimization tolerance to $f_{\rm thresh}=10^{-3}$, (c) using all $N_{\rm opt} = N_s$ sampling points and dropping the optimization step entirely. This affords us confidence that our optimization algorithm is accurate (and motivates us to use reduced tolerances below). 

\subsubsection{Constant Non-Gaussianity}

\noindent Before proceeding to the more exotic templates, we briefly validate the constant non-Gaussianity estimators. This corresponds to the schematic model:
\beq
    T_\zeta \sim g_{\rm NL}^{\rm con}\left[P_\zeta(k_1)P_\zeta(k_2)P_\zeta(k_3)P_\zeta(k_3)\right]^{3/4},
\eeq
which was designed to have a contact-type shape with features arising only from the CMB transfer functions \citep{Fergusson:2010gn}. Whilst it is phenomenologically interesting, this model has not been used to create non-Gaussian simulations and has been rarely studied since its introduction. To validate the estimator we (a) check that it does not lead to false detections of non-Gaussianity, (b) compare the Monte Carlo Fisher matrix to the analytic (unmasked) prediction, and (c) compare the empirical and theoretical variances using $100$ Gaussian simulations. We additionally note that the \polyspec $g_{\rm NL}^{\rm con}$ estimator is a simpler cousin of the $\gnl$ form, featuring additional symmetry. For this analysis, we adopt the same set-up as for $\gnl$, using $N_{\rm disc}=100$ to remove disconnected contributions.

\begin{figure}
    \centering
    \begin{minipage}[t]{0.4\linewidth}
        \centering
        \includegraphics[width=\linewidth]{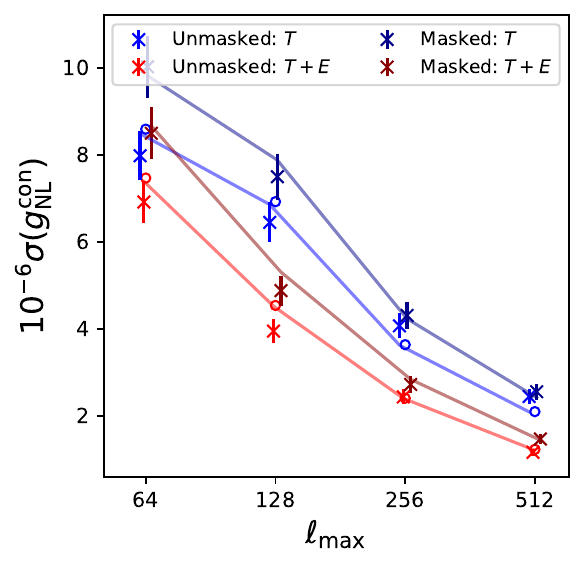}
    \caption{Optimality test for the $g_{\rm NL}^{\rm con}$ estimator, as in Fig.\,\ref{fig: g-err}. We find good agreement between the empirical variances (errorbars), Fisher matrix predictions (lines) and, where relevant, ideal analytic calculations (circles), implying that the estimator is unbiased and close-to-optimal. The errorbars scale roughly as $1/\ell_{\rm max}$ and shrink significantly when polarization is included.}
    \label{fig: con-err}
    \end{minipage}
    \hfill
    \begin{minipage}[t]{0.55\linewidth}
        \centering
        \includegraphics[width=\linewidth]{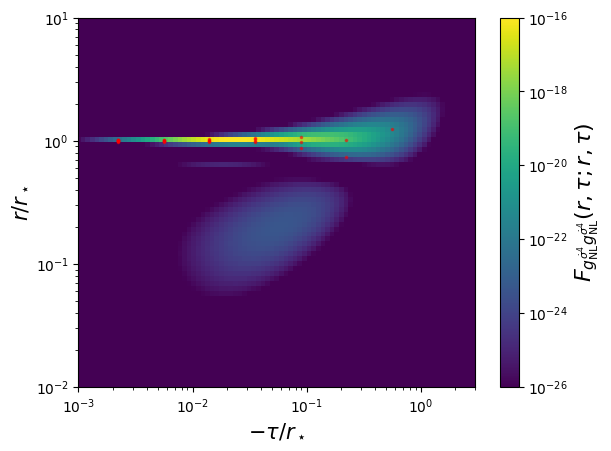}
    \caption{Contributions to the ideal $\gnldotdot$ Fisher matrix diagonal as a function of the integration variables $r,\tau$ (generalizing \resub{the left panel of Fig.\,\ref{fig: optim1d}}). The integrand is sharply peaked around $r=r_\star$ with $-\tau\lesssim r_\star$, and its full structure can be approximated using $35$ points, shown in red dots. The horizontal stripe around $r\approx 0.7r_\star$ corresponds to reionization. This is computed for using $T$- and $E$-modes for $\ell\in[2,512]$. Results for the other EFTI templates, $\gnldotdel$ and $\gnldeldel$, are analogous.}
    \label{fig: optim2d}
    \end{minipage}
\end{figure}

In Fig.\,\ref{fig: con-err}, we plot the three sets of errorbars and their dependence on $\ell_{\rm max}$. As before, we find good agreement between the empirical variances and $\F^{-1}$, implying that the estimator is close to optimal. Coupled with the fact that the Monte Carlo Fisher matrix closely matches the analytic calculation, this indicates that the overall estimator in unbiased. In this case, we find that the inclusion of polarization yields significantly tighter errorbars, with up to $70\%$ improvements for large $\ell_{\rm max}$. Moreover, we find $\sigma(g_{\rm NL}^{\rm con})\sim \ell_{\rm max}^{-1}$, though the scaling is somewhat weaker in temperature-only analyses. As expected, we find consistent results when a \textit{Planck}-like mask is included, which both validates our estimator and implies that our choice of weighting scheme, $\Si$, is appropriate. 

\subsection{EFTI Non-Gaussianity}\label{subsec: valid-efti}
\noindent We now consider the EFTI templates. These are contact trispectra that can arise in both single- and multi-field models of inflation and take the schematic form
\beq
    T_\zeta \sim \gnldotdot\Delta^4_\zeta\frac{f}{k_1k_2k_3k_4k_T^5} + \gnldotdel\Delta^4_\zeta\frac{g(k_3,k_4,k_T)}{k_1k_2k_3k_4k_T^5}(\hk_3\cdot\hk_4) + \gnldotdel\Delta^4_\zeta\frac{h(k_1,k_2,k_3,k_4,k_T)}{k_1k_2k_3k_4k_T^5}(\hk_1\cdot\hk_2)(\hk_3\cdot\hk_4),
\eeq
where $k_T = k_1+k_2+k_3+k_4$, $\Delta_\zeta^2$ is the power spectrum amplitude and $f$, $g$, and $h$ are some dimensionless separable polynomials specified in \paperone. These are much more difficult to implement than for $g_{\rm NL}^{\rm loc, con}$ due to the $1/k_T$ factors; as discussed in \citep{2015arXiv150200635S}, efficient estimators for $\gnldotdot$, $\gnldotdel$ and $\gnldeldel$ can be obtained by introducing an integral over conformal time $\tau$. The temperature signatures of these templates have been searched for in both WMAP and \textit{Planck} data \citep{2015arXiv150200635S,Planck:2015zfm,Planck:2019kim}, but the polarization signal has yet to be explored. Given the complexity of the above form, non-Gaussian simulations have not been generated, thus we must validate our estimators in the Gaussian limit.

\subsubsection{Optimization}
\noindent First, we demonstrate the efficacy of the optimization scheme discussed in \S\ref{sec: estimators}. Despite the additional $\tau$ integral, the optimization proceeds similarly to that of the $\gnl$ template, starting from a finely-spaced array of $N_s = 4161$ $(r_i,\tau_i)$ points, as described in \S\ref{sec: code}. In Fig.\,\ref{fig: optim2d}, we show the diagonal of the ideal $\gnldotdot$ Fisher derivative as a function of $r,\tau$, \textit{i.e.}\ the quantity integrated to obtain the full Fisher matrix. This is sharply peaked around $r\approx r_\star$, but has support for a fairly wide range of $\tau$ values with $\tau\lesssim -r_\star$ (which is expected, since $\tau$ is the inflationary conformal time). For a temperature-plus-polarization analysis with $\ell_{\rm max}=512$, we find that the Fisher matrix structure can be reproduced to within $f_{\rm thresh}=10^{-4}$ with just $N_{\rm opt}=35$ sampling points, with a further $4$ required for $\gnldotdel$ and $\gnldeldel$; this reduce the computation time by a factor of $100$. As seen in the figure, the optimized points cluster closely around $r\approx r_\star$, but feature a wider range in $\tau$. Our optimization routine is highly stable: the means (errorbars) of all EFTI results presented in this section are unchanged to $0.01\sigma$ ($0.3\%$) when doubling $N_s$ or increasing the tolerance to $f_{\rm thresh}=10^{-3}$.

\subsubsection{Performance on Gaussian Simulations}
\noindent The three EFTI estimators can be validated using a large suite of Gaussian simulations, generated at the \textit{Planck}-like signal-plus-noise power spectrum described in \S\ref{subsec: setup}. By default, we use the hyperparameter set $\{N_{\rm disc}=50, N_{\rm fish}=50, \ell_{\rm min}=2, f_{\rm thresh} = 10^{-4}, N_{\rm side}=256\}$, analyzing $50$ simulations. We use a lower $N_{\rm fish}$ than for $\gnl$ due to the fast convergence found previously; this choice will be validated below. In all analyses, we simultaneously estimate $\gnldotdot, \gnldotdel$ and $\gnldeldel$; if we set the off-diagonal elements of the Fisher matrix to zero, this dataset be used to obtain independent constraints on the three models (ignoring their correlations). 
%We consider results both with and without a mask, with and without polarization data, and for $\ell_{\rm max}\in\{64,128,256,512\}$ (noting that our conclusions naturally extend to larger $\ell_{\rm max}$, but are more expensive to obtain therein).

Regardless of the choice of mask, $\ell_{\rm max}$, or choice of fields, we find unbiased results from each estimator when applied to $50$ Gaussian simulations. For example, in the most realistic analysis (temperature-plus-polarization at $\ell_{\rm max}=512$), we obtain $10^{-5}\gnldotdot = -3.7\pm 2.3$ ($-0.9\pm3.0$), $10^{-5}\gnldotdel = -3.0\pm2.3$ ($-0.47\pm3.2$), $10^{-5}\gnldeldel = -0.56\pm0.56$ ($0.12\pm0.77$) without (with) a mask, all of which are consistent with zero within $1.6\sigma_{\rm mean}$. This validates that (a) the Gaussian biases are being subtracted appropriately, and (b) the number of numerator simulations, $N_{\rm disc}$, is appropriate.

\begin{figure}[!t]
    \centering
    \includegraphics[width=0.9\linewidth]{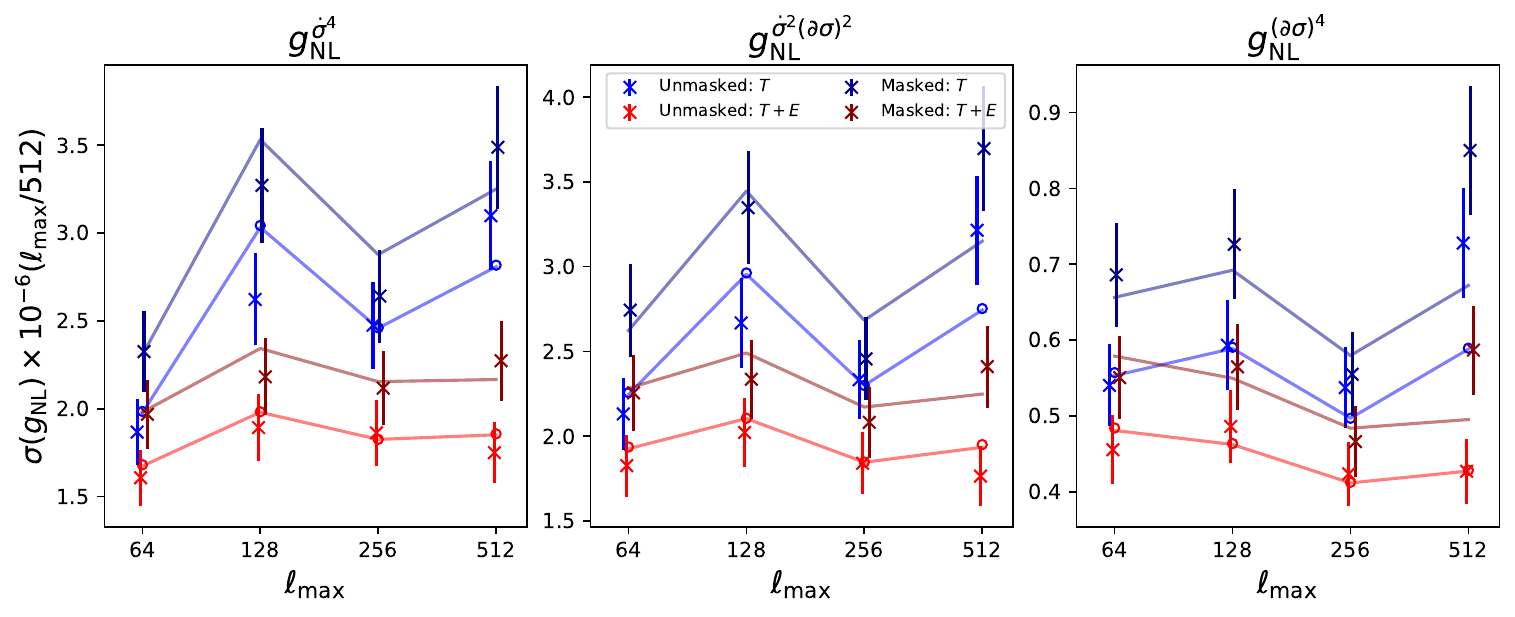}
    \caption{Optimality test for the three EFTI estimators, as in Fig.\,\ref{fig: g-err}. For all templates and choices of $\ell_{\rm max}$, the empirical Gaussian errors from $50$ simulations (crosses) are consistent with the Fisher matrix predictions (lines) and, in the ideal limit, the analytic calculations (circles). As for $\gnl$, errors scale as $1/\ell_{\rm max}$, and we find $30-50\%$ tighter errorbars when including polarization. Here, we analyze each template independently; their correlation is assessed in Fig.\,\ref{fig: efti-corner}. Note that all data-points use the same simulations, leading to correlated noise.}
    \label{fig: efti-err}
\end{figure}

\begin{figure}
    \centering
    \includegraphics[width=0.6\linewidth]{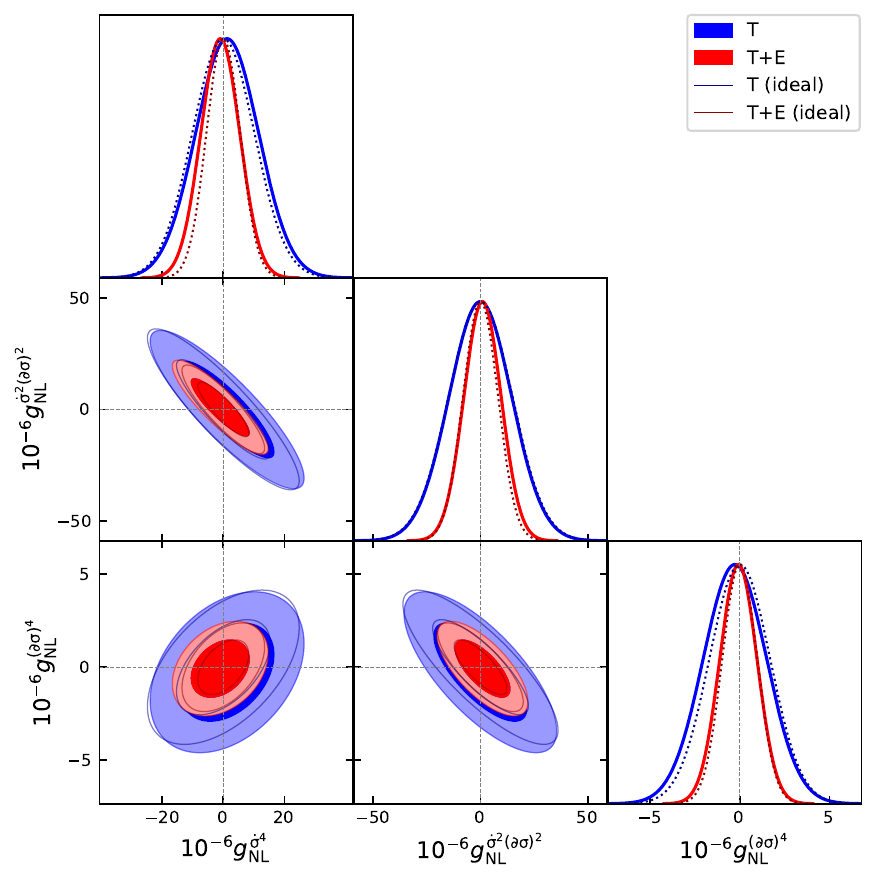}
    \caption{Joint analysis of the three EFTI templates, as applied to the mean of $50$ Gaussian simulations. The blue (red) contours show the empirical posteriors from a temperature-only (temperature-plus-polarization) analysis at $\ell_{\rm max}=512$ without a mask, and the dashed lines give the theoretical predictions (setting the mean to zero and the covariance to $\F^{-1}$). We find good agreement between theory and simulation, and observe strong correlations between the templates \citep[cf.][]{2015arXiv150200635S,Philcox4pt1}, particularly those with $\gnldotdel$. The constraints benefit strongly from $E$-modes; these have not been included in previous analyses.}
    \label{fig: efti-corner}
\end{figure}

In Fig.\,\ref{fig: efti-err}, we compare the empirical, numerical (Fisher) and analytical (ideal Fisher) variances of the three estimators, ignoring their correlations. In the idealized case, we find excellent agreement between the exact and Monte Carlo estimates of $\F$, validating our Fisher matrix computation. In all cases, the empirical variances are consistent with the theoretical forms (though with a possible slight excess in masked data for $\ell_{\rm max}=512$) -- this implies that the estimator is close to minimum variance and, coupled with the consistency in $\F$ , unbiased. We observe an approximate $\sigma(g_{\rm NL})\sim \ell_{\rm max}^{-1}$ scaling, as expected for equilateral templates \citep{Kalaja:2020mkq}, and find a $30-50\%$ gain from the inclusion of polarization on small scales.

Next, we perform a joint Gaussian analysis of the three EFTI templates at $\ell_{\rm max}=512$, with results displayed in Fig.\,\ref{fig: efti-corner}. As before, we find excellent agreement between the empirical and theoretical distributions, and significantly narrower contours in the joint temperature-plus-polarization analysis. The three templates are highly correlated, with the marginalized constraints weaker than the unmarginalized forms by a factor of $2.4-4.9$. This agrees with previous work including \citep{2015arXiv150200635S} and \paperone, and can be ameliorated by analyzing only $\gnldotdot$ and $\gnldeldel$ \citep[cf.,][]{Planck:2015zfm,Planck:2019kim}.
%, which forecasted a $99\%$ correlation between $\gnldotdel$ and some linear combination of $\gnldotdot$ and $\gnldeldel$, and \paperone, which found a $95\%$ degeneracy between $\gnldotdel$ and $\gnldotdot$ in primordial-space. To ameliorate this, the \textit{Planck} papers analyzed only $\gnldotdot$ and $\gnldeldel$ \citep{Planck:2015zfm,Planck:2019kim}. 
We can also reduce degeneracies by projecting onto specific inflationary models, such as the Lorentz Invariant and DBI actions discussed in \paperone.
%: for example, recasting the $T+E$ constraints of Fig.\,\ref{fig: efti-corner} onto the Lorentz Invariant and DBI models discussed in \paperone gives the 95\% one-dimensional posteriors $c_s>0.014$, $-0.23<(H/\Lambda_1)^4<0.19$, where $c_s$ and $\Lambda_1$ are the microphysical parameters of the models. 
Such transformations will be applied in \paperthree to constrain microphysical parameters using \textit{Planck} data.

\subsubsection{Consistency Tests}
\noindent Finally, we test the dependence of our results on hyperparameter choices, focusing on the masked temperature-plus-polarization analysis at $\ell_{\rm max}=512$. Halving $N_{\rm fish}$ induces an error below $0.15\%$ in $1/\sqrt{\F}$, corresponding to $<0.25\%$ in the errorbars and $<0.001\sigma$ in the mean, implying that our Monte Carlo Fisher matrix is highly converged.\footnote{Whilst the variance in $\F$ always scales as $1/N_{\rm fish}$, the prefactor varies between templates (and was much larger for $\taunl$). If we assemble the Fisher matrix in series, the convergence rate can be easily assessed on the fly.} Reducing $N_{\rm disc}$ to $25$ has a bigger effect, leading to a bias of up to $0.34\sigma$ in the mean and a $3\%$ change in the empirical errorbars. As discussed above, using a large value of $N_{\rm disc}$ is important to eliminate any residual disconnected contributions to the trispectrum numerator. 
%Next, we increase $N_{\rm side}$ to $512$ (which can have a non-trivial effect for masked data), this is sources a $\lesssim 0.2\sigma$ variation in the mean and $10\%$ in the errors (but sub-percent in the Fisher matrix); that said, this may be not correspond to a physical shift since the precise realizations of the numerator simulations are different in this case.\footnote{\polyspec generates Gaussian random simulations at fixed $\ell_{\rm max}=3N_{\rm side}$, which we use for removing disconnected contributions biases. For the tests considered herein, we fix the realizations for all analysis at the same $N_{\rm side}$ to allow detailed comparison without sample variance.} 
Finally, we double the $k$-space resolution by increasing $N_k$ to $6877$; this yields consistent results within $0.002\sigma$ in the mean, and $0.1\%$ in the errors and Fisher matrix, implying that the fiducial setting is appropriate.

\subsection{Direction-Dependent Non-Gaussianity}\label{subsec: valid-direc}
\noindent Having validated all of the contact trispectrum estimators (and $\widehat{\tau}_{\rm NL}^{\rm loc}$), we now move to the exchange sector. All of the remaining primordial templates are generalizations of the $\taunl$ shape, designed to explore additional physical and phenomenological regimes. Here, we consider the `direction-dependent' non-Gaussianity templates introduced in \paperone (building on \citep{Shiraishi:2013oqa,Shiraishi:2016mok}), which have the schematic form
\beq
    T_\zeta \sim \sum_{n_1n_3n}\tau_{\rm NL}^{n_1n_3n}P_\zeta(k_1)P_\zeta(k_3)P_\zeta(K)\sum_{m_1m_3m}\tj{n_1}{n_3}{n}{m_1}{m_3}{m}Y_{n_1m_1}(\hk_1)Y_{n_3m_3}(\hk_3)Y_{nm}(\hK),
\eeq
where $Y_{nm}(\hk)$ is a spherical harmonic and the parentheses indicate a Wigner $3j$ symbol, encoding a tripolar spherical harmonic (which represents the most general angular dependence, assuming rotation invariance). These are related to the parity-even and parity-odd templates proposed in \citep{Shiraishi:2013oqa,Shiraishi:2016mok}:
\beq
    T_\zeta &\sim& \sum_{n}\left[\tau_{\rm NL}^{n,\rm even}+\tau_{\rm NL}^{n,\rm odd}(\hk_1\cdot\hk_3\times\hK)\right]P_\zeta(k_1)P_\zeta(k_3)P_\zeta(K)\left[\mathcal{L}_n(\hk_1\cdot\hk_3)+\mathcal{L}_n(\hk_1\cdot\hK)+\mathcal{L}_n(\hk_3\cdot\hK)\right]\\\nonumber
\eeq
for Legendre polynomial $\mathcal{L}_n$, which are even and odd under $\vk\to-\vk$, and map to $\tau_{\rm NL}^{n_1n_3n}$ shapes with even and odd $n_1+n_3+n$ respectively. \polyspec includes estimators for $\tau_{\rm NL}^{n_1n_3n}$, $\tau_{\rm NL}^{n,\rm even}$ and $\tau_{\rm NL}^{n,\rm odd}$ for (small) positive integer $n_i$, which we validate below using Gaussian simulations. We assume the fiducial parameters $\{N_{\rm disc}=50, N_{\rm fish}=50, \ell_{\rm min}=2, L_{\rm min}=2, L_{\rm max}=10, f_{\rm thresh} = 10^{-3}, N_{\rm side}=256\}$ (which are validated below), using a weaker optimization threshold than before since (a) the direction-dependent templates are more expensive to optimize and (b) we previously found excellent convergence at $f_{\rm thresh}=10^{-3}$.

\subsubsection{General Direction-Dependence}
\noindent We first validate the $\tau_{\rm NL}^{n_1n_3n}$ estimator by comparing its empirical variance to the Fisher matrix prediction, using $50$ Gaussian simulations. Noting that computational costs scale quadratically with $n_{\rm max}$, we restrict to $n_i\in\{0,1,2\}$, but consider each non-trivial combination of $n_1,n_3,n$ obeying the triangle conditions, split into even and odd $n_1+n_3+n$, which are approximately uncorrelated. Given the eight even and three odd templates, we perform optimization as discussed in \S\ref{sec: code}, using the same integration points for all templates.
%attempting to approximate the two-dimensional $r,r'$ integrals with a small set of sampling points (shared across templates) and associated (template-dependent) weights. 
The algorithm is slow to converge for templates with odd $n_1$ and/or $n_3$ (which we purposefully optimize last), requiring $N_{\rm opt}\approx 100$ basis points to reach $f_{\rm thresh}=10^{-3}$ instead of the $N_{\rm opt}\approx 20$ required for the other templates.%\footnote{To guard against non-convergence, the \polyspec optimization routine terminates if the score has failed to improve by at least $5\%$ across the addition of the last $5$ basis points. We do not require this feature in this work.}

\begin{figure}
    \begin{minipage}[t]{0.49\linewidth}
    %\vspace{0pt} 
    \includegraphics[width=\linewidth]{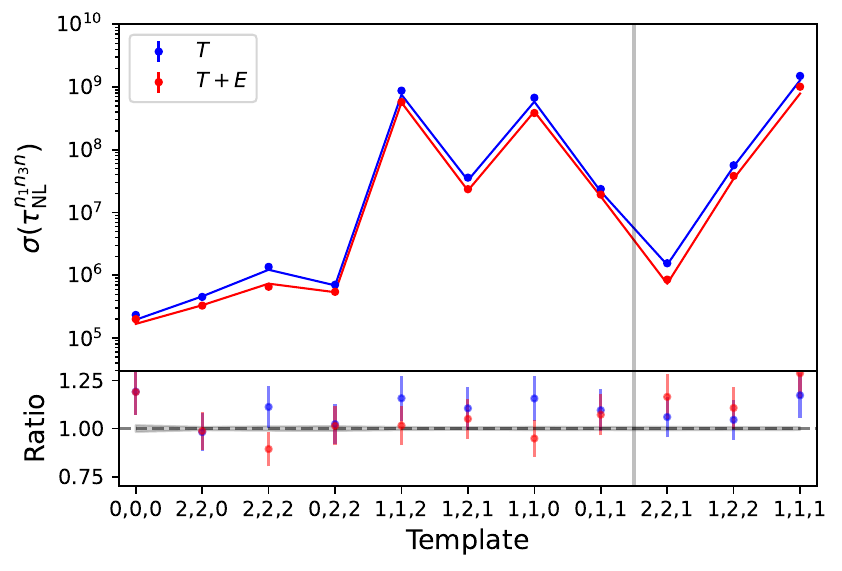}
    \caption{Optimality of the $\tau_{\rm NL}^{n_1n_3n}$ direction-dependent templates. We show results for eight parity-even (left) and three parity-odd (right) templates, analyzed separately, comparing the empirical errors from $50$ unmasked Gaussian simulations (points) to the Fisher matrix prediction (lines). The bottom panel shows the ratio, with the faint gray band indicating the scatter due to noise in $\F$. Whilst we find agreement between simulations and theory in all cases (up to noise), we caution that templates with odd $n_1,n_3$ do not pass all our consistency tests (and are extremely poorly constrained from the data). Correlations between templates are shown in Fig.\,\ref{fig: direc-errs}.}
    \label{fig: direc-errs}
    \end{minipage}
    \hfill
    \begin{minipage}[t]{0.49\linewidth}
    %\vspace{0pt}\raggedright
    \includegraphics[width=1.0\linewidth]{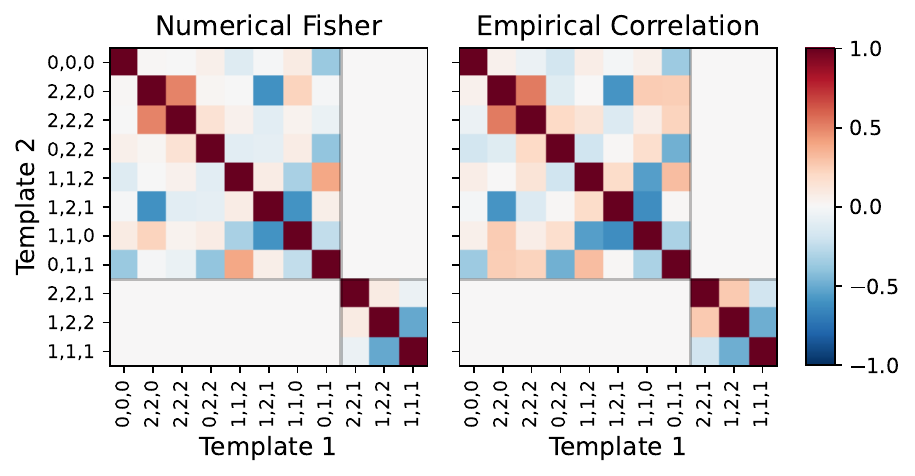}
    \caption{Correlations between the parity-even (top left) and parity-odd (bottom right) $\tau_{\rm NL}^{n_1n_3n}$ direction-dependent trispectrum templates, obtained from a joint analysis of unmasked data at $\ell_{\rm max}=512$, using both $T$- and $E$-modes. We show the Fisher matrix (left) as well as the empirical correlation from  50 Gaussian simulations, which are in good agreement. In order to directly test the Fisher matrix, we do not normalize by $\F^{-1}$; the full correlation matrix is the inverse of the above. Most templates exhibit only weak correlations implying that they can be separately constrained from data.}
    \label{fig: direc-corr}
    \end{minipage}
\end{figure}

Our main results are shown in Fig.\,\ref{fig: direc-errs}, assuming $\ell_{\rm max}=512$ and $L_{\rm max}=10$. Across all of the eleven templates, we find excellent agreement between the empirical and theoretical $\tau_{\rm NL}^{n_1n_3n}$ variances, with a maximal deviation of $1.4\sigma_{\rm mean}$ in the joint temperature-plus-polarization dataset (or $2.1\sigma_{\rm mean}$ from temperature-only analyses). Furthermore, we find that $E$-modes reduce $\sigma(\tau_{\rm NL}^{n_1n_3n})$ by $20-70\%$, depending on $n_1,n_3,n$. Although the various templates are phenomenologically similar, their detectability varies dramatically. For example, the $1\sigma$ constraint on $\tau_{\rm NL}^{000}\equiv (4\pi)^{3/2}\taunl$ is $2000\times$ stronger than $\tau_{\rm NL}^{110}$ (which includes the angular factor $\hk_1\cdot\hk_3$) and $6000\times$ stronger than $\tau_{\rm NL}^{111}$ (which contains $\hk_1\cdot\hk_3\times \hK$). Constraints on templates with even $n_1,n_3$ are much more consistent, varying only by a factor of four across $\tau_{\rm NL}^{000},\tau_{\rm NL}^{220},\tau_{\rm NL}^{222},\tau_{\rm NL}^{022}$ and $\tau_{\rm NL}^{221}$. This arises from geometric restrictions; templates with odd $n_1,n_3$ vanish in the exact squeezed limit. Moreover, the poorly constrained templates are precisely those that were difficult to optimize. 

In Fig.\,\ref{fig: direc-corr}, we test the Fisher matrix by comparing it to the empirical correlation obtained from the numerators of 50 Gaussian simulations. Up to noise fluctuations, the matrices are in good agreement, implying that the inverse of $\F$ can correctly capture the structure of the covariance as well as its diagonal. Focusing on the well-behaved templates (even $n_1,n_3$), we find generally weak ($<10\%$) correlations between shapes, except for a $56\%$ anti-correlation between $\tau_{\rm NL}^{220}$ and $\tau_{\rm NL}^{222}$ (obtained from the inverse of Fig.\,\ref{fig: direc-corr}). This implies that the amplitudes can be independently extracted from data (as well as used to constrain specific model amplitudes, including microphysical parameters of solid inflation and $U(1)$ gauge fields \citep[e.g.,][]{Shiraishi:2016mok,Bartolo:2014xfa}). The odd-$n_1,n_3$ shapes have similar correlations, with $\approx 50\%$ correlation obtained for the $(\tau_{\rm NL}^{1,2,1},\tau_{\rm NL}^{1,1,0})$ and $(\tau_{\rm NL}^{1,2,2},\tau_{\rm NL}^{1,1,1})$ pairs.

Lastly, we perform hyperparameter tests. Running the optimization algorithm using $N_{\rm fish}^{\rm opt}=1$ Monte Carlo realizations to compute the idealized Fisher matrix (instead of the fiducial $N_{\rm fish}^{\rm opt}=5$) leads to a $\approx 0.4\%$ ($<0.1\%$) variation in the Fisher matrix and a $0.03\sigma$ ($0.01\sigma$) shift in the estimator mean across all (even $n_1,n_3$) templates; moreover, doubling the number of points in the initial radial grid leads to a shift below $1.6\%$ ($1.2\%$) in $\F$ or $0.09\sigma$ in the mean ($0.02\sigma$). This indicates that our optimization procedure is converged. Furthermore, the mean and Fisher matrices vary by just $0.006\sigma$ ($0.006\sigma$) and $0.9\%$ ($0.9\%$) when reducing to $N_{\rm fish}=25$. Finally, we double the $k$-space resolution by increasing $N_k$ to $6877$. Templates with even $n_1,n_3$ are highly stable, with maximal shifts of $0.02\sigma$ in the mean and $1.6\%$ in the Fisher matrix. For odd $n_1,n_3$ templates, we find variation in $\F$ by up to $19\%$
%(though $N_{\rm opt}$ is reduced by a factor of two), 
and we do not find convergence when increasing to $N_k=13805$.
% (except at the $\mathcal{O}(1)$ level).
This indicates that templates with odd $n_1,n_3$ cannot be meaningfully constrained from the CMB 
%(with constraints non-trivially biased by sampling noise in the integrals), 
and justifies our previous difficulties in finding optimized representations, as well as the large variances seen in Fig.\,\ref{fig: direc-errs}. For these reasons, we will not attempt to constrain such templates in \paperthree. 

\subsubsection{Even \& Odd Templates}

\noindent As discussed in \paperone, the optimal estimators for $\tau_{\rm NL}^{n,\rm even}$ and $\tau_{\rm NL}^{n,\rm odd}$ can be formed as a linear combination of the $\tau_{\rm NL}^{n_1n_3n}$, and thus are implicitly validated by the above tests. For completeness, we perform additional tests in this section, which allow us to verify the scalings with $\ell_{\rm max}$ discussed in the literature. In the previous section, we found that $\tau_{\rm NL}^{n_1n_3n}$ amplitudes with odd $n_1,n_3$ are difficult to constrain; these correspond to $\tau_{\rm NL}^{n,\rm even}$ with odd $n$ and $\tau_{\rm NL}^{n,\rm odd}$ with even $n$. Here, we estimate $\tau_{\rm NL}^{n,\rm even}$ with $n\in\{0,1,2,3\}$ and $\tau_{\rm NL}^{n,\rm odd}$ with $n\in\{0,1,2\}$, extending beyond the previous analysis by including order-three spherical harmonics.
%(noting that $\tau_{\rm NL}^{\rm odd}$ contains $\tau_{\rm NL}^{n_1n_3n}$ with $n_{1,3}^{\rm max}=n+1$).
As expected, the well-behaved even (odd) templates with even (odd) $n$ are easy to optimize, requiring $N_{\rm opt}\approx 15-30$ terms, whilst the other templates require $N_{\rm opt}\approx 100$. We find a similar dependence on $k$-resolution to before: doubling $N_k$ leads to consistent results within $0.01\sigma$ (mean) and $0.1\%$ (Fisher) for the well-behaved templates, but $\approx 0.2\sigma$ and $10-20\%$ variation for the other templates, indicating lack of convergence.

\begin{figure}
\begin{minipage}[t]{0.46\linewidth}
    %\vspace{0pt}
    \centering
    \includegraphics[width=\linewidth]{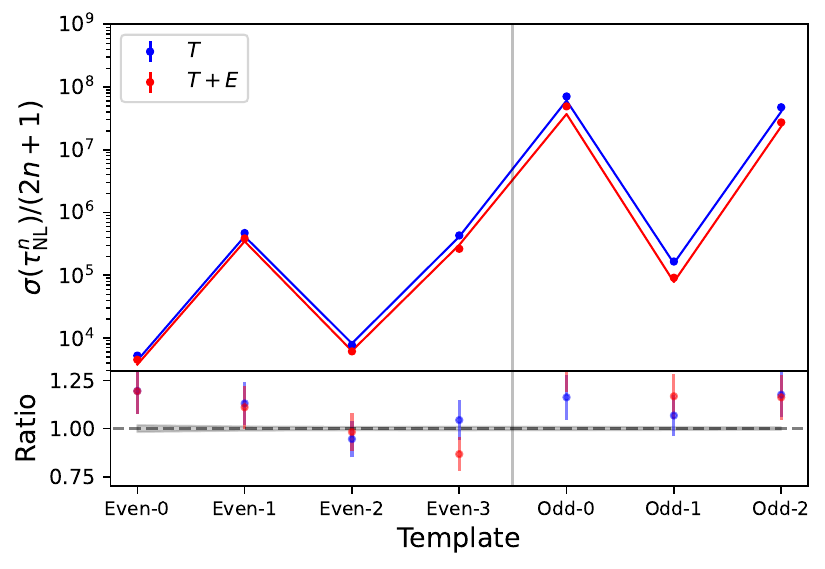}
    \caption{Optimality of the $\tau_{\rm NL}^{n,\rm even}$ and $\tau_{\rm NL}^{n,\rm odd}$ estimators, as in Fig.\,\ref{fig: direc-errs}. The left (right) panel shows even (odd) templates, analyzed separately, as a function of $n$, and we normalize by $1/(2n+1)$ to facilitate comparison with Fig.\,\ref{fig: direc-errs}. Only even templates with even $n$ and odd templates with odd $n$ can be meaningfully constrained from the data. All results are obtained at $\ell_{\rm max}=512$ without a mask.}
    \label{fig: err-even-odd}
\end{minipage}
\hfill
\begin{minipage}[t]{0.52\linewidth}
    %\vspace{0pt}\raggedright
    \centering
    \includegraphics[width=\linewidth]{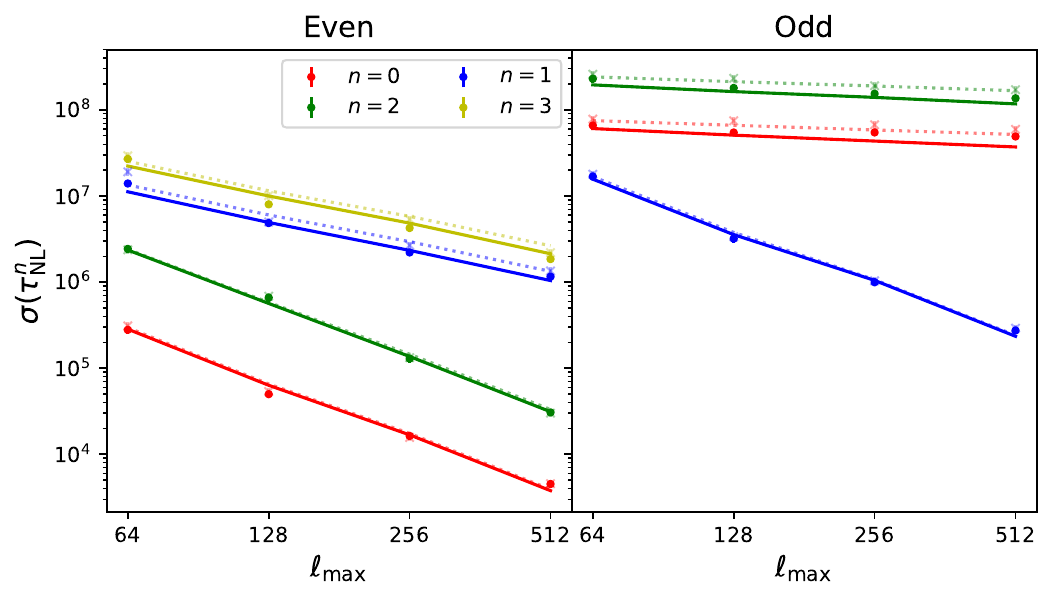}
    \caption{Scaling of the parity-even and parity-odd $\tau_{\rm NL}$ constraints with $\ell_{\rm max}$, assuming an unmasked temperature-plus-polarization analysis. Solid and dashed lines indicate results computed using $L_{\rm max} = 10$ and $5$ respectively, and errorbars show the results from 50 Monte Carlo simulations. The $\ell_{\rm max}=512, L_{\rm max}=10$ are equivalent to those of Fig.\,\ref{fig: err-even-odd}. We find a strong $1/\ell_{\rm max}^2$ scaling for the `well-behaved' templates (parity-even with even $n$ and parity-odd with odd $n$), but a much weaker scaling for the other terms.}
    \label{fig: err-even-odd-scaling}
\end{minipage}
\end{figure}

In Fig.\,\ref{fig: err-even-odd}, we compare the variances of each template using the fiducial scale-cuts of $\ell_{\rm max}=512$ and $L_{\rm max}=10$. We find similar conclusions to above: the empirical and theoretical standard deviations agree (with at most $1.6\sigma_{\rm mean}$ discrepancy across the well-behaved templates), implying that the estimator is close-to-optimal. As expected, the well-behaved terms $\tau_{\rm NL}^{0,\rm even},\tau_{\rm NL}^{2,\rm even},\tau_{\rm NL}^{1,\rm odd}$ are the best constrained, with errors on the other amplitudes larger by several orders of magnitude.\footnote{Note that the normalizations of $\tau_{\rm NL}^{n_1n_3n}$ and $\tau_{\rm NL}^{n,\rm even/odd}$ differ, with, for example, $\tau_{\rm NL}^{n,\rm even}$ containing an extra factor of $1/(2n+1)$ relative to $\tau_{\rm NL}^{nn0}$.} The utility of $E$-modes varies between templates; in general we find a larger impact for parity-odd templates than for parity-even shapes ($60-90\%$ improvement in $\sigma(\tau_{\rm NL})$, compared to $15-35\%$).
%), with the $\tau_{\rm NL}^{0,\rm even}$ result matching that found earlier for $\taunl$.

Fig.\,\ref{fig: err-even-odd-scaling} shows the dependence of the errorbars on scale-cuts. We find a clear bifurcation: well-behaved templates exhibit $1/\ell_{\rm max}^2$ scalings (as for $\taunl$), whilst the other terms show weaker dependence on resolution, approximately $1/\ell_{\rm max}$ for $\tau_{\rm NL}^{n,\rm even}$ and closer to constant for $\tau_{\rm NL}^{n,\rm odd}$. This occurs due to the cancellations inherent in such templates in the squeezed limit
%(and through the projection integrals) 
and is in agreement with the analytic results of \citep{Shiraishi:2013oqa,Shiraishi:2016mok}. Changing from $L_{\rm max}=10$ to $L_{\rm max}=5$ increases the well-behaved errorbars by $\lesssim 5\%$, but leads to $\approx 20-100\%$ inflation for the other shapes, implying that they have significant signal-to-noise outside the collapsed limit, again matching \citep{Shiraishi:2013oqa,Shiraishi:2016mok}.\footnote{Whilst one could use a larger $L_{\rm max}$ to constrain these shapes (e.g., $L_{\rm max}=\ell_{\rm max}$); however, this departs from the spirit of our direction-dependent templates and is expensive to optimize.} Our conclusion is that only the parity-even templates with even $n$ and parity-odd templates with odd $n$ can be meaningfully constrained from the collapsed CMB trispectrum.%, though there is little hope for constraining the other templates in this manner (as expected from \citep{Shiraishi:2013oqa,Shiraishi:2016mok}).

\subsection{Collider Non-Gaussianity}
\noindent Our final set of primordial templates correspond to massive particles in inflation, modeled using the `cosmological collider' formalism. Despite extensive theoretical work on their inflationary signatures, these models have seen almost no application to data, except for the first bispectrum analyses in \citep{Cabass:2024wob,Sohn:2024xzd} and a primitive study of the galaxy trispectrum in \citep{Cabass:2022oap}. We consider two types of model: \textbf{light} -- intermediate and light particles with $m\lesssim H$ (formally the complementary series); \textbf{heavy} -- high-mass particles with $m\gtrsim H$ (formally the principal series), for inflationary Hubble scale $H$.\footnote{The midpoint occurs at $m=3H/2$ (spin-zero) or $m = (s-1/2)H$ (spin-$s$), which corresponds to the conformally-coupled regime.} In the collapsed limit, spin-zero particles correspond to the following schematic trispectra:
\beq\label{eq: template-heavy}
    T_\zeta &\sim & \tau_{\rm NL}^{\rm light}(0,\nu_0)\left(\frac{K^2}{k_1k_3}\right)^{3/2-\nu_0}P_\zeta(k_1)P_\zeta(k_3)P_\zeta(K)\\\nonumber
    &\sim& \tau_{\rm NL}^{\rm heavy}(0,\mu_0)\left(\frac{K^2}{k_1k_3}\right)^{3/2}\cos\left[\mu_0\log\frac{K^2}{k_1k_3}+\phi_0\right]P_\zeta(k_1)P_\zeta(k_3)P_\zeta(K),
\eeq
where $\nu_0 = i\mu_0 = \sqrt{9/4-m^2/H^2}$ is the mass parameter and $\phi_0$ is some model-dependent phase. These generalize the local exchange model featuring additional scale-dependence and oscillations. Higher-spin particles, defined by mass parameter $\nu_s = i\mu_s = \sqrt{(s-1/2)^2-m^2/H^2}$, generate similar trispectra, but with additional angular factors \citep{Philcox4pt1,Arkani-Hamed:2015bza}.

\polyspec includes estimators for both $\tau_{\rm NL}^{\rm light}$ and $\tau_{\rm NL}^{\rm heavy}$ for positive integer spin $s$ and arbitrary $\nu_s, \mu_s$ (subject to appropriate physical bounds). These are similar to the $\taunl$ estimator, but incorporate angular factors analogous to the direction-dependent trispectra, and, for heavy particles, complex fields.
%(since we expand $2\cos\mu\log \kappa = \kappa^{i\mu}+\text{c.c.}$). 
As discussed in \paperone, the theoretical trispectra are valid only in the collapsed limit, thus we must restrict to $K\lesssim  k$. This is achieved by imposing the separable condition $k>k_{\rm coll}, K<k_{\rm coll}$ for some appropriate $k_{\rm coll}$, whose value is discussed below. To further limit contamination from non-collapsed regimes, we marginalize over the three EFTI templates and carefully assess the dependence of our results on $L$-space cuts (noting that $L_{\rm max}\sim k_{\rm max}\chi_\star$).
%Importantly, this is separable in $k_i$, thus our estimators can be efficiently implemented. 

% In the subsections below, we validate the collider estimators, focusing first on the spin-zero estimator, then generalizing to higher spin. %, testing the estimators against Gaussian random fields using a variety of hyperparameters and scale cuts, as before. 
% Given the direction-dependent results above (and the theoretical discussions in \citep[e.g.,][]{MoradinezhadDizgah:2018ssw,Arkani-Hamed:2015bza}), we expect that even-spin particles will be easier to constrain than those of odd-spin. 

\subsubsection{\texorpdfstring{Dependence on $k_{\rm coll}$}{Dependence on Cut-Off Scale}}

\begin{figure}
    \centering
    \includegraphics[width=0.8\linewidth]{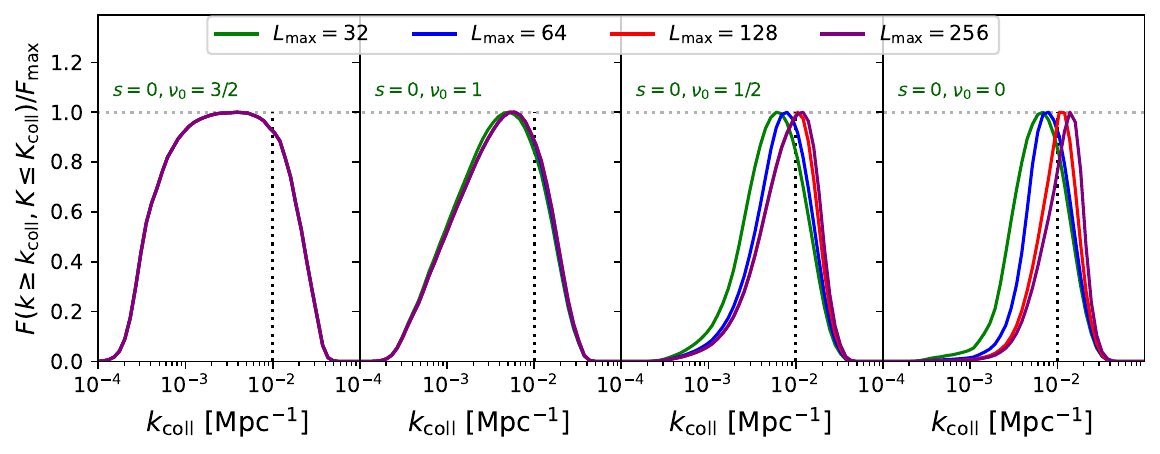}\\
    \includegraphics[width=0.8\linewidth]{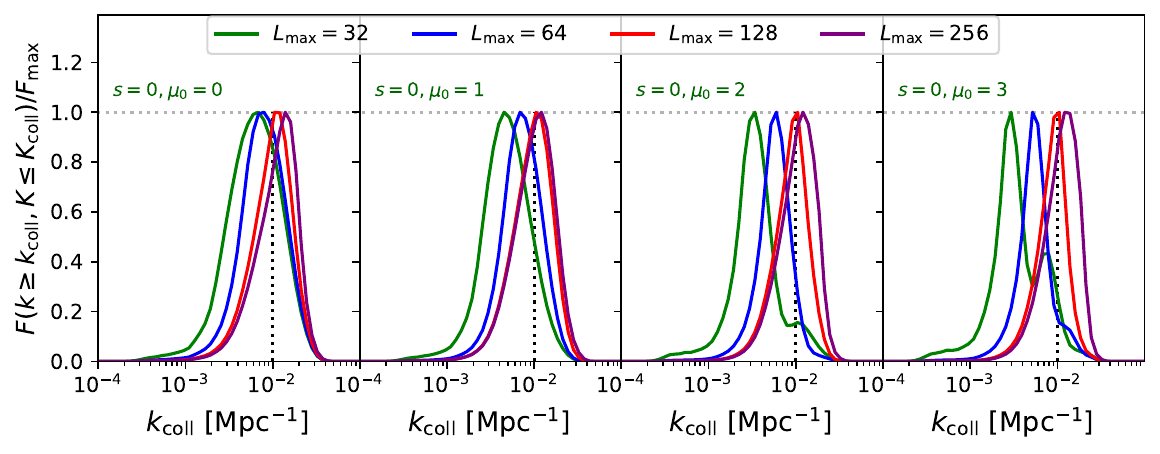}
    \caption{Dependence of the light (top) and heavy (bottom) spin-zero cosmological collider constraints on the truncation scale $k_{\rm coll}$. We display CMB Fisher forecasts for the information content truncated to modes with $k>k_{\rm coll}$ and $K<k_{\rm coll}$ (corresponding to collapsed configurations), for various choices of mass parameters (marked in green). We normalize to the maximum Fisher matrix obtained from these tests, and give results for various choices of $L_{\rm max}$ (indicated by colors). All results are obtained from a temperature-plus-polarization analysis at $\ell_{\rm max}=512$, and projected out the three EFTI shapes. The vertical dashed line shows the adopted value of $k_{\rm coll}$, which captures the majority of the signal-to-noise for all values of $\nu_0,\mu_0$.}
    \label{fig: kcoll-dependence}
\end{figure}

\noindent Choosing $k_{\rm coll}$ requires balancing two constraints: (1) low $k_{\rm coll}$ leads to less contamination from equilateral regimes (since the external legs are dominated by large $\ell \sim k\chi_\ast$); (2) high $k_{\rm coll}$ leads to increased signal-to-noise. To assess this, we perform a suite of spin-zero Fisher forecasts (\textit{i.e.}\ we compute $\F$) for various choices of $k_{\rm coll}$, template, and $L_{\rm max}$, as shown in Fig.\,\ref{fig: kcoll-dependence}.
% obtaining the results ts for both light and heavy spin-zero templates, for various values of $\nu_0,\mu_0$ and $L_{\rm max}$ (which sets the squeezing of the tetrahedron in harmonic-space), comparing the Fisher matrix at a given $k_{\rm coll}$ to the maximal value. 
For light particles with $m\to 0, \nu_0\to 3/2$ (which asymptote to $\taunl$), we find that the constraints are broadly insensitive to $k_{\rm coll}$. This is expected: the collapsed limit scales as $(K^2/k_1k_3)^{3/2-\nu_0}$ \eqref{eq: template-heavy}, thus the signal-to-noise is dominated by the lowest $K$-modes and highest $k$-modes.\footnote{Even stronger scalings can be obtained if one allows for inflationary ``tachyonic'' particles with $\nu_0<0$ \citep{McCulloch:2024hiz}.} As $\nu_0$ decreases, the function becomes more sharply peaked, indicating that the signal-to-noise is principally concentrated in ``not-too-collapsed'' regimes (though still non-equilateral, since we never allow for $K\gtrsim k$). This story is corroborated by the $L_{\rm max}$-dependence: whilst the results are invariant at large $\nu_0$, the peak $k_{\rm coll}$ shift upwards as $L_{\rm max}$ increases, since signal-to-noise can be extracted from smaller scales.

Due to the $(K^2/k_1k_3)^{3/2}$ scaling in \eqref{eq: template-heavy}, results for heavy particles are similar to the light-particle results with $\nu_0\to 0$. 
%This is expected since the amplitude of the signal scales as $(K^2/k_1k_3)^{3/2}$ in all cases, which is not strongly peaked at $K\ll k$ unlike for $\nu_0\to 3/2$. 
Noting that our restrictions on $k,K$ already restrict to `quasi-collapsed' regimes (particularly given that we set $L_{\rm max}<\ell_{\rm max}$ and marginalize over the EFTI shapes), our goal is to find a value of $k_{\rm coll}$ that preserves most of the information content across the various templates. Motivated by Fig.\,\ref{fig: kcoll-dependence}, we use $k_{\rm coll}=0.01\mathrm{Mpc}^{-1}$ for the remainder of this work, which captures the majority ($\gtrsim 90\%$) of the Fisher information for all templates with $L_{\rm max}\gtrsim \ell_{\rm max}/4$. 
%We additionally marginalize over the three EFTI shapes \S\ref{subsec: valid-efti} to remove any residual equilateral behavior, which is not specific to the collider phenomenology. 
As shown in Fig.\,\ref{fig: collider-ratio-Lmax}, none of the results are highly sensitive to $L_{\rm max}$, implying that our scale-cuts are appropriate. 

\subsubsection{Spin-Zero Results}

\begin{figure}
    \centering
    \includegraphics[width=0.6\linewidth]{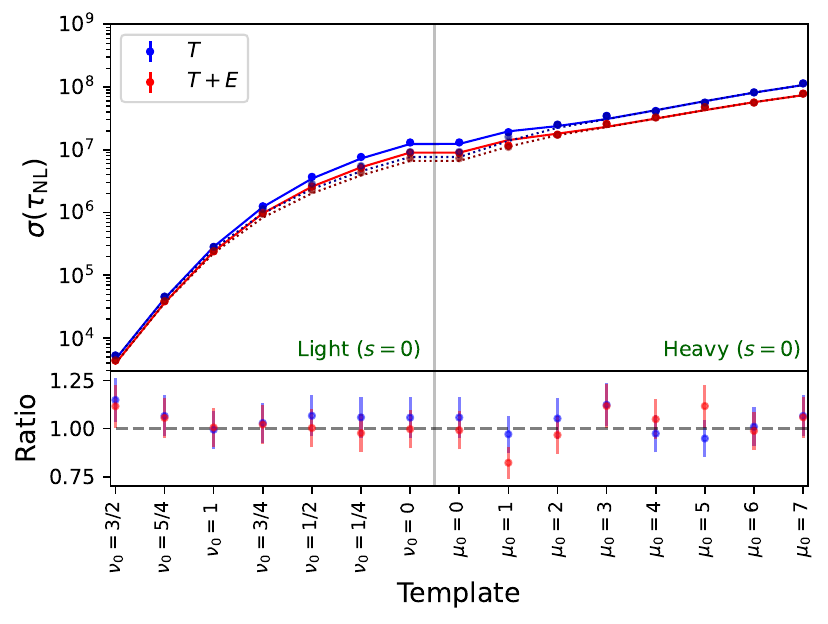}
    \caption{Optimality of the spin-zero cosmological collider estimators, as in Fig.\,\ref{fig: direc-errs}. We show results for light ($m\leq 3H/2$, $\nu_0 = \sqrt{9/4-m^2/H^2}$) and heavy ($m\geq 3H/2$, $\mu_0 = \sqrt{m^2/H^2-9/4}$) particles on the left and right respectively, analyzing each template independently. The two templates are coincident for conformal particles (with $m=3H/2$, $\nu_0 = \mu_0 = 0$). All results are computed for unmasked data at $\ell_{\rm max}=512$ and $L_{\rm max}=256$. The dashed lines indicate constraints without marginalization over the EFTI shapes -- this predominantly impacts shapes with $m\approx 3H/2$. The bottom panel shows the ratio of empirical to theoretical errors; these are consistent with unity showing that our estimators are close to optimal. 
    %The strong scaling of $\sigma(\tau_{\rm NL}^{\rm light})$ with $\nu_0$ occurs is partly caused by our (equilateral-focused) normalization scheme.
    }
    \label{fig: collider-errs}
\end{figure}

\noindent We now validate the spin-zero estimators using Gaussian simulations. Our key results are shown in Fig.\,\ref{fig: collider-errs}, comparing the empirical and theoretical errorbars on $\tau_{\rm NL}^{\rm light, heavy}$ across a wide range of mass parameters $\nu_0$ and $\mu_0$. The two sets of errorbars agree within $1.8\sigma$ (accounting for the expected scatter), validating our estimators. The addition of polarization leads to significantly tighter constraints, with $\sigma(\tau_{\rm NL})$ reducing by $20\%$ for local-like templates \resub{($\nu_0\gtrsim 3/4$)} and $40\%$ for the remaining templates.

We find strong variation with mass: at low-$\nu_0$, the collapsed limit strongly diverges (particularly for $\nu_0>3/4$ \citep[cf.][]{Kalaja:2020mkq,Bordin:2019tyb}), leading to tight constraints on $\tau_{\rm NL}^{\rm light}$, whilst the $(K^2/k_1k_3)^{3/2}$ scaling for $\nu_0\to0$ gives a much weaker constraint on Hubble-scale particles. Constraints on heavy particles vary more slowly with mass, though there is some loss of constraining power at high-$\mu_0$ due to the `wash-out' of high-frequency oscillations.
%. This occurs since the oscillations are `averaged out' due to the high frequency (scaling as $\mu_s\log(K^2/k_1k_3)$), given the limited $k$-resolution of the experiment. 
Omitting the marginalization over the equilateral EFTI templates, constraints on conformally coupled particles ($\nu_0,\mu_0\to 0$) tighten by $40\%$; this matches expectations since these collider shapes are featureless and not strongly divergent. In contrast, EFTI marginalization in low-mass ($\nu_0\to 3/2$) or high-mass ($\mu_0\gg 0$) regimes, due to the minimal overlap of these templates with the equilateral shapes.
These results agree schematically with the idealized CMB and 21cm forecasts of \citep{Kalaja:2020mkq,Floss:2022grj}.
%., as well as the 21cm (and three-dimensional) forecasts given in \citep{Floss:2022grj}. 
%weakened by marginalization over the equilateral EFTI templates (implying that this helps to remove any non-collapsed signatures from the templates), with an inflation of up to $40\%$ in $\sigma(\tau_{\rm NL})$, but with negligible impact on low-mass ($\nu\to0$) or very high-mass $\mu\gg 0$) templates. Once again this matches our expectation: low-mass templates features divergences and high-mass templates feature fast oscillations, neither of which lead to significant overlap with the equilateral shapes.

\begin{figure}[!t]
    \centering
    \includegraphics[width=0.7\linewidth]{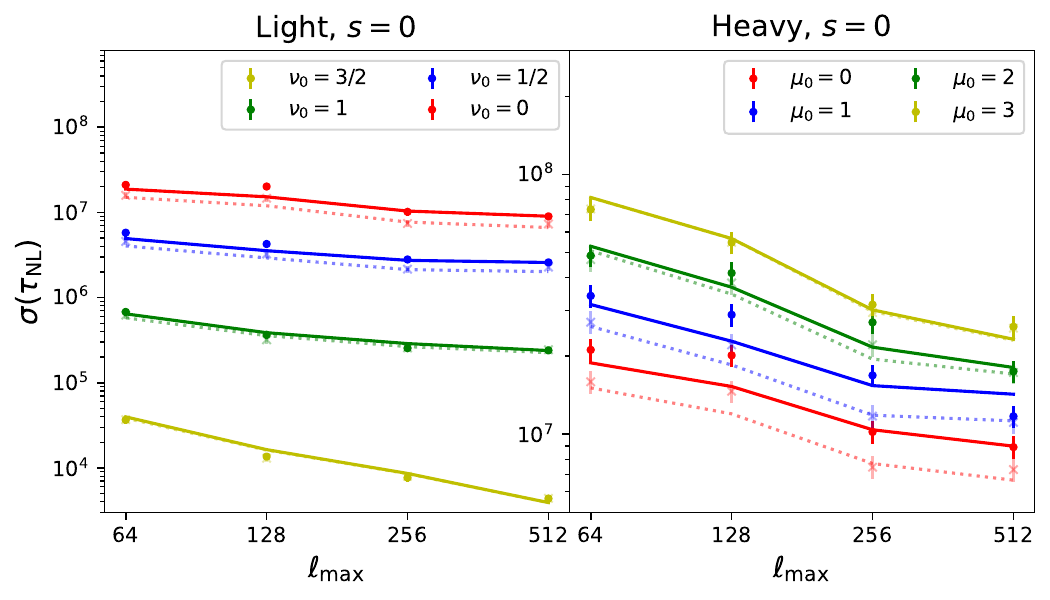}
    \caption{Dependence of the spin-zero collider non-Gaussianity constraints on the maximum external mode, $\ell_{\rm max}$, fixing the maximum internal leg to $L_{\rm max}=\ell_{\rm max}/2$. As in Fig.\,\ref{fig: collider-errs}, we show both empirical (points) and theoretical (lines) variances computed from unmasked temperature-plus-polarization data, with the EFTI-unmarginalized results shown in crosses and dashed lines. For low-mass particles with $\nu_0\to3/2$, the errorbars scale as $\ell_{\rm max}^{-2}$, but we find a weaker scaling for higher-mass particles, approaching $\sigma(\tau_{\rm NL})\sim \ell_{\rm max}^{-1}$ in the conformal limit ($\nu_0=\mu_0=0$). Note that the two panels have different vertical scales.}
    \label{fig: collider-errs-lmax}
\end{figure}

In Fig.\,\ref{fig: collider-errs-lmax}, we demonstrate the dependence of the constraints on $\ell_{\rm max}$. The $\nu_0 = 3/2$ scalings match those of $\taunl$, with $\sigma(\tau_{\rm NL})\sim \ell_{\rm max}^{-2}$ due to the enhanced collapsed limit. As $\nu_0$ decreases, this becomes more muted (matching \citep{Kalaja:2020mkq,Bordin:2019tyb}), and approaches $\sigma(\tau_{\rm NL})\sim \ell_{\rm max}^{-1}$ scaling in the conformal limit. For heavy templates, we find a similar scaling to that of $\nu_0=0$ (expected given the divergence properties), though some enhancement for the highly oscillatory templates with large $\mu_0$.% for heavy templates is similar to the light conformal template, as expected (noting that these have the same shapes, just with additional oscillations). 

\begin{figure}
    \begin{minipage}{0.49\linewidth}
    \centering
    \includegraphics[width=\linewidth]{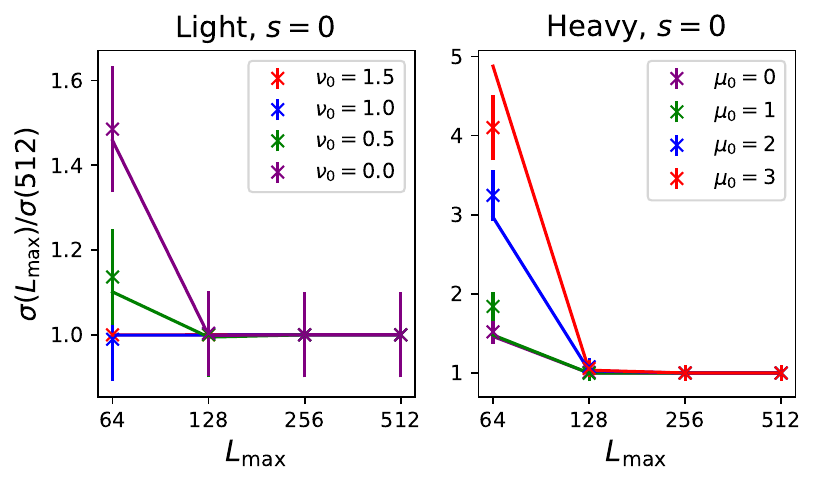}
    \caption{Dependence of the spin-zero collider constraints on the maximum internal mode, $L_{\rm max}$. We plot the ratio of the empirical (points) and theoretical (lines) errors as a function of $L_{\rm max}$ for fixed $\ell_{\rm max}=512$, using the same dataset as in Fig.\,\ref{fig: collider-errs-lmax} (without accounting for sample variance cancellation). In the conformal limit of \resub{$\nu_0\to0$ for light particles, as well as for heavy particles at all $\mu_0$,} %,\mu_0\to 0$, 
    restricting to $L_{\rm max}<128$ causes significant loss of information, but we find negligible changes at larger $L_{\rm max}$. This implies that our $k$-space truncation restricts the analysis to collapsed configurations, as desired.}
    \label{fig: collider-ratio-Lmax}
    \end{minipage}
    \hfill
    \begin{minipage}{0.49\linewidth}
    \centering
    \includegraphics[width=\linewidth]{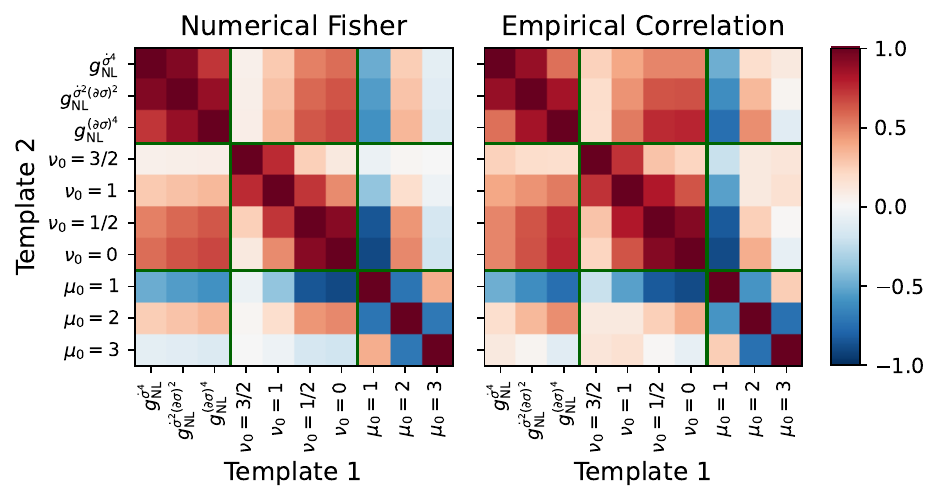}
    \caption{Correlations between the spin-zero cosmological collider and EFTI amplitudes. We compare the numerical Fisher matrix (left) to the empirical correlation obtained from the estimator numerators; as in Fig.\,\ref{fig: direc-corr}, the full correlation matrix is the inverse of this. The EFTI, light collider and heavy collider templates are shown in the top left, middle, and bottom right respectively. We find good agremeent between theoretical and empirical matrices, and note 
    % uses the same results as in Fig.\,\ref{fig: collider-errs-lmax}, but features a joint analysis of all templates, including the equilateral EFTI shapes (top left), light collider shapes (middle) and heavy collider shapes (bottom right). As in Fig.\,\ref{fig: direc-corr}, we show both theoretical and empirical Fisher matrices (showcasing the structure of the inverse covariance), which are in good agreement. We find 
    significant positive (negative) correlations between shapes with similar $\nu_0$ (similar $\mu_0$).}
    \label{fig: collider-corr-plot}
    \end{minipage}
\end{figure}

Fig.\,\ref{fig: collider-ratio-Lmax} shows the corresponding dependence on the internal scale-cut. For light templates with $\nu_0\gtrsim 1$, we find negligible dependence on $L_{\rm max}$; this matches the conclusion of Fig.\,\ref{fig: tau-err}. As the mass increases, larger $L$-modes become progressively more important, as seen from the ratio of the $L_{\rm max}=64$ results to obtained with $L_{\rm max}=512$. This is particularly true for the heavy collider, with the constraints on $\mu_0\gtrsim1$ dominated by $L\gtrsim 64$. %Whilst this demonstrates that 
%hilst changing from $L_{\rm max}=128$ to $L_{\rm max}=512$ negligibly impacts our constraints, we find a fairly large inflation in $\sigma(\tau_{\rm NL})$ when setting $L_{\rm max}=64$. when restricting to $L_{\rm max}$ for the high-mass shapes ($\nu\lesssim 0.5$, as well as the heavy templates). The lack of $L_{\rm max}$-dependence for light $\nu_0$ templates implies that the shapes are dominated by the most collapsed configurations (large $\ell$, small $L$), whilst it is clear that 
%the massive particle analyses extract information from modes beyond the largest scales, 
In all cases, however, we find stable constraints for $L_{\rm max}\geq 128$ ($=\ell_{\rm max}/4$), within $0.4\%$ for $\mu_0<2$, and $8\%$ else. This indicates that our $k$-space restrictions are working as expected, nulling contributions from strongly non-collapsed tetrahedra. This validates the approach discussed in \paperone: we are able to probe collider physics by searching for its distinctive features in quasi-collapsed regimes.
% outside the $L$ templates gain some constraining power from modes outside this regime. However, the fact that $\sigma(\tau_{\rm NL})$ is stable at $L_{\rm max}\geq 128=\ell_{\rm max}/4$ (to within $0.4\%$ for $\mu<2$, and $8\%$ else) for all templates indicates that our 

Next, we consider the correlation between each of the spin-zero trispectrum templates. As shown in Fig.\,\ref{fig: collider-corr-plot}, we find good agreement between the numerical Fisher matrix and the empirical prediction (from the estimator numerator), further validating our approach. The precise correlations match the earlier discussion (see also \citep{Floss:2022grj}): the three EFTI shapes are strongly correlated, and the EFTI shapes exhibit $\gtrsim50\%$ degeneracies with the collider templates around the conformal limit, but only weak correlations in the high- and low-mass limits. As predicted in \paperone, particles with similar mass are hard to distinguish; however, the correlations drop to $\lesssim 30\%$ for $|\nu_1-\nu_2|\gtrsim 1/2$. Heavy and light particles are only weakly correlated (except for $\nu_0\to 0$), and neighboring heavy templates are anticorrelated, due antiphase oscillations.
%in the templates, and may promote a slightly narrower sampling in $\mu_0$ (since the correlation must fall to zero at some $|\mu_1-\mu_2|$). 
Whilst this demonstrates that performing joint analyses of multiple templates at similar $\nu_0,\mu_0$ will lead to a significant loss of constraining power (and motivates a more Bayesian analysis scheme), it is clear that CMB trispectra can be used to probe distinct regimes (e.g., massless, conformal, and highly massive particles). As we shown in \paperthree, these correlations are reduced at higher $\ell_{\rm max}$, further aiding practical analyses.% can be separately probed in teh CMB statistics.% remains hope to distinguishing the CMB signatures of particles across the mass spectrum allowed by inflation. 

As a final test of the spin-zero results, we perform consistency checks as described above. Reducing to $N_{\rm fish}$ yields negligible shifts in the mean of 50 simulations and the Fisher matrix, with maximal deviations of $0.1\sigma$ and $0.8\%$ respectively. Setting $N_{\rm disc}=25$ has a larger effect, yielding shifts up to $0.25\sigma$ in the mean -- this re-emphasizes the importance of carefully subtracting the disconnected term. By reducing $f_{\rm thresh}$ to $10^{-4}$ and doubling the number of initial radial integration points, we find that the optimization algorithm is well-converged: all results are stable within $0.03\sigma$ (mean) and $0.3\%$ (Fisher). A similar conclusion holds for the $k$-resolution, with $0.1\sigma$ and $1.2\%$ consistency found when doubling $N_k$.

% In each of the below tests, we report the maximal deviation in the mean in $\sigma$ units (averaged across $50$ realizations) and the Fisher matrix in $\%$ for light (heavy) templates. Reducing to $N_{\rm fish}=25$ gives negligible shifts of $0.006\sigma$ \& $0.8\%$ ($0.00\sigma$ \& $0.07\%$), whilst reducing to $N_{\rm disc}=25$ leads to a larger shift of $0.24\sigma$ ($0.25\sigma$), implying that it is important to carefully subtract off the Gaussian contributions (though this shift is $\sqrt{50}$ times smaller in a single data analysis). Reducing the optimization tolerance to $f_{\rm thresh}=10^{-4}$ affects results by at most $0.01\sigma$ \& $0.06\%$ ($0.01\sigma$, $0.2\%$), whilst doubling the number of initial radial points gives $0.03\sigma$ \& $0.3\%$ ($0.02\sigma, 0.1\%$) agreement, showing that our optimization procedure is highly stable. Finally, doubling $N_k$ gives consistent results to $0.01\sigma$ \& $0.4\%$ ($0.14\sigma$ \& 1.2\%), indicating convergence.

\subsubsection{Higher-Spin Results}
\noindent The most complex templates included in \polyspec are the light and heavy collider templates with spin $s>0$. Due to the Higuchi bound, $m^2/H^2\geq s(s-1)$ \citep{Higuchi:1986py}, which implies that the light template is restricted to $\nu_s\in[0,1/2]$,
%\equiv \sqrt{(s-1/2)^2-m^2/H^2}\in[0,1/2]$, which 
avoiding the local-type divergence at $\nu_0\to 3/2$. As such, we expect that higher-spin trispectra will be harder to constrain; moreover, odd spin particles have a cancellation in the collapsed limit, which suppresses their collapsed-limit scaling by a factor of $K^2/k_1k_3$ \citep{Arkani-Hamed:2015bza,Lee:2016vti,MoradinezhadDizgah:2018ssw}.

\begin{figure}[!t]
    \begin{minipage}{0.49\linewidth}
    \centering
    \includegraphics[width=\linewidth]{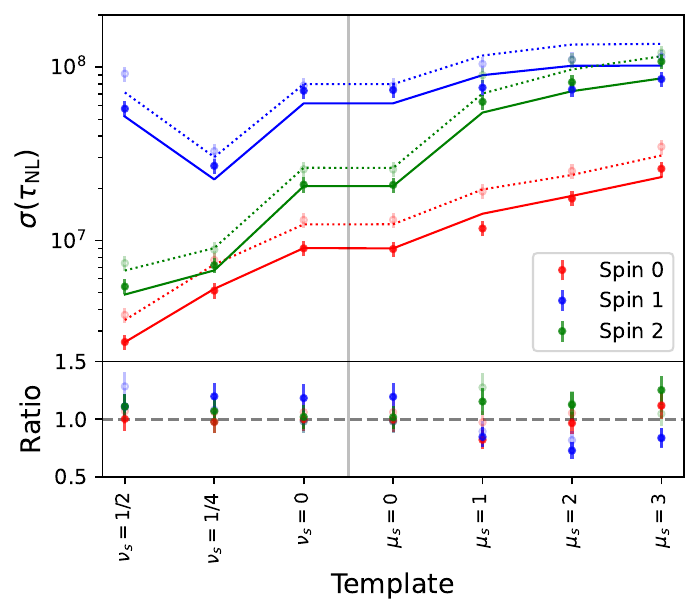}
    \caption{Optimality of the higher-spin cosmological collider estimators, as in Fig.\,\ref{fig: collider-errs}. The solid (dashed) lines indicate analyses including (excluding) $E$-modes and we show results for spin-$0,1,2$, indicated by color. All results include marginalization over the three EFTI templates. We find good agreement of empirical and predicted variances in all cases, and note that $E$-modes tighten the bounds by $\approx 30-40\%$. Though the constraints on spin-two and spin-zero models are similar, light spin-one particles are more difficult to detect due to cancellations in the collapsed limit. At large masses, we find comparable constraints on all spins, differing by at most a factor of five.% high masses ($\nu_0\to 0$ and $\mu_0$), we find only weak dependence of $\sigma(\tau_{\rm NL})$ on the mass parameter, with higher spins being suppressed by a factor $\approx 5\times$ relative to spin-$0$.
    }
    \label{fig: collider-spin-err}
    \end{minipage}
    \hfill
    \begin{minipage}{0.49\linewidth}
    \centering
    \includegraphics[width=\linewidth]{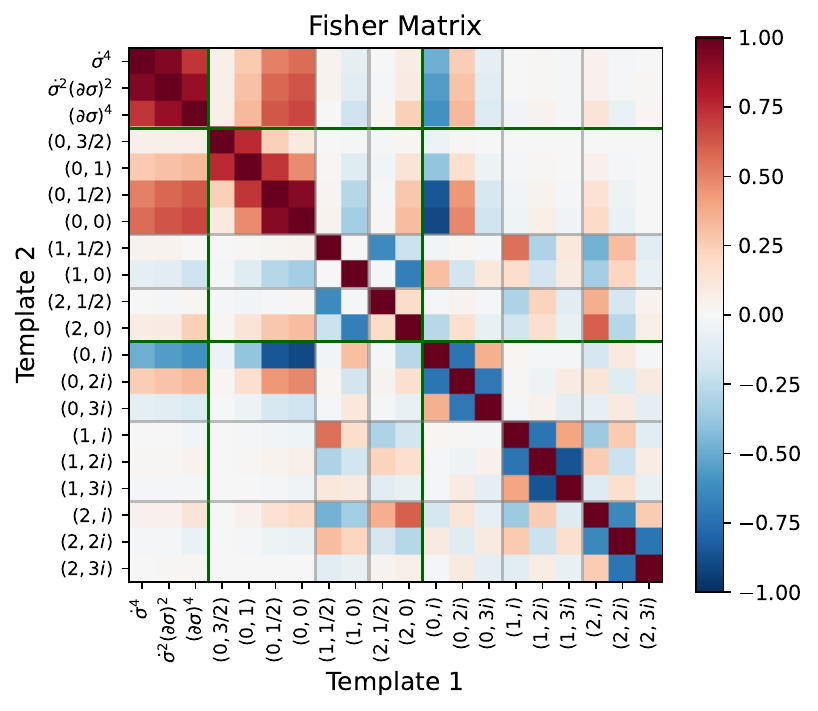}
    \caption{Fisher matrix for the EFTI templates (top left), light collider templates (middle) and heavy collider templates (bottom right). This matches Fig.\,\ref{fig: collider-corr-plot} but includes higher spins, with the captions indicating $(s,\nu_s)$. All results are obtained at $\ell_{\rm max}=512$, $L_{\rm max}=256$ using the numerical Fisher matrix. Whilst we find strong correlations between templates of the same spin (along the diagonal), correlations between different spins are weak, thus that they can be separately constrained from data. The higher-spin templates are not appreciably correlated with the EFTI shapes, thus previous trispectrum constraints cannot rule out primordial collider physics.}
    \label{fig: collider-spin-correlation}
\end{minipage}
\end{figure}

In Fig.\,\ref{fig: collider-spin-err}, we compare the empirical and predicted $\tau_{\rm NL}$ variances for spins $s\in\{0,1,2\}$, finding excellent agreement in all cases. 
%As expected, the two $\sigma(\tau_{\rm NL})$ estimates are in good agreement, implying that our estimator is correctly implemented (noting also that it exactly reproduces the direction-dependent estimator in certain, albeit non-physical, parameter regimes).
%More interesting is the dependence of the constraints on the spin and mass parameters. 
For the even-spin templates, we find tightest constraints at large $\nu_s$ (which feature the strongest collapsed scaling), finding a similar behavior for $s = 0$ and $s = 2$. The light spin-$1$ constraints are much weaker, however, and roughly mass independent; this occurs due to the lack of a collapsed-limit divergence. % and indicates that the information content is sourced by not-quite-collapsed' modes (noting that we remove any equilateral modes via our $k$-space filtering). 
For heavy templates, the constraints vary only weakly with $\mu_s$ (though somewhat more strongly for $s=2$), with comparable constraining power for all spins (within $\lesssim 5\times$, \citep[cf.,][]{Bordin:2019tyb}).
%with higher spin-templates being somewhat harder to detect than their spin-$0$ brethren, but only by a factor of $\lesssim 5$. 
This indicates that such models can be meaningfully constrained with observational data. Moreover, we find that adding $E$-modes significantly tightens the constraints, with $\sigma(\tau_{\rm NL})$ reduced by $\approx(30-40)\%$.

To assess the distinguishability of the various templates, we perform a joint analysis of $17$ collider templates and the $3$ EFTI shapes, obtaining the Fisher matrix shown in Fig.\,\ref{fig: collider-spin-correlation}. As for spin-zero, we find anti-correlations between the heavy-spin templates with similar $\mu_s$; however, the light $\nu_s = 1/2$ and $\nu_s = 0$ shapes are almost uncorrelated for $s=1,2$. Importantly, we find only weak correlations between different spins and with the EFTI templates. This indicates that (a) CMB data can meaningfully constrain higher-spin particles, and (b) the cosmological collider templates are not well described by the local shape (\textit{i.e.\ }$s=0,\nu_0 = 3/2$) or the EFTI templates. This provides strong motivation for \paperthree; previous \textit{Planck} trispectrum constraints cannot rule out the signals induced by inflationary particle exchange.

Finally, we assess the convergence of our results. Since the higher-spin collider templates are expensive to analyze (and our estimators have been heavily validated by this point), we focus on the most pertinent test: convergence of the $k$-integrals (recalling the non-convergence found for some direction-dependent templates). 
%For the light templates with $s\in\{0,1,2\}$, doubling $N_k$ shifts the mean of 50 simulations by $\{0.003\sigma, 0.01\sigma, 0.02\sigma\}$ and the Fisher matrix by $\{0.6\%, 0.6\%, 0.7\%\}$. For the heavy templates, we find a similarly small shift: $\{0.01\sigma, 0.007\sigma, 0.02\sigma\}$ in the mean and $\{1.2\%, 0.5\%, 1.7\%\}$ in the Fisher matrix. As desired, this variation is small, indicating that the results are converged and our analysis is accurate.
For the light (heavy) templates with $s\in\{0,1,2\}$, doubling $N_k$ shifts the mean of 50 simulations by at most $0.02\sigma$ ($0.02\sigma$) and the Fisher matrix by $0.7\%$ ($1.7\%$). As desired, this is a small variation, indicating that our results are converged and accurate.\footnote{This is in contrast to the results found for $\tau_{\rm NL}^{n_1n_3n}$ for odd $n_1,n_3$. Whilst the angular dependence of the collider shapes is similar to the direction-dependent templates, we note that the collider shapes have (a) different dependence on the tetrahedron momentum ratios, (b) a much larger $L_{\rm max}$, (c) a physical cut in $k,K$-space, and (d) a differnet combination of spherical harmonics. As such, the aforementioned difficulties for odd-$n_1,n_3$ templates do not preclude convergence for the spin-$1$ collider shapes.}

\subsection{CMB Lensing}
\noindent Finally, \polyspec includes estimators for two late-time effects: point-sources and gravitational lensing.\footnote{\resub{\polyspec can additionally estimate trispectra arising from cross-correlations of CMB lensing and the integrated Sachs-Wolfe effect; this is discussed in detail in \citep{Philcox:2025lxt}.}} Although they do not encode inflationary physics, these effects are important contaminants to primordial non-Gaussianity searches and can be accounted for using joint analyses \citep[e.g.,][]{Planck:2019kim,Philcox4pt3}. The point-source effect is sourced by an uncorrelated Poisson distribution of sources, which produces an angle-independent trispectrum with amplitude $t_{\rm ps}$. As shown in \paperone, the corresponding estimator is trivial (simply a quartic product of the inverse-variance-filtered data) and \resub{will be discussed further in this work}.  
% CMB lensing  
% Although this trispectrum is not sourced by primordial physics, it is an important contaminant for  and must be carefully modeled to avoid false detections
% of new physics. In \paperthree, this is achieved by estimating the four-point lensing amplitude, here denoted $A_{\rm lens}$ ($\propto C_L^{\phi\phi}$) jointly with the trispectrum shapes of interest, optionally with a tight prior around the expected value of unity. 
The \polyspec lensing estimator takes a similar form to the $\taunl$ estimator (involving the `exchange' of a lensing field), and optimally estimates the lensing amplitude $A_{\rm lens}$ ($\propto C_L^{\phi\phi}$), using $T$-, $E$-, and $B$-modes. As described in \paperone, this is analogous to the conventional quadratic lensing estimators \citep[e.g.,][]{Okamoto:2003zw,Planck:2018lbu,Lewis:2006fu,Carron:2022edh}, but naturally incorporates the $\mathrm{N}^{(1)}$ bias, first-principles `realization-dependent-noise' \citep{Namikawa:2012pe}, optimal polarization correlations \citep[cf.][]{Maniyar:2021msb}, and mask- and weighting-dependent normalization. %, through our Fisher matrix $\F$. 
%Whilst a similar approach could also be used to estimate the lensing amplitude in $L$-bins, though \polyspec does not yet include such an implementation.

\begin{figure}[!t]
    \centering
    \includegraphics[width=0.55\linewidth]{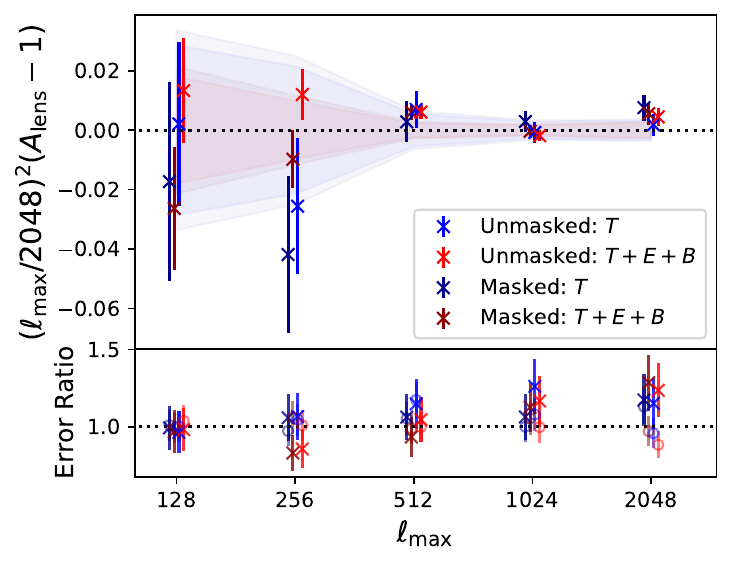}
    \caption{Validation of the \polyspec lensing estimator. The top panel shows the constraints on the lensing amplitude $A_{\rm lens}$ obtained from $50$ lensed simulations (crosses) using the \polyspec temperature (blue) or temperature-plus-polarization (red) estimators. The solid bands give theoretical predictions obtained from the numerical Fisher matrix, and we normalize all results by $\ell_{\rm max}^2$. We find excellent agreement: the mean of the unmasked (masked) $T+E+B$ results yield $\widehat{A}_{\rm lens} = 1.0044 \pm  0.0030$ ($\widehat{A}_{\rm lens} = 1.0055\pm0.0036$), which are consistent with the fiducial value within $1.5\sigma_{\rm mean}$. The bottom panel shows the ratio of the empirical and theoretical errors, which are broadly consistent with unity. The unfilled circles show the variances obtained from unlensed simulations (where our estimator should be optimal); we find some slight evidence for suboptimality at high $\ell_{\rm max}$ due to the additional lensing covariance.}
    \label{fig: lensing}
\end{figure}

To validate the lensing estimator, we employ $150$ Gaussian realizations of temperature, polarization and lensing potential.\footnote{Specifically, we use the FFP10 realizations \citep[e.g.,][]{Planck:2018lkk}, which are publicly available on \textsc{nersc}.} Using the \textsc{lenspyx} code \citep{Reinecke:2023gtp}, we generate non-Gaussian simulations with a known lensing amplitude $A_{\rm lens}$, which are supplemented with Gaussian noise and a \textit{Planck}-mask, as before. 
%are used both for in the estimator numerator and as mock data and numerator simulations). We then add noise and optionally a \textit{Planck}-like mask, as above, and apply the \polyspec estimator. 
Here, we use the hyperparameter set $\{N_{\rm disc}=100, N_{\rm fish}=50, \ell_{\rm min}=2, L_{\rm min}=2, N_{\rm side}=512\}$, fixing $L_{\rm max}=\ell_{\rm max}$ and using lensed $TT, TE, EE$ and $BB$ spectra where necessary \citep{Hanson:2010rp}, noting that the lensing estimators do not require optimization (since they do not contain radial integrals). 
%and the (non-linear) $\phi\phi$ lensing spectra, which are required to form the close-to-optimal lensing estimators 
%When analyzing only exchange-templates, one can fix $L^{(\rm lens)}_{\rm max} = L_{\rm max}^{(\tau_{\rm NL})}$ (since we only need to account for the lensing signatures from correlated modes); the contact-templates include information from all $L$ however, so we should set $L_{\rm max}^{(\rm lens)} = 2\ell_{\rm max}$. 
Given that the fiducial lensing amplitude is non-zero, it is important to validate \polyspec at high resolution, thus we perform mock analyses up to $\ell_{\rm max}=2048$ (increasing $N_{\rm side}$ to $1024$ where necessary).\footnote{As discussed above, it is safe to validate the other estimators at a reduced $\ell_{\rm max}$ since the errors scale with $\sigma(A)$ in the absence of a signal (assuming that the disconnected contributions are correctly subtracted).}

In Fig.\,\ref{fig: lensing}, we show constraints on the lensing amplitude as a function of $\ell_{\rm max}$ for both unmasked and masked datasets. We find sharp dependence on scale-cuts with $\sigma(A_{\rm lens})\sim \ell_{\rm max}^{-2}$ at high $\ell_{\rm max}$, and even stronger dependencies on large scales. At $\ell_{\rm max}=2048$ (the fiducial value used in \paperthree), we find $2.1\%$ ($2.6\%$) constraints on $A_{\rm lens}$ from the temperature-plus-polarization (temperature-only) dataset, which reduce to $2.6\%$ ($3.1\%$) when a mask is included. Furthermore, we recover the input value of $A_{\rm lens}$ within $0.7\%$ or $1.5\sigma_{\rm mean}$. These results are consistent with the $\approx 40\sigma$ ($30\sigma$) detection of lensing in \textit{Planck} PR3 with (without) polarization \citep{Planck:2018lbu}. Furthermore, we find good agreement between the empirical and theoretical errors at low $\ell_{\rm max}$, but weak evidence for an excess at $\ell_{\rm max}=2048$. Rerunning the pipeline using $100$ unlensed simulations yields variances consistent with theory (see Fig.\,\ref{fig: lensing}), suggesting that this excess is caused by non-Gaussian contributions to the estimator variance. This is not unexpected: the \polyspec estimators only satisfy the Cram\'er-Rao bound if the data is Gaussian. 

Finally, we can assess the dependence on hyperparameters (fixing $\ell_{\rm max}=2048$). %This is simpler than before since the lensing templates do not require any optimization routines or numerical integrals. 
We find excellent convergence in the Fisher matrix, with a shift in the mean (Fisher matrix) of at most $0.04\sigma$ ($0.05\%$) induced by reducing $N_{\rm fish}$ to $25$; in fact, the Fisher matrix is converged to $1.5\%$ accuracy with just one iteration. In contrast, reducing $N_{\rm disc}$ from $100$ to $50$ leads to a $0.18\sigma$ variation in $A_{\rm lens}$, representing the dominant systematic uncertainty in the estimator. Whilst this is still comparatively small, it can appear significant when one averages across multiple simulations, using the same disconnected realizations for each, and informs the fiducial choice of $N_{\rm disc}$ used in \paperthree.%. In \paperthree, we will use at least $N_{\rm disc}=50$ to avoid any potential biases (though these are of less relevance to our study, since we estimate the lensing amplitude only to debias the primordial templates).

\section{Timing \& Scalings}\label{sec: timings}
\noindent We now discuss the computational scalings of \polyspec and demonstrate its performance at scale. Since the properties are fairly generic,  we will principally restrict to the local templates ($\gnl$ and $\taunl$), though we compare the runtime of each estimator in \S\ref{subsec: timings-templates}. In all cases, we compute a single Fisher matrix realization (\textit{i.e.}\ set $N_{\rm fish}=1$) and analyze a single dataset, ignoring the disconnected contributions; iteration over both simulations and Fisher matrix realizations can be embarrasingly parallelized, thus this does not affect the scalings. Unless otherwise specified, we set $N_{\rm CPU}=64$, $N_{\rm side}=512$, $\ell_{\rm max}=512$, $L_{\rm max}=10$ and $f_{\rm thresh}=10^{-4}$.

%To build intuition for the results below, it is useful to give a rough overview of the computations involved. To build the $\gnl$ numerator we (a) filter the data by $\Si$ and transform to harmonic-space, (b) apply $\ell$-space filters for each radial point and return to map-space, (c) multiply the four copies of the maps in pixel-space and sum over $r$. The first two steps of the $\taunl$ numerator are the same, following which we (c) multiply a pair of maps in pixel-space for each $r$ and transform to harmonic-space, (d) sum over two copies of the vector of harmonic-space, taking the inner product with respect to $r,r'$. To form the $\gnl$ Fisher matrix, we proceed similarly, but finish with (c) multiplying three copies of the maps in pixel-space, returning to harmonic-space and applying an $\ell$-space filter. Finally the $\taunl$ Fisher matrix 

\begin{figure}
    \centering
    \includegraphics[width=0.9\linewidth]{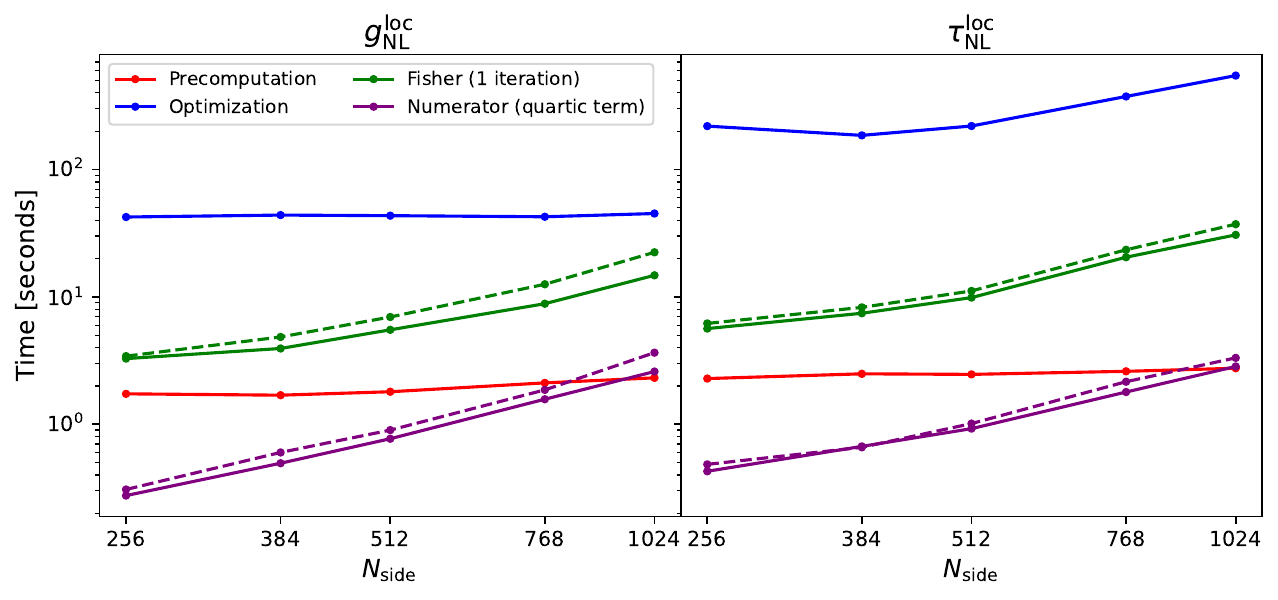}
     \includegraphics[width=0.9\linewidth]{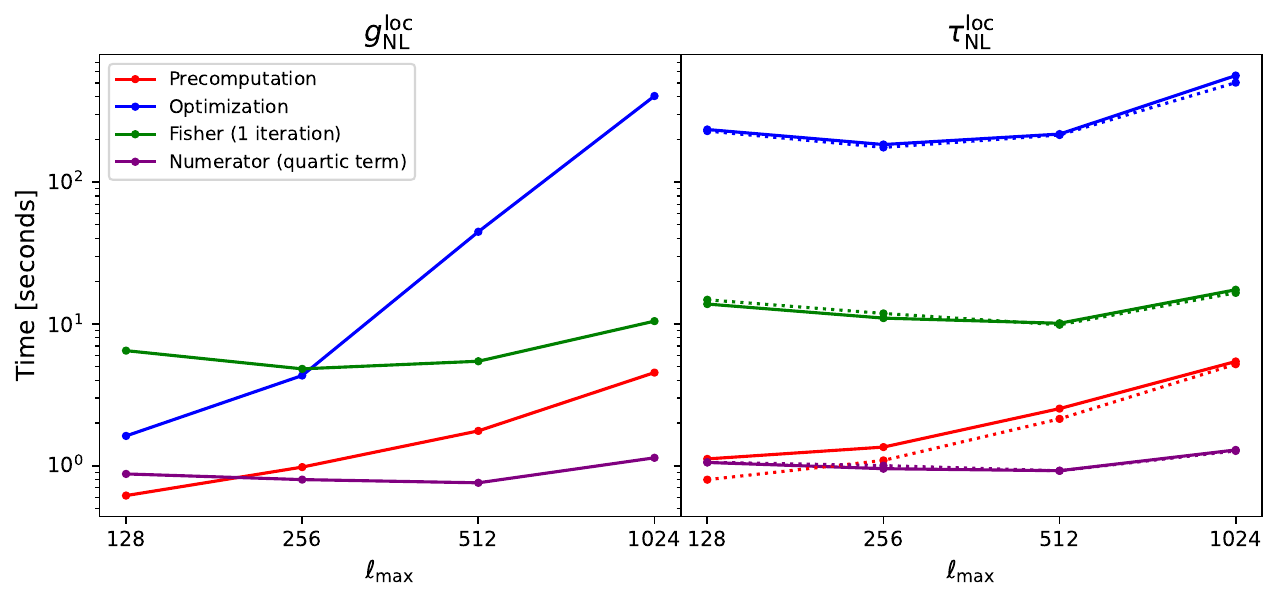}
    \caption{Runtime of the \polyspec estimator applied to the $\gnl$ (left) and $\taunl$ (right) templates. We give the total runtime of each section of the code (see \S\ref{sec: code}) a function of $N_{\rm side}$ (top) and $\ell_{\rm max}$ (bottom), restricting to a single dataset and Fisher matrix realization (dropping the disconnected terms). By default, we assume $\{\ell_{\rm max},L_{\rm max},N_{\rm side},f_{\rm thresh}\} = \{512,10,512,10^{-4}\}$, and analyze unmasked temperature and polarization data. In the top panels, the dashed lines indicate timings using masked data, whilst the dotted lines in the bottom panel indicate results with $L_{\rm max}=5$. At low $\ell_{\rm max}$, the optimization algorithm is less efficient leading to larger $N_{\rm opt}$ (since the integrand is less sharply peaked) and the code is limited by memory allocation.% thus the numerator and Fisher matrix timings plateau.
    }
    \label{fig: timings}
\end{figure}

\subsection{\texorpdfstring{Dependence on $N_{\rm side}$}{Dependence on HEALPix Resolution}}
\noindent A key hyperparameter in \polyspec is the \textsc{healpix} resolution, $N_{\rm side}$. This sets three quantities: (1) the number of pixels (via $N_{\rm pix}=12N_{\rm side}^2$); (2) the maximum multipole (requiring $\ell_{\rm max}\leq \ell_{\rm max}^{\rm healpix}\equiv 3N_{\rm side}-1$); (3) the speed of spherical harmonic transforms (which scale asymptotically as $\mathcal{O}(N_{\rm side}\ell_{\rm max}^2)$ \citep{Reinecke_2013}). At fixed $\ell_{\rm max}$, we expect both pixel-space summations and harmonic transforms to scale linearly with $N_{\rm side}$. 

The top panel of Fig.\,\ref{fig: timings} shows the runtime of each segment of the $\gnl$ and $\taunl$ estimators as a function of $N_{\rm side}$ (as defined in \S\ref{sec: code}). For $\gnl$, both the precomputation and optimization steps are independent of $N_{\rm side}$; these involve only harmonic-space operations (including computation of the Bessel functions, the transfer function integrals, and the analytic Fisher matrix). As outlined in \S\ref{sec: code}, computing the numerator and Fisher matrix of both templates requires SHTs and pixel-space summation: this explains the dependencies seen in Fig.\,\ref{fig: timings}.\footnote{In practice, we find a slightly stronger scaling, since the SHT precomputation steps dominate if $\ell_{\rm max}\ll \ell_{\rm max}^{\rm healpix}$} When analyzing masked data, we find reduced scalings due to the inpainting procedure contained within the $\Si$ filtering. For $\taunl$, the optimization algorithm exhibits a weak dependence on $N_{\rm side}$, due to the grid-based computation of the Fisher matrix, though the runtime is dominated by other contributions at low $N_{\rm side}$. %, such as memory allocation and matrix inversions. 
In general, the runtime is dominated by optimization and computation of the Fisher matrix -- this is not a significant limitation given that these steps are data independent and can be precomputed.
% these do not depend on the observed data (and the former does not depend on the mask or $\Si$ weighting), thus they can be computed once and stored. 
Furthermore, since the optimization is performed under idealized conditions, we can use a comparatively low $N_{\rm side}\lesssim \ell_{\rm max}/3$. Computing the $g_{\rm NL}$ and $\tau_{\rm NL}$ numerators is very fast: these take only $\sim 2$s at $N_{\rm side}=1024$ (though the runtime increases when including disconnected contributions).

\subsection{\texorpdfstring{Dependence on $\ell_{\rm max}$}{Dependence on Scale Cuts}}
\noindent The bottom panel of Fig.\,\ref{fig: timings} gives the scalings of \polyspec with $\ell_{\rm max}$ (assuming fixed $N_{\rm side}=512$). The precomputation step depends strongly on $\ell_{\rm max}$: this occurs since we must compute the Bessel functions and $\ell$-space filters for all $\ell\in[\ell_{\rm min}, \ell_{\rm max}]$. For $\taunl$, there is a slight dependence on $L_{\rm max}$, since \polyspec also computes the quadratic coupling $F_L(r,r')$ for all $L\in[L_{\rm min},L_{\rm max}]$, though this is comparatively cheap. The scaling of the numerator and Fisher matrix is more complex, and, at first glance, counterintuitive. This is driven by an interplay of three factors: (1) the number of points in the optimized integral representation, $N_{\rm opt}$, which is largest at small $\ell_{\rm max}$ (since the Fisher matrix integrand is less sharply peaked); (2) memory allocation, which scales as $N_{\rm pix}N_{\rm opt}$; (3) SHTs, which scale as $\ell_{\rm max}^2$ for fixed $N_{\rm side}$. At small $\ell_{\rm max}$, the rate-limiting step is allocating memory to the vector of $N_{\rm opt}$ real-space maps,\footnote{For contact trispectra, we could reduce the algorithm's memory footprint by analyzing each radial point in turn rather than storing all $N_{\rm opt}$ pixel-space maps; for the exchange diagrams, this is not possible since the estimators correlate pairs of maps at different radii.} and the timings are $\ell_{\rm max}$-independent; as $\ell_{\rm max}$ increases, the other processes become increasingly important, but are mollified by the reduction in $N_{\rm opt}$. For $\tau_{\rm NL}$, we find similar scalings for the optimization and Fisher matrix computation, which is as expected, since the former involves the latter.
%optimization we find similar scalings as for the Fisher matrix: this is as expected, since it involves the latter process (and the requisite memory allocations scale as $N_{\rm pix}N_{\rm opt}$, which is less sensitive to $\ell_{\rm max}$). 
In contrast, optimization of the $\gnl$ Fisher matrix is a strong function of $\ell_{\rm max}$, since it involves $\mathcal{O}(\ell_{\rm max}^2)$ analytic computation of the idealized Fisher matrix \citep[cf.][]{2015arXiv150200635S}.

\subsection{\texorpdfstring{Dependence on $N_{\rm CPU}$}{Dependence on Number of CPUs}}

\begin{figure}[!t]
    \centering
    \includegraphics[width=0.9\linewidth]{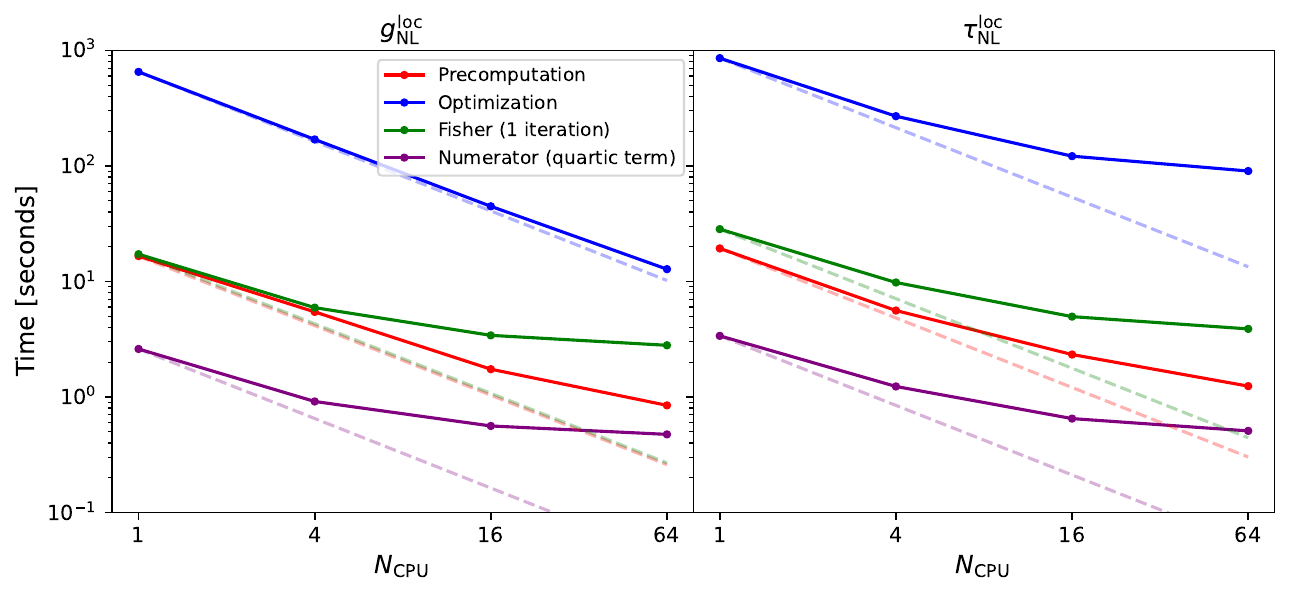}
    \caption{Scaling of the \polyspec estimator with the total number of CPUs, $N_{\rm CPU}$. The dashed lines show the idealized scalings $\propto N_{\rm CPU}^{-1}$. %The rate-limiting steps of the algorithm are parallelized using \textsc{OpenMP}, though many steps become memory limited at large $N_{\rm CPU}$.
    We find fairly good scalings for $N_{\rm CPU}\lesssim 16$, but a rough plateau thereafter; this indicates that non-parallelized processes such as memory allocation eventually become limiting. Whilst improved scalings could potentially obtained by rewriting the algorithm in \textsc{c} (instead of \textsc{cython}/\textsc{python}) this would lead to a loss of flexibilty.
    %  the intrinsic flexibility of \textsc{python}, which allows for a flexible and unified analysis of many types of templates (each of which requires slightly different algorithms).
    %This is an intrinsic limitation of writing \polyspec in \textsc{python}, although it does allow for much greater flexibility. 
    In $\gnl$, analytic calculation of the idealized Fisher matrix is rate-limiting, but exhibits almost perfect scaling with $N_{\rm CPU}$.}
    \label{fig: cpus}
\end{figure}

\noindent To perform fast analyses of high-resolution datasets it is vital to take full advantage of the available computational resources, including through parallelization.\footnote{We do not attempt to include MPI parallelism in \polyspec. Whilst this would allow distributed computing across nodes, it is more efficient to use such nodes to run the estimator on different templates, different Fisher realizations, or different simulations.} In \polyspec, we compute SHTs using the \textsc{ducc} code, which is heavily parallelized \citep[cf.][]{Reinecke_2013}; however, the full computation involves much more than SHTs. For efficiency, we have written the rate-limiting steps of \polyspec in \textsc{cython},\footnote{These include: multiplication and summation of pixel-space maps, assembly of the Fisher matrices via outer products, (pre-)computation of the spherical Bessel functions, computation of all $\ell$- and $L$-space filters via $k$- and $K$-space integrals, shifting harmonic space maps from $a_{\ell m}\to a_{(\ell+\delta\ell)(m+\delta m)}$ (required for direction-dependent and spin-$s>0$ estimators), and computation of the ideal $g_{\rm NL}$ Fisher matrices.} parallelized with \textsc{OpenMP} (typically distributing across the $N_{\rm opt}$ radial integration points), using \textsc{python} for interfacing, memory access, and overall structure. Whilst a full \textsc{c}-based code would be more efficient for simple template analyses, this is less flexible, particularly given the large number of templates that are included in our analysis (and the even larger number that could be included in the future). In typical runs, we find that the code exhibits $70-80\%$ CPU efficiency, implying that most of the CPUs are being utilized most of the time.

In Fig.\,\ref{fig: cpus}, we show the explicit scalings of \polyspec with the CPU count, $N_{\rm CPU}$. For most sections of the code, we find a similar behavior: there is a close to optimal scaling at low $N_{\rm CPU}$ but a gradual plateau by $N_{\rm CPU}\approx 16$. In this limit, we are hampered by (a) having more CPUs than independent processes (\textit{i.e.}\ $N_{\rm CPU}>N_{\rm opt}$) and (b) memory management. We find somewhat improved scalings for the precomputation step, where computation of the $\ell$-space filters can be parallelized over all threads with limited overheads, and near-perfect parallelization for the $\gnl$ optimization step, since the entire operation can be performed in \textsc{c}, without intermediate memory management or SHTs. Overall, our parallelism induces a speed-up of $\approx 10\times$ relative to the na\"ive single-thread implementation; whilst a sufficiently adept coder could likely optimize \polyspec further (and is welcome to do so!), we find this sufficient for the \textit{Planck} analysis presented in \paperthree.

\subsection{Dependence on Templates}\label{subsec: timings-templates}

\begin{figure}
    \centering
    \includegraphics[width=0.65\linewidth]{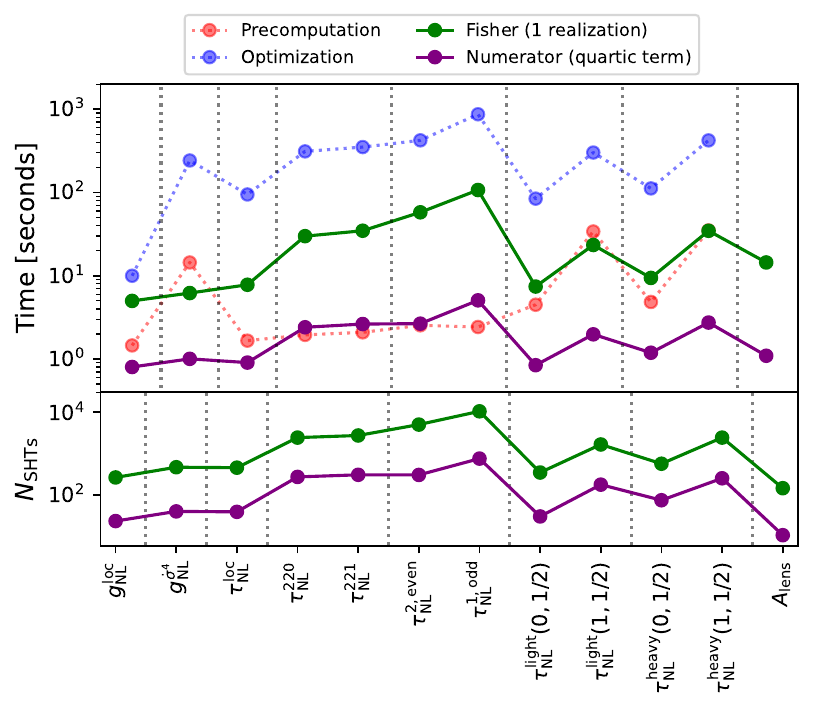}
    \caption{Dependence of \polyspec on the template shape. We show the time required to analyze each template independently (top panel), as well as the total number of SHTs required (bottom panel), working with unmasked data at $\ell_{\rm max}=512$, $N_{\rm side}=512$, and $f_{\rm thresh}=10^{-3}$ in all cases (which leads to a different $N_{\rm opt}$ for each analysis). We chose a value of $L_{\rm max}$ appropriate for each type of analysis: $L_{\rm max}=10$ for direction-dependent and local templates, $L_{\rm max}=256$ for the collider shape and $L_{\rm max}=512$ for lensing. Whilst the computation time depends on a number of effects (described in the text), we find that the overall variation in the numerator and Fisher matrix steps are within an order of magnitude or so, implying that \polyspec can compute both simple and complex templates in reasonable computation times.}
    \label{fig: template-timings}
\end{figure}

\noindent The above sections have focused only on the simplest trispectrum templates: $\gnl$ and $\taunl$. Here, we consider the dependence of the runtime on the type of template, considering representative examples across all the trispectrum types validated in \S\ref{sec: validation}. In general, the runtimes can vary due to a number of factors, including (a) the extra SHTs and summations required to implement exchange templates, (b) the two-dimensional radial integration required for the EFTI shapes, (c) the additional operations required to add angular dependence (d) the use of spin-weighted harmonic transforms in the lensing and EFTI estimators, (e) the need for complex fields in the heavy templates, and (f) variations in $N_{\rm opt}$.% (particularly for the EFTI shapes), which perform a joint optimization in the $(r,\tau)$ plane). 

Fig.\,\ref{fig: template-timings} shows the result of this complex interplay of factors. Regardless of the type of template considered, the runtimes of the key algorithm steps (the numerator and Fisher matrix) are consistent to within a factor of around ten. This is an excellent outcome, since it implies that the more complex estimators can be applied to high-resolution \textit{Planck} data prohibitive computational costs (though the analyses of \paperthree still required $\mathcal{O}(10^5)$ CPU-hours). For $\gnldotdot$, we find a similar runtime to $\gnl$, although the precomputation and optimization steps are more expensive (due to the additional $\tau$ integral and the more complex analytic Fisher matrix). The runtime of the direction-dependent $\tau_{\rm NL}^{n_1n_3n}$ templates is $\approx 3\times$ larger than for $\taunl$: this is due to the additional harmonic transforms and Gaunt factor summations, which affects all components except for precomputation. The even and odd estimators are a sum over several $\tau_{\rm NL}^{n_1n_3n}$ forms; as such, their runtime is similar (though $\tau_{\rm NL}^{1,\rm odd}$ requires $50\%$ larger $N_{\rm opt}$). The light spin-zero and spin-one collider shapes are analogous to $\taunl$ and $\tau_{\rm NL}^{n_1n_3n}$, with the $k,K$-truncation causing only a slight increase in the precomputation step (since the $K$-space integrals are computed numerically instead of analytically). The heavy collider is somewhat more expensive: this is due to the addition of complex maps, which require twice the number of SHTs. Finally, the lensing estimator is cheap since it does not require optimization or $r$-space summation, though we do require spin-three SHTs when polarization is included. %In brief, the conclusion of this exercise is that the runtime of the estimators scales as expected.

Fig.\,\ref{fig: template-timings} also shows the number of SHTs required to analyze each template. For the numerators, this varies between $N_{\rm SHTs}=11$ for lensing to $N_{\rm SHTs}=760$ for $\tau_{\rm NL}^{\rm odd,1}$; in contrast, computation of a single Fisher matrix realization can require $N_{\rm SHTs}=10^4$. %Whilst these numbers are large, they are not unexpected. 
These match expectations, with simple estimators requiring $\mathcal{O}(N_{\rm opt})$ SHTs (in addition to those required to implement the $\Si$ filtering), which increases by a factor of $(2n_{\rm max}-1)$ for the direction-dependent and spin-exchange forms. Of course, these values scale with $N_{\rm fish}$ and $N_{\rm disc}$; a typical analysis could thus require $10^5-10^6$ SHTs,\footnote{This number does not scale with the number of simulations analyzed: the Fisher matrix is data-independent, and many of the disconnected terms can be stored if multiple datasets are analyzed in succession.} clearly demonstrating the utility of fast SHT algorithms such as those provided by \textsc{ducc}.

\subsection{High-Resolution Testing \& Breakdown}

\begin{figure}
    \centering
    \includegraphics[width=0.8\linewidth]{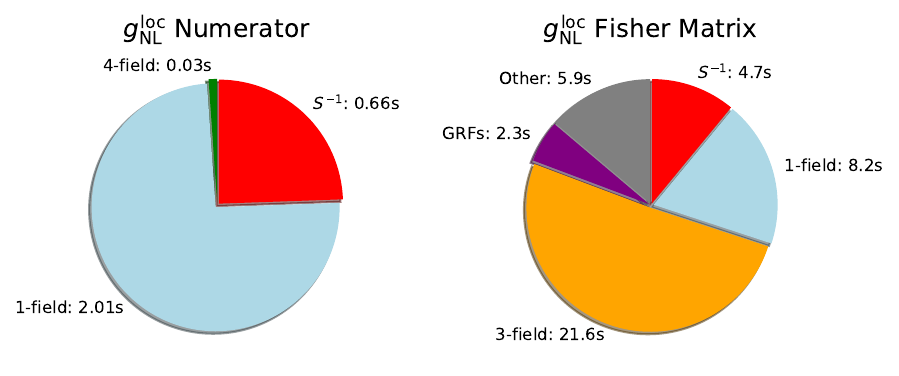}
    \includegraphics[width=0.8\linewidth]{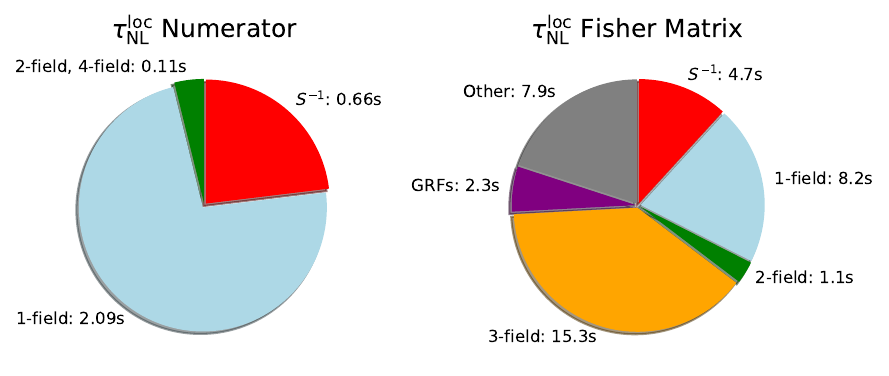}
    \vskip2em
    \includegraphics[width=0.8\linewidth]{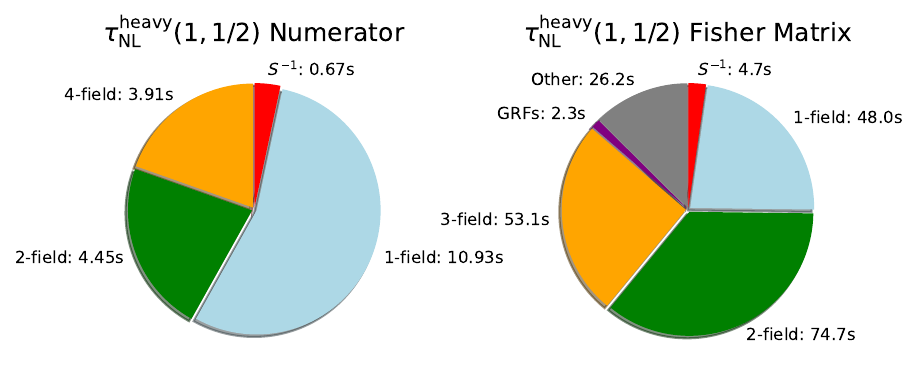}
    \caption{Breakdown of computation time for analyzing the numerator (left) and Fisher matrix (right) of three trispectrum templates at \textit{Planck}-like precision settings $\ell_{\rm max}=2048$, $L_{\rm max}=1024$ and $N_{\rm side}=1024$, using masked temperature and polarization data. We split the numerator computation into various parts: applying the $\Si$ filters ($S^{-1}$), constructing transfer-function-weighted maps using the data (1-field), computing pairwise products of these maps (2-field), and summing over four copies of the maps to form the output (4-field). The Fisher matrix involves additional steps: computing cubic products of the maps (3-field), creating Gaussian random field realizations (GRFs), and various additional processes including memory management (other).}
    \label{fig: pie-charts}
\end{figure}

\noindent For our final test, we apply \polyspec at scale, computing the numerator and Fisher matrices of three representative templates using similar precision settings to \paperthree. In particular, we analyze $\{\gnl, \taunl, \tau_{\rm NL}^{\rm heavy}(1,1/2)\}$ (which span the full gamut of contact shapes, exchange shapes, angular-dependence and oscillatory shapes), using masked temperature and polarization data with $\ell\in[2,2048]$, $L\in[2,1024]$, fixing $N_{\rm side}=1024$ and using $N_{\rm opt}=30$ integration points for each template (to allow fair comparison). %This emulates the analyses performed in \paperthree.

Fig.\,\ref{fig: pie-charts} shows the time required to analyze a single numerator and Fisher realization for each template, breaking down each segment into various aspects. Even at high-precision, \polyspec is relatively fast: computation of the three numerators requires 2.7, 2.9, and 20 node-seconds, whilst the Fisher matrix require 43, 39, and 210 node-seconds. Approximately half of this time is spent performing SHTs; the rest involves pixel- and harmonic-space weightings and summation. 
%This makes it feasible to run \textit{Planck}-scale analyses of high-order templates. 
Notably, a significant chunk of the computation time is spent applying the quasi-optimal weighting scheme $\Si$ to the data (or, for the Fisher matrices, random fields); this is slow since it includes a non-parallelized inpainting scheme (which could be optimized). In practice, one may wish to use a more complex $\Si$ filtering, such as a conjugate gradient scheme (see \paperthree); this will inevitably lead to slower computation, though we note that the filtering only has to be applied only once (in the numerator) or three times (for the Fisher matrix).\footnote{The time required to evaluate $\Si$ depends on whether we require the inputs and outputs in pixel- or harmonic-space, since \polyspec is configured to use as few SHTs as possible.} 

For the local templates, numerator computation is dominated by the `1-field' operations, whence the dataset is filtered by $\ell$- and $r$-dependent weights and transformed to pixel-space (as in \eqref{eq: g-r-def}), creating a vector of size $N_{\rm opt}\times N_{\rm pix}$. Since this involves SHTs, it is far more expensive than the eventual summation in pixel- or harmonic-space (the `4-field' contribution). For the spin-1 collider template, the `1-field' operations are around five times slower due to the requisite angular couplings and Gaunt functions, which require creation of $(2s+1)N_{\rm opt}$ filtered maps.
% (encoding the $Y_{s\lambda}$ harmonics with $|\lambda|<s$, noting that $\lambda \to -\lambda$ is not a symmetry, given the complex radial scalings). 
Furthermore, the exchange estimators involve an additional $N_{\rm opt}$ harmonic transforms on products of the filtered maps: this is the `2-field' term, which becomes expensive for the collider shape, again due to the additional spherical harmonic couplings. Finally, we note that `4-field' summation in $\tau_{\rm NL}^{\rm heavy}$ has an $\mathcal{O}(N_{\rm opt}^2)$ scaling, though does not involve SHTs. Although the various components differ between estimators, the overall runtimes are broadly consistent (as in Fig.\,\ref{fig: template-timings}) -- this is a result of careful code optimization and similar asymptotic scalings.

The breakdown of the Fisher matrix is somewhat more complex. As well as construction of the `1-field' filters (which takes $\approx 4\times$ longer, since we require a pair of random fields and two filtering schemes),
%($\Ai$ and $\Si$), 
construction of the $Q$ derivative discussed in \S\ref{subsec: sep-impl} requires cubic products of the maps. For local templates, this is the rate-limiting step, and involves $N_{\rm opt}$ pixel-space multiplications, $N_{\rm opt}$ SHTs, and finally a summation over all harmonic-space maps, weighted by some filter. This is more expensive for the collider templates due to additional permutations, Gaunt functions and the presence of complex maps. We additionally require generation of two sets of Gaussian random fields which must be multiplied by the mask -- this has a non-negligible runtime, scaling as $\left(\ell_{\rm max}^{\rm healpix}\right)^3\propto N_{\rm side}^3$. % (since we cannot filter to $\ell\leq\ell_{\rm max}$ before applying the mask). 
Finally, we find some timing residual (``other''); this represents the time spent allocating arrays and memory in \textsc{python} (and could probably be reduced with a more careful implementation). Whilst computation time is shared between a number of different steps, the slowest steps are precisely those expected from the theoretical scalings -- as such, we conclude that our implementation is efficient and ready to be applied to observational data.

\section{Summary \& Next Steps}\label{sec: summary}
\noindent In this work, we have introduced the \polyspec code -- a fast and flexible \textsc{python} package for estimating a wide variety of primordial trispectrum amplitudes from CMB data. By necessity, this is a complex code: the full trispectrum estimators involve more than $4000$ lines of \textsc{python} and $2000$ lines of \textsc{cython}. In addition to an outline of the code structure and usage, this paper has performed an extensive battery of tests, focusing on the following properties:
\begin{itemize}
    \item \textbf{Bias}: When possible, we have applied \polyspec to non-Gaussian simulations and demonstrated that it recovers the input non-Gaussianity parameter. These tests are performed for the $\gnl, \taunl, A_{\rm lens}$ templates.  We further validate that every estimator returns zero when applied to Gaussian simulations.
    \item \textbf{Variance}: For Gaussian input data, the estimator covariance should be equal to the inverse normalization, $\F^{-1}$, up to sample variance. This has been validated for all templates. Furthermore, we have validated that the stochastically computed normalization agrees with the analytic predictions for contact estimators.
    \item \textbf{Scalings}: The detectability of many trispectrum templates have been considered previously in the literature. Where possible, we have compared the scalings of our errorbars with $\ell_{\rm max}$ to the published forecasts, finding excellent agreement.
    \item \textbf{Stability}: \polyspec depends on a number of hyperparameters, which set the accuracy of the numerical integrations and summation (summarized in Tab.\,\ref{tab: hyperparameters}). We have ensured that the estimators are stable with respect to these choices, \textit{i.e.}\ that the results are converged. All templates have passed these checks except for certain direction-dependent $\tau_{\rm NL}$ shapes, which previous forecasts suggested would be hard to constrain -- these will be excluded from the \textit{Planck} analysis in \paperthree.
    \item \textbf{Efficiency}: We have verified that the runtime of the code scales as expected with various resolution choices. Furthermore, we find that the code is parallelized effectively at low $N_{\rm CPU}$, though becomes limited by memory at high $N_{\rm CPU}$. The performance could be improved with more efficient memory allocation and harmonic transforms, such as through GPU acceleration \citep[e.g.,][]{Belkner:2024vor}.
\end{itemize}
The combination of the above results gives us confidence that the \polyspec estimators are efficient and free from malignant human and numerical errors. 

In addition to validating our implementation, the numerical tests in this paper allow for several interesting conclusions. Firstly, we find that the time required to estimate trispectrum amplitudes is reasonable at $\ell_{\rm max}=2048$, regardless of the type of template. The estimator numerators require around $3-30$s to compute (per simulation, with $\mathcal{O}(100)$ required to remove disconnected contributions), whilst computation of the Fisher matrices requires around $30-300$s for each of $\mathcal{O}(20)$ Monte Carlo realizations. The most expensive templates are those involving direction-dependence or higher-spin, with a spin-$s$ template requiring $\mathcal{O}(2s+1)$ more SHTs than the local estimator. 

Next, we note that constraints on non-Gaussianity amplitudes are strong functions of $\ell_{\rm max}$, and improve by up to $40\%$ when $E$-modes are included in the analysis. This implies that the official \textit{Planck} bounds on local non-Gaussianity can be surpassed, and gives hope for future higher-resolution surveys. % particularly for constraining $\tau_{\rm NL}$-type non-Gaussianity, which has a strong divergence in the collapsed limit. 
Finally, and perhaps most importantly, we find that the correlations between templates are generally weak (with some exceptions, such as the three EFTI shapes, as well as collider signatures with similar mass). This is particularly apparent for collider non-Gaussianity, as shown in Fig.\,\ref{fig: collider-spin-correlation}. This implies that (a) the various can be independently constrained from CMB data and (b) most templates considered in this work have not been indirectly constrained by the previous non-detection of local and EFTI non-Gaussianity, \textit{i.e.}\ searching for their signatures in \textit{Planck} data is a worthwhile exercise. 

We close by elucidating our next steps. Up to this point, we have presented the inflationary templates and their optimal estimators (\paperone) and built an efficient code to search for such signatures (this paper). In \paperthree, we will search for non-Gaussian signatures using \textit{Planck} data. %This will be aided by the validation tests presented in \S\ref{sec: validation}, which specify the precision settings we will require for unbiased analyses.
Motivated by the above results, we will perform both single-template and joint analyses of a range of inflationary templates, including all of the shapes considered in this work except for the poorly constrained direction-dependent forms with odd $n_1,n_3$ (cf.\,\S\ref{subsec: valid-direc}). We will further account for CMB lensing contamination using our $A_{\rm lens}$ template, and perform both temperature-only and joint temperature-and-polarization analyses. The result will be strong constraints on a wide variety of inflationary phenomena; these can be further improved using future CMB or large-scale structure datasets \citep[e.g.,][]{Cabass:2022epm,Sailer:2021yzm,Floss:2022grj}.

\acknowledgments
{\small
\noindent We thank Giovanni Cabass, William Coulton, Adriaan Duivenvoorden, Sam Goldstein, Colin Hill and Maresuke Shiraishi for insightful discussions. \resub{We additionally thank the anonymous referee for their careful reading of the manuscript and useful comments.} \begingroup\hypersetup{hidelinks}OHEP is a Junior Fellow of the Simons Society of Fellows, and thanks
\href{https://www.flickr.com/photos/198816819@N07/54308122315/}{Lludwig van Beethoven} for auditory assistance.\endgroup OHEP would also like to thank the Center for Computational Astrophysics for their hospitality across the multiple years this set of papers took to write. The computations in this work were run at facilities supported by the Scientific Computing Core at the Flatiron Institute, a division of the Simons Foundation.
}

\appendix

\bibliographystyle{apsrev4-1}
\bibliography{refs}% Produces the bibliography via BibTeX.

\end{document}